\documentclass[amsfonts,amsmath,amssymb,prb,twocolumn]{revtex4}
\usepackage{graphicx}
\usepackage{color}
\usepackage{bm}
\usepackage{bbm}
\usepackage{dsfont}
\usepackage{mathbbol}
\usepackage{multirow}

\newcommand{\e}{\begin{equation}}
\newcommand{\be}{\begin{equation}}
\newcommand{\ee}{\end{equation}}
\newcommand{\bea}{\begin{eqnarray}}
\newcommand{\one}{\mathbbm{1}}
\newcommand{\Tr}{\mbox{Tr}}

\newcommand{\ea}{\begin{eqnarray}}
\newcommand{\eea}{\end{eqnarray}}

\newcommand{\vectornorm}[1]{\left|\left|#1\right|\right|}
\newcommand{\ket}[1]{\left. | #1 \right\rangle }

\newcommand{\bra}[1]{\left\langle #1 |\right. }

\newcommand{\einheit}[1]{\hspace{1mm}\mbox{#1}}
\newsavebox{\mysquare}
\savebox{\mysquare}{\textcolor{black}{\rule{2.5mm}{2.5mm}}}
\usepackage{float}

\begin{document}

\title{Optimal Control for Open Quantum Systems: Qubits and Quantum Gates}
\author{R. Roloff}
 \email{robert.roloff@uni-graz.at}
\author{M. Wenin}%
 \email{markus.wenin@uni-graz.at}
\author{W. P\"otz}%
 \email{walter.poetz@uni-graz.at}

\affiliation{Institut f\"ur Physik, Theory Division,\\ Karl Franzens Universit\"at Graz, Universit\"atsplatz 5, 8010 Graz, Austria}

\date{\today}

\begin{abstract}
This article provides a review of recent developments in the formulation and execution of optimal control strategies 
for the dynamics of quantum systems.   A brief introduction to the concept of optimal control, the dynamics of of open quantum systems, and quantum information processing  is followed by a presentation of recent developments regarding the two main tasks in this context: state--specific and state--independent optimal control.    For the former, we present an extension of conventional theory (Pontryagin's principle) to quantum systems which undergo a non--Markovian time--evolution.   Owing to its importance for the realization of quantum information processing, the main body of the review, however, is devoted to state--independent optimal control.  Here, we address three different approaches: an approach which treats dissipative effects from the environment in lowest--order perturbation theory, 
a general method based on the time--evolution superoperator concept, as well as one based on the Kraus representation of  the time--evolution superoperator.  Applications which illustrate these new methods focus on single and double qubits (quantum gates) whereby the environment is modeled either within  the Lindblad equation or a bath of bosons (spin--boson model).  
While these approaches are widely applicable, we shall focus our attention to solid--state based physical realizations, such as semiconductor-- and superconductor--based systems. 
While an attempt is made to reference relevant and representative work throughout the community, the exposition will focus mainly on work which has emerged from our own group.
\end{abstract}
\maketitle

\tableofcontents

\section{Introduction}\label{INT}

\subsection{Preliminaries and overview}\label{PREL}

Recent developments throughout modern  nanophysics,  both regarding physical phenomena and technological  applications, have pushed the development of theoretical approaches for the optimal control of the dynamics of open quantum systems.   While the field of quantum information processing is still open for new ideas, solid--state based realizations, such as semiconductor quantum--dot systems and Josephson--junction based  quantum gates, have  emerged as major candidates. Appealing features are the high levels of technological abilities regarding design and fabrication of these systems, as well as their scalability into large arrays.~\cite{DiVi98,Burk99,Mak01,Nak99,Yam03,Clarke,Nakahara,Mooij,Spoerl07}
Current drawbacks are isolation problems of solid--state quantum systems from their environment on time--scales where external control can be administered,  the complexity of controlled fabrication of these artificially grown structures, as well as the generation and precise application of sufficiently strong control fields.  Since quantum interference, generally,  is a frail effect on a mesoscopic length scale, it stands to reason to apply optimal control theory to maximize ones ability to steer nanostructures in quantum--coherent fashion. In this article we shall review  recent progress in this direction, focusing on recent progress within our group.  

Optimal control theory (OCT) generally concerns itself with algorithms for finding  control fields which minimize or maximize a given performance index, often called \textit{cost functional}. As such, OCT represents an example for an inverse problem. The fundamental prerequisite to formulate and apply OCT is the ability to express the control objective, as a quantitative mathematical criterion, \textit{i.e.}, in form of a cost functional. Typical is also the presence of additional constraints.  
In fact, there is an enormous number of problems for which the latter can be formulated in form of differential equations. Probably the most famous 
example in physics is provided by classical mechanics, however, this type of problem can be found throughout  quantitative science.

The basic mathematical description for an optimal control problem can be given as follows.~\cite{Krotov96} Suppose we have a set $\mathbb{M}$ and a functional $J(v)$, with $v\in \mathbb{M}$. $\mathbb{M}$ corresponds to the space of solutions and $v$ is an arbitrary particular solution to the problem. The functional $J$ maps $\mathbb{M}$ onto the set of Real numbers, \textit{i.e.}, $J: \mathbb{M} \rightarrow \mathbb{R}$, and characterizes the quality of $v$ so that if solution $v_1$ is better than $v_2$, then $J(v_1)<J(v_2)$. The best, and therefore called \textit{optimal} solution is defined as $v^*=\operatorname{arg\,min}_{v\in \mathbb{M}} J(v)$. The pair $(\mathbb{M},J)$ is a mathematical model of our control problem. Much of this work deals with proper modeling of dynamical processes.
Here, the cost functional $J$ usually depends on a set of state variables, whose dynamics is governed by (integro--) differential equations. By proper tuning of the controls it is possible to alter the evolution of these state variables and to find an extremum of the cost functional.\par
OCT has its origins in the calculus of variations, especially in curve minimization problems to which considerable attention has been paid at the end of the 17th century.~\cite{Suss97} Since then, there has been an ongoing advancement in the field of optimization which culminated in the work of Pontryagin and Bellman in the 1950s. Nowadays OCT is used in many different areas including engineering, finance, economics and physics. Below, we shall take a closer look on the latter, namely optimal control of dissipative or \textit{open} quantum systems with a particular focus on quantum information processing (QIP).   We start out in Sec. \ref{OPT} with a brief review of 
standard optimal control theory in the presence of constraints in form of differential equations  (kinetic equations), associated numerical approaches, the dynamics of open quantum systems, and models for dissipation and decoherence. Sec. \ref{APP} gives a brief account of quantum subsystems. The main part of the paper discusses  
state--selective optimal control for Markovian and non--Markovian quantum systems in Sec. \ref{SD}, and state--independent optimal control in Sec. \ref{SID}.   
Relevant physical examples are given in the respective chapters. Sec. \ref{SUM} gives a summary and an outlook.
\par
\subsection{The optimality system }\label{OPT}

We confine ourselves to a brief introduction to standard optimization theory for continuous systems.  Detailed expositions may be found in the literature.\cite{Bryson75,Krotov96}

\subsubsection{The cost functional}
We consider a  continuous system and denote its state vector  by $\boldsymbol x(t)$ and the control by $\varepsilon(t)$. In general, both, state vector and the control, will be multi--dimensional. The former is an element of a linear vector space.  The most general case of dynamics of the state vector $\boldsymbol x(t)$ we are considering is described by an integro--differential equation,
\begin{equation}
\frac{d}{d t}\boldsymbol{x}(t)=\int\limits_{t_0}^{t}{d t' \boldsymbol f(\boldsymbol x(t'),\varepsilon(t'),t,t') },\label{opt1}
\end{equation}
which may be nonlinear in $\boldsymbol x$. The integral kernel $\boldsymbol f(\boldsymbol x(t'),\varepsilon(t'),t,t')$ depends on two times:  the current time $t$ and a time $t'$ accounting for the past $t'< t$.
The cost functional can be written as,~\cite{Hanson07b,Bryson75}
\begin{equation}
J[\boldsymbol{x},\varepsilon](\boldsymbol{x}_0,t_0,t_f)=\int\limits_{t_0}^{t_f}{dt L(\boldsymbol{x}(t),\varepsilon(t),t)}+\Phi(\boldsymbol{x}(t_f),t_f), \label{cost1}
\end{equation}
where $L(\boldsymbol{x}(t),\varepsilon(t),t)$ is often called the \textit{running cost} or the \textit{Lagrangean} and $\Phi(\boldsymbol{x}(t_f),t_f)$ is the \textit{terminal cost}. Eq.~\eqref{cost1} is referred to as \textit{Bolza} type, whereas cost functionals containing only the terminal or running penalty are called \textit{Mayer} or \textit{Lagrange} type, respectively.~\cite{Stengel94,Hanson07b}\par
The choice of the particular form of the cost functional reflects the desired objective to be achieved. If we want to steer our system into a given final state $\boldsymbol{x}_f$, the straightforward choice is to set $L=0$ and $\Phi(\boldsymbol{x}(t_f),t_f)=-\left\langle \boldsymbol{x}(t_f),\boldsymbol{x}_f \right\rangle$, where $\left\langle .,.\right\rangle $ denotes a real--valued scalar product defined in the linear state--vector space.  Another common choice for the Lagrangean is, $L(\boldsymbol{x}(t),\varepsilon(t),t)=-\left\langle \boldsymbol{x}_D(t),\boldsymbol{x}(t)\right\rangle$. Here we want the system's state variable to follow a given desired trajectory $\boldsymbol{x}_D(t)$. If we set $\boldsymbol{x}_D(t)=\boldsymbol{x}_D=const.$, the system, when subjected to the corresponding optimal solution, approaches the desired state $\boldsymbol{x}_D$ as fast as possible and tries to stay in that state, which is often called ``state trapping''.\par

The dependence of the Lagrangean on $\varepsilon$ and $\dot\varepsilon$ (the latter was not explicitly included above) allows the implementation of additional constraints on the control, such as shape, duration rate of change, or intensity. Due to physical considerations it is sometimes reasonable to include a constraint imposed on the control intensities, \textit{i.e.} to choose a Lagrangean of the form $L(\boldsymbol{x}(t),\varepsilon(t),t)= L'(\boldsymbol{x}(t),t)+\alpha \left|\varepsilon(t)\right|^2$, which penalizes large control field intensities, where $\alpha$ characterizes the degree of penalty. In fact, such a constraint may be mandatory to render the optimization problem well--defined mathematically. (See  Sec. \ref{SD} and Ref.~\onlinecite{Krotov96}). 
 If one desires  control fields which vanish at $t_0$ and $t_f$ it is convenient to use a penalty function which depends on time, \textit{i.e.}, $\alpha \rightarrow \alpha(t)$ and which takes on large values near initial and final time. The time derivative ${\dot \varepsilon(t)}$ can be included in the Lagrangean to introduce a means for suppressing unphysically rapid variations in the control field in the cost functional explicitly, as well as to preserve an analogy to the formalism of classical mechanics in form of a velocity--dependent Lagrangean. 

\subsubsection{Optimality conditions}
The optimal control field is defined by,
\begin{equation}
\varepsilon^*(\boldsymbol{x}_0,t_0,t_f)=\underset{\varepsilon\in {\cal L}^2 [t_0,t_f]}{\operatorname{arg\,min}}\left\lbrace J[\boldsymbol{x},\varepsilon](\boldsymbol{x}_0,t_0,t_f)\right\rbrace.
\end{equation}
A necessary condition for an optimal point is,
\begin{eqnarray}
&&\left.\frac{\delta J}{\delta \varepsilon}\right|_{\varepsilon^*}=0, \quad \frac{\delta J}{\delta \varepsilon}=\left.\left\langle  \frac{ \delta \Phi(\boldsymbol{x}(t),t_f)}{\delta \boldsymbol x(t)}, \frac{\delta \boldsymbol x(t)}{\delta \varepsilon}\right\rangle\right|_{t=t_f} +\nonumber \\
&& \int\limits_{t_0}^{t_f}{dt \left[ \left\langle \frac{\delta L(\boldsymbol{x},\varepsilon,t)}{\delta \boldsymbol x}, \frac{\delta \boldsymbol x}{\delta \varepsilon}\right\rangle +\frac{\delta L(\boldsymbol{x},\varepsilon(t),t)}{\delta \varepsilon}\right] }. \label{dJdE}
\end{eqnarray}
However, at least for analytical investigations of the optimality system, Eq.~\eqref{dJdE} is not very useful because $\boldsymbol{x}(t)$ depends implicitly on $\varepsilon(t)$. The variation of $\boldsymbol x$ with respect to $\varepsilon$,  $\frac{\delta \boldsymbol x}{\delta \varepsilon}$, may be complicated (nonlocal in time).  Most formulations of the optimality system which circumvent this problem are based on Lagrangean multipliers. The next  subsections will consider the formulation of such an approach for Markovian systems.
Application to non--Markovian quantum systems will be given in Sec.~\ref{NMARK}.

\subsubsection{Markovian kinetic equation}\label{MARK}
If the Kernel $\boldsymbol{f}$ is local in time, Eq.~\eqref{opt1} reduces from an integro--differential equation to a differential equation. Then, adjoining the system differential equations by the use of Lagrangean multipliers $\boldsymbol\lambda(t)$, often called \textit{co--state}, one may formulate a new cost functional,
\begin{eqnarray}
\hat J&=&\int\limits_{t_0}^{t_f}{dt \left\lbrace L(\boldsymbol{x},\varepsilon,t)+\boldsymbol \left\langle \boldsymbol\lambda(t), \dot{\boldsymbol x}(t)-\boldsymbol{f}(\boldsymbol{x},\varepsilon,t)\right\rangle  \right\rbrace }\nonumber \\
&&+\Phi(\boldsymbol{x}(t_f),t_f).
\end{eqnarray}
Now the differential equation constraint has been incorporated. By proper choice of the multipliers $\boldsymbol \lambda(t)$, (see Eq.~\eqref{dLdX}), we can eliminate the dependence of $\frac{\delta \hat J}{\delta \varepsilon}$ on  $\frac{\delta \boldsymbol x}{\delta \varepsilon}$.
If we define the Hamiltonian,
\begin{equation}
H(\boldsymbol{x},\varepsilon,\boldsymbol\lambda,t)=L(\boldsymbol{x},\varepsilon,t)+ \left\langle \boldsymbol\lambda,\boldsymbol{f}( \boldsymbol{x},\varepsilon,t)\right\rangle ,
\end{equation}
and apply the calculus of variations, the necessary conditions for an optimal point can be derived (see Ref.~\onlinecite{Bryson75}),
\begin{eqnarray}
\frac{\partial H}{\partial \boldsymbol x}&=&-\frac{d}{dt}\boldsymbol\lambda(t)=\left( \frac{\partial L}{\partial \boldsymbol x}+\frac{\partial}{\partial \boldsymbol x} \left\langle \boldsymbol{f},\boldsymbol\lambda\right\rangle \right), \label{dLdX} \\\nonumber
\boldsymbol\lambda(t_f)&=& \left.\frac{\partial \Phi}{\partial \boldsymbol x}\right|_{t=t_f},\nonumber \\
\frac{\partial H}{\partial \boldsymbol\lambda}&=&\frac{d}{dt}\boldsymbol x(t)=\boldsymbol f( \boldsymbol x,\varepsilon,t),\nonumber \\
\frac{\partial H}{\partial \varepsilon}&=& 0 = \left( \frac{\partial L}{\partial \varepsilon}+\left\langle  \boldsymbol\lambda(t), \frac{\partial}{\partial \varepsilon}\boldsymbol{f}( \boldsymbol{x},\varepsilon,t)\right\rangle\right). \nonumber
\end{eqnarray}
These equations take the form of Hamilton's equations of motion of classical mechanics and are referred to as Pontryagin's minimum principle. They are simultaneously satisfied for an optimal trajectory.~\cite{Grei03,Krotov96}  

One of the first applications of OCT to quantum systems has been the theoretical examination of how to control the final state of a diatomic molecule.~\cite{Peirce88} A similar approach based on a variational principle has been used to maximize the probability of a certain pathway in a chemical reaction by using coherent two--photon processes.~\cite{Tann85}
The use of Lagrangean multipliers and the similar Krotov method is quite common.~\cite{Sola98,Palao03} These techniques have been applied to a broad range of quantum mechanical problems.~\cite{Palao02,Grace07,Borzi02,Hohen07,Poetz2,Poetz3} A comparison between the Krotov method and gradient methods can be found in Ref.~\onlinecite{Sola98}.

When OCT is applied to quantum mechanics, typical examples for state vectors $\boldsymbol{x}(t)$ are wave wave functionsfunctions $\ket{\psi(t)}$~\cite{Borzi02,Tesch02} or density matrices $\rho(t)$.~\cite{Rol07,Poetz3,Poetz2,Wenin06} 
In the quantum computation context, the elements of the unitary time evolution operators $U(t)$ are a common choice for ${\boldsymbol x}$ because initial--state--independent optimization schemes are necessary in order to optimize quantum gates.\cite{Palao02,Palao03,Grace07,Monta07} Recently,  time--evolution superoperator--based formulations have been proposed.~\cite{Wenin08c,Rol08,Schulte06}
\par

\subsection{Numerical aspects }
 
The variational calculus  provides the necessary conditions for an extremum of $J$ in form of gradients which may be used as input to a broad range of numerical  schemes which search for minima of a function for which both function and gradients are available analytically.\cite{Press92}  
However, deriving the co--state equations  for an open quantum system often is  a tedious task and one may want to use a solely numerical method to compute the gradients. One possibility is to discretize the control field $\varepsilon(t)$ in time,
\begin{eqnarray}
t &\rightarrow & t_n=n h, \; \mbox{with} \; n\in[0,N], \; t_f=Nh+t_0, \nonumber \\
\varepsilon(t) &\rightarrow & \varepsilon_n=\varepsilon(t_n), \nonumber \\
h&...&\mbox{grid spacing}, \nonumber
\end{eqnarray} 
and to compute the gradient of the cost functional directly via finite differences,
\begin{eqnarray}
\frac{\delta J}{\delta \varepsilon} &\rightarrow& \left( \frac{\delta J}{\delta \varepsilon_0},\frac{\delta J}{\delta \varepsilon_1},...,\frac{\delta J}{\delta \varepsilon_N}\right), \nonumber \\
\frac{\delta J}{\delta \varepsilon_n}&=&\frac{J\left(\varepsilon_0,...,\varepsilon_n + \Delta \varepsilon_n,... \right) -J\left(  \varepsilon_0,...,\varepsilon_n,...\right) }{\Delta \varepsilon_n}. \nonumber
\end{eqnarray}
The price paid when using finite differences is a significant loss in numerical stability relative to the indirect variational method introducing a co--state and care must be taken in identifying the parameter range over which meaningful results are obtained so that convergence can be reached.\cite{Poetz3} Usually, the extra effort spent in deriving and evaluating co--state equations pays dividends when performing the optimization numerically. 

Using the gradient $\frac{\delta J}{\delta \varepsilon}$, one can utilize e.g. a conjugate gradient method to search for a minimum of the cost functional.\cite{Press92} However, such a method is prone to get stuck in local minima and convergence may be slow when the initial guess is poor. Instead one can use global search algorithms, the most prominent being stochastic function minimizers, like genetic and differential evolution algorithms or simulated annealing.~\cite{Schmitt01,Storn97,Kirk83,Grace07,Amstrup93,Poetz06} These methods have the advantage that no calculation of the gradient is needed and that it is more likely to find a global minimum of the cost functional. On the other hand, these algorithms usually need lots of cost functional evaluations which may be computationally expensive.  \par

For all numerical implementations the control scheme has to be executed on a time grid.  The upper limit in grid size is usually determined by the numerical requirements 
posed by the differential equations for state and co--state.  In principle one may discretize the control using the same grid.\cite{Poetz06}   This quickly leads to a control field vector of high dimension and an according number of field gradients which makes computation time--consuming.  
Frequently it is advantageous when physical intuition or experimental limitations narrow down the solution space.  In particular, parameterising  the control has been shown to provide significant speedup and reduction in numerical complexity.

\section{Application of optimal control theory to quantum information processing}\label{APP}
\subsection{Introduction to quantum information processing}
In 1982 Feynman published a paper in which he discusses the question of whether it is possible to simulate quantum mechanics effectively using a classical (probabilistic) computer.~\cite{Feyn82} ``Effective'' here means that the computational resources, (\textit{i.e.}, computation time and memory) scale  polynomially, as opposed to exponentially,  with  the size of the physical system  to be simulated.  He also introduced the concept of a quantum computer as a universal quantum simulator which uses ``quantum elements'' in order to simulate another quantum system. For a quantum computer, such a ``quantum element'' is the \textit{quantum bit} or \textit{qubit}, which can be seen as the quantum mechanical analogue to the classical bit. The difference with respect to the classical bit, which is either in the state 0 or 1, is that a qubit can be in a superposition state . If we denote the computational basis states of the quantum two level system by $\ket{0}$ and $\ket{1}$, the pure state $c_0\ket{0} +c_1\ket{1}$, with $c_i\in \mathbb{C}$ and $\left|c_0\right|^2+\left|c_1\right|^2=1$, is also a valid qubit state. Another difference arises if we examine $n$--partite systems, e.g. a two--partite system. For a classical 2--bit system it is always possible to assign a definite state to each of it is components whereas for a two--qubit system this is not always possible. If two qubits are entangled, e.g. if they are in the pure state $\frac{1}{\sqrt{2}}\left( \ket{00}+\ket{11}\right)$, only the composite system is in a definite state.  

\begin{figure}[h]
\includegraphics[width=5cm]{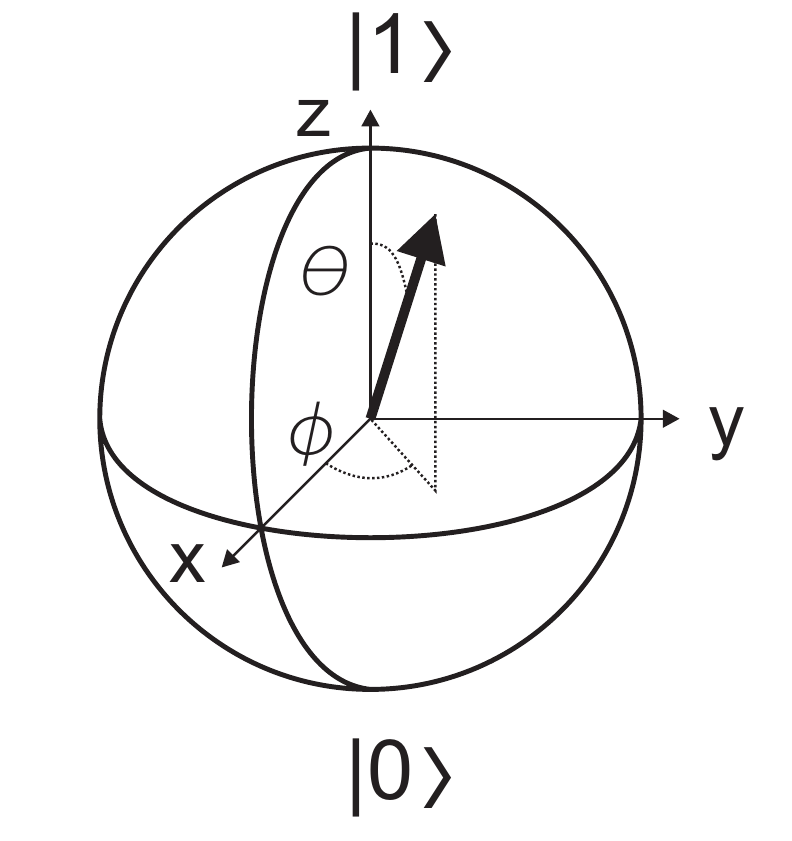}
\caption{\label{bloch} Bloch sphere and Bloch vector.}
\end{figure}

In addition to pure states, a quantum system can also be in a mixed state represented by a density operator $\rho$.\cite{Fick90} 
To visualize the state of a single qubit one often uses the so called \textit{Bloch sphere} and \textit{Bloch vector}. 
Any qubit state $\rho$, pure or mixed, can be written as,
 \begin{equation}\label{Blochvec}
 \rho=\frac{1}{2}(\mathbbm{1}+\vec{R}\cdot\vec{\sigma}),
 \end{equation}
 where $\vec{R}=(x,y,z)$ is the real Bloch vector, $|\vec{R}| \leq 1$, and
 $\vec{\sigma}$ is the spin--vector, containing the
 Pauli--matrices $\sigma_{i}$, $i=x,y,z$.
By rewriting the pure state of a qubit, $\ket{\psi}=c_0\ket{0} +c_1\ket{1}$ into $\ket{\psi}=\sin{\frac{\theta}{2}}\ket{0}+e^{i\phi}\cos{\frac{\theta}{2}}\ket{1}$, we may use the angles ${\theta,\phi}$ to represent $\ket{\psi}$ by a vector (the Bloch vector),  with length $1$ for pure states, which points to a specific point on the surface of a unit sphere (the Bloch sphere), see Fig.~\ref{bloch}.  
\par
\begin{figure}[h]
\begin{tabular}{c}
(a) \includegraphics[width=7.5cm]{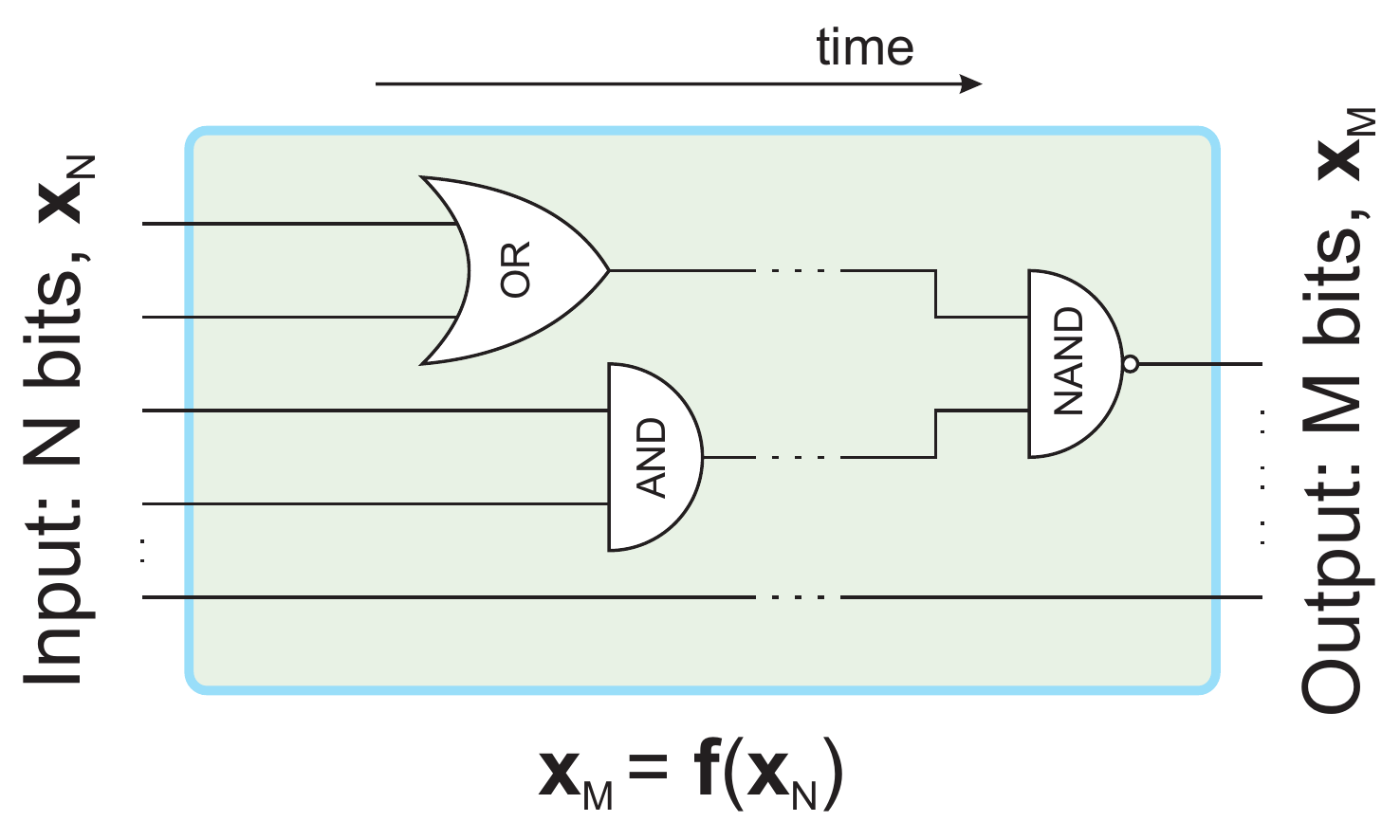} \\
(b) \includegraphics[width=7.5cm]{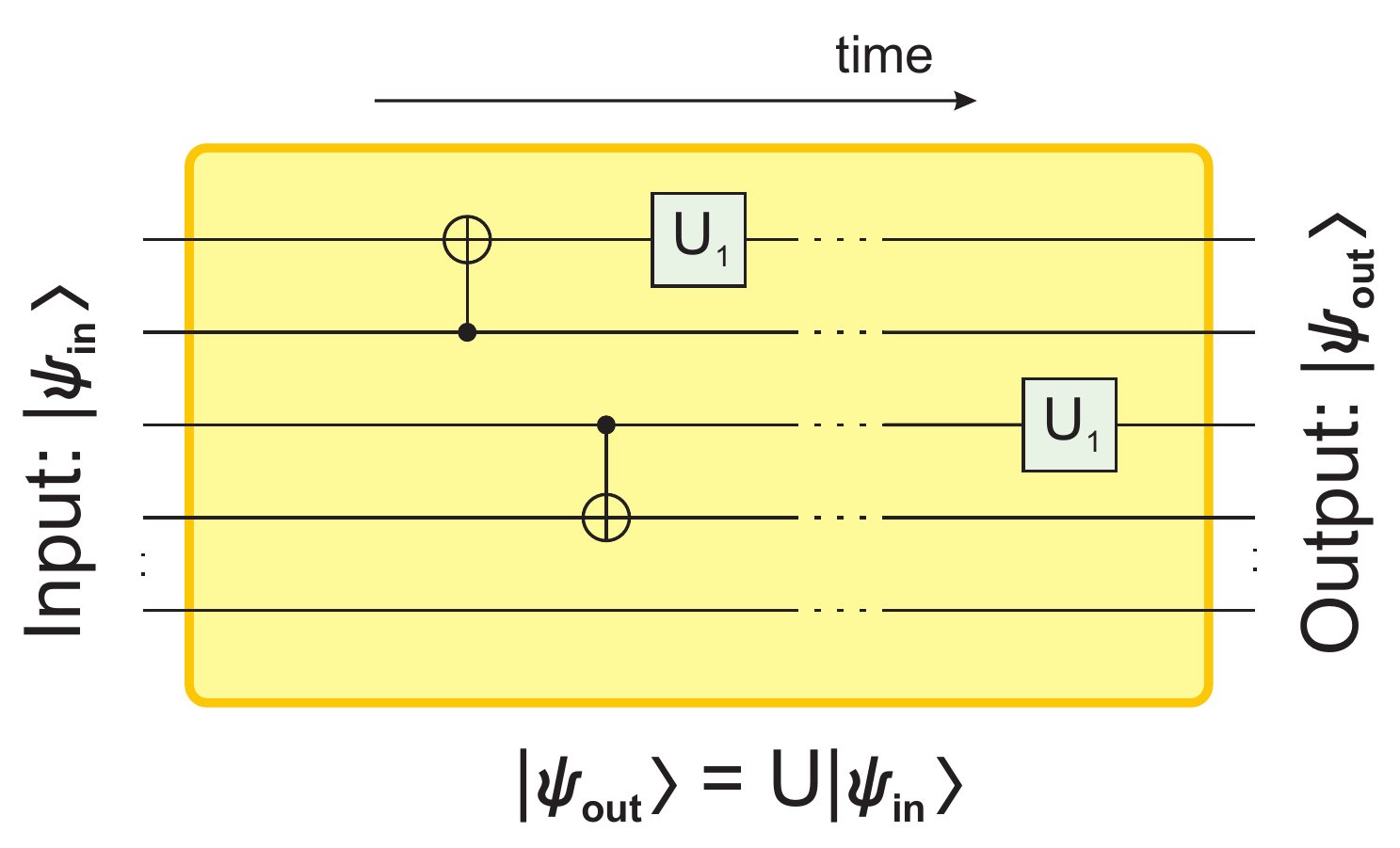}
\end{tabular}
\caption{\label{circuit} (a) Decomposition of a classical algorithm into elementary gates. Universal gate: e.g. NAND. (b) Decomposition of a quantum algorithm. Universal gates: e.g. one qubit operations, $U_1$, together with the controlled NOT (CNOT).}
\end{figure}

Within the circuit model of quantum computing, every quantum algorithm can be decomposed into elementary operations which correspond to unitary transformations on the qubits, see Fig.~\ref{circuit}.~\cite{Niel02,Zeil00} In analogy to classical computing one can identify universal quantum gates. It has been shown that one--qubit operations together with the so called \textit{controlled NOT} (CNOT) gate, which is a operation on two qubits, is universal for quantum computing.~\cite{Bar95} However, the particular choice of the CNOT gate is not mandatory. It has been proven that almost every gate that operates on two or more qubits represents an universal gate.~\cite{Deu95}

\subsection{Dynamics of quantum systems}
\subsubsection{Unitary and non--unitary time--evolution}
When a quantum system  is perfectly isolated from its environment the dynamics is governed by the unitary time--evolution operator, $U(t)\equiv U(t,0)$,
\begin{equation}
U(t)=\operatorname{T}\exp {\left\lbrace-\frac{i}{\hbar}\int\limits_{0}^{t}{ H(t') dt'}\right\rbrace },
\label{PROP}
\end{equation}
with the system Hamiltonian,
\be\label{HSYS}
H(t)=H_0+H_c(t),
\ee
where $H_0$ denotes the Hamiltonian of the intrinsic quantum system and $H_c$ the control part. The wave function evolves accordingly,
\begin{equation}
\ket{\psi(t)}=U(t)\ket{\psi(0)}.
\end{equation}
The basic equation of motion for an isolated quantum system in a mixed state $\rho$ is the von Neumann 
equation,~\cite{Fick90}
\begin{equation}
\frac{d}{dt}\rho(t)=-\frac{i}{\hbar}\left[ H(t), \rho(t)\right]. \label{von Neumann}
\end{equation}

If we are dealing with open quantum systems, e.g., a system $S$ which is in contact with an environment $B$, the resulting time--evolution for subsystem $S$ is non--unitary in general. This means a description based upon the Schr\"odinger or von Neumann equation is no longer appropriate for $S$. 
In order to compute the dynamics of open quantum systems one can use a stochastic Schr\"odinger equation.~\cite{Gard00} 
However, the general approach to deal with the non--unitary time evolution of a subsystem is to start from a suitably enlarged composite quantum system 
which obeys the von Neumann equation Eq. \eqref{von Neumann}, followed by a reduction to the degrees of freedom of the subsystem.
The form of the composite system (system S and bath B) Hamiltonian is,
\begin{equation}
H(t)=H_S(t) \otimes \openone_B + \openone_S \otimes H_B + H_{SB},
\end{equation}
where $H_S$, $H_B$ and $H_{SB}$ denote the Hamiltonians of system, environment and the interaction between system and environment, respectively.
By tracing out the environmental degrees of freedom in Eq.~\eqref{von Neumann}, one can deduce the differential equation for the density matrix of subsystem $S$,
\begin{equation}
\frac{d}{dt}\rho_S(t)=-\frac{i}{\hbar}\operatorname{tr}_B\left\lbrace \left[ H(t), \rho(t)\right] \right\rbrace .
\end{equation}
Calculation of the dynamics of the reduced system is, except for very few simple examples, a demanding task. Despite its simple appearance, it is, in fact, often difficult to cast the above equation in a form in which numerical evaluation is tractable. 
For Markovian processes one may employ the Lindblad master equation approach which will be described in the next section. To obtain kinetic equations within  microscopic quantum mechanical models one can use a perturbative expansion in the system--environment coupling, non--perturbative resummation techniques, or projection operator techniques. For a detailed description of common methods see Ref.~\onlinecite{Breu03}.
\par
\begin{figure}[h]
\includegraphics[width=8.5cm]{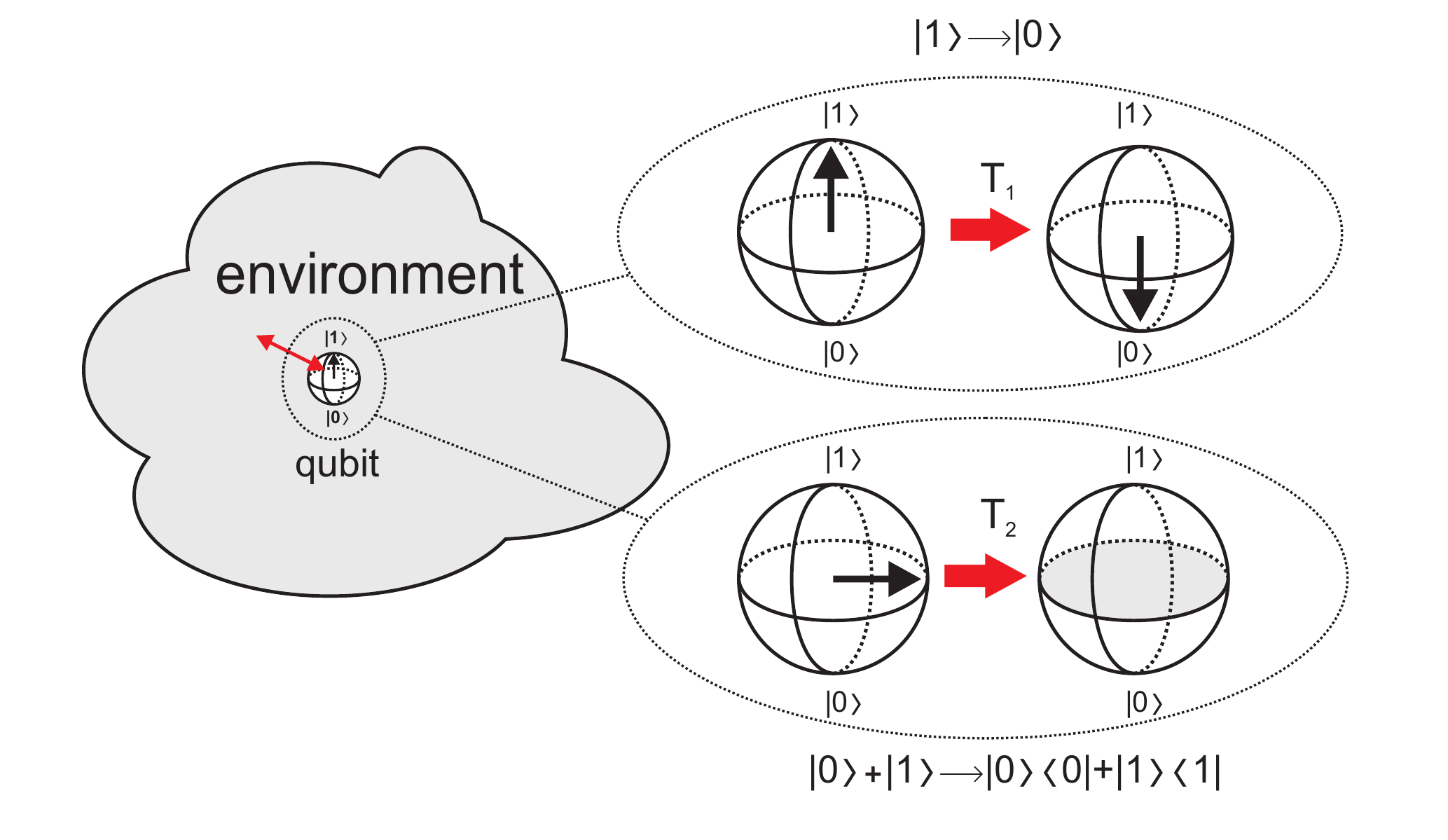}
\caption{\label{decoherence} Relaxation and dephasing (\textit{i.e.}, decoherence) due to unwanted environmental interactions within the Bloch sphere picture. $T_1$ denotes the relaxation and $T_2$ the decoherence time.}
\end{figure}

State superposition and entanglement are key--ingredients for quantum information processing. Making the qubit to perform a specific unitary transformation is done by proper tuning of external controls interacting with the dynamics of the quantum system. However, via the same channels by which one couples to the qubit, as well as by additional  sources (``the environment'') over which one has no direct control, noise can enter the system. These unwanted perturbations, in general, lead to decoherence (destruction of state--superposition and entanglement) and/or dissipation (\textit{i.e.}, state relaxation), both being detrimental for quantum computation, see Fig.~\ref{decoherence}.

\subsubsection{Models of dissipation and decoherence}\label{DISSI}
The \textit{Lindblad equation} is the most general Markovian  differential equation which is of first order in time and which preserves positivity and trace=1 for the system's density matrix. It can be viewed as a Markovian  extension of the von Neumann equation to open quantum systems by adding a dissipator to the equation.   In its general form,~\cite{Breu03}
 \begin{equation}\label{MEQ}
 i\hbar\dot{\rho_S}=[H_S(t),\rho_S]+D[\rho_S],\hspace{5mm}
 \rho_S(0)=\rho_0,
 \end{equation}
with
\begin{eqnarray}\label{LIND}
D[\rho_S] & = & i\hbar\sum\limits_{\mu=1}^{N^2-1}{\gamma_\mu\left(\vphantom{\frac{1}{2}} L_\mu\rho_S L_\mu^\dag - \right.} \nonumber \\
&&\left. \frac{1}{2}L_\mu^\dag L_\mu \rho_S - \frac{1}{2}\rho_S L_\mu^\dag L_\mu\right),
\end{eqnarray}
the  dimensionless operators $L_\mu$ contained in the dissipator  describe the different decay and decoherence channels of the quantum system which are opened by its interaction with the environment.  The quantities $\gamma_\mu$ are effective relaxation rates which  may be set phenomenologically 
or be computed on basis of a microscopic model. Since the Lindblad structure is the most general of a Markovian master equation for $\rho_S$,  it serves as reference for Markovian master equations which are derived from microscopic models within approximations.
Optimal control schemes have been used to optimize the dynamics of quantum systems which are described by a Lindblad equation.~\cite{Schulte06, Poetz3,Wenin06, Wenin08,Wenin08b} The use of Lindblad operators with constant $\gamma_\mu$'s is phenomenological and lacks details about the quantum mechanical interaction between sub--system and environment. Therefore, controllability of the system is usually poor when this model is appropriate.\par

Among the elementary microscopical models for a system--bath interaction the most prominent example is probably the \textit{spin--boson model}.~\cite{Leg87,Weiss99}. In its basic form, a spin $\frac{1}{2}$--particle couples linearly to the oscillator bath polarization. The Hamiltonian for bath and system--bath interaction, respectively, are usually written as,
\begin{eqnarray}
{H_B} &=& \sum\limits_k{\hbar \omega_k b_k^{\dag}b_k}, \nonumber \\
H_{SB} &=& {\hat S} \otimes\Gamma, \; \Gamma=\hbar\sum\limits_k{g_k\left( b_k+b_k^\dag\right) },\label{SB}
\end{eqnarray}
where $b_k^{(\dag)}$ is the bosonic (creation) annihilation operator for mode $\omega_k$ and $g_k$ is the effective coupling strength of the $k$th mode to the spin ${\hat S}\in\left\lbrace S_x,S_y,S_z \right\rbrace $. $\Gamma$ denotes the bath polarization. In QIP one focuses mainly on the case of weak coupling of the bath to the two level system (interpreted as the qubit). For this case perturbative methods, such as the Born approximation and the Bloch--Redfield approach, are best suited to describe the reduced dynamics of the system.~\cite{Car02,Breu03} If one is interested in the strong coupling regime, techniques like the polaron transformation~\cite{Mahan} or path--integral approaches [e.g. the non--interacting blip approximation (NIBA)] are available. For a method which deals with both regimes see Ref.~\onlinecite{Nesi07}. In general, the validity of each of the mentioned approximation schemes also depends on the bath temperature and/or on other bath--spectral--density specific characteristics (e.g. the cutoff frequency).~\cite{Hart00,Nesi07,Grif96} Optimal control of a qubit system subjected to a polaron transformation and a subsequent approximation by a second order expansion in the tunneling parameter $\Delta$ [which is equivalent to the NIBA approximation, see Ref.~\onlinecite{Aslan86}] has been performed in Refs.~\onlinecite{Poetz06,Poetz2}. For longitudinal couplings, \textit{i.e.}, $S=S_z$, and $H_S \propto S_z$ analytical solutions are available.~\cite{Mahan, Rei02} Because control within $H_S$ is restricted within these cases due to lack of control with respect to orthogonal directions, they are of minor importance for QIP applications.\par

Another microscopic model which has been employed to take into account the effects of an environment is the \textit{spin bath}. For a review see Ref.~\onlinecite{Prokof00}.  In Ref.~\onlinecite{Grace07}, optimal control techniques have been applied in order to obtain high--fidelity one-- and two--qubit gates in the presence of coupling to ``environmental'' two level systems, which can be interpreted as spin--$\frac{1}{2}$ particles. The Hamiltonian for $m$ qubits and $N$ spin--$\frac{1}{2}$ particles is of Heisenberg form,
\begin{equation}
H=\sum\limits_{i=1}^{N}{\omega_i S_{z,i}} - \sum\limits_{i=1}^{m}{\mu_i C(t) S_{x,i}}+ \sum\limits_{i=1}^{N-1}{\sum\limits_{j>i}^{N}{\gamma_{ij}\boldsymbol S_i \cdot \boldsymbol S_j}},
\end{equation}
where $\left\lbrace \boldsymbol S_i=(S_{x,i},S_{y,i},S_{z,i}) \right\rbrace $ denotes the (pseudo) spin--operator and  $\omega_i$ the (pseudo) Zeeman splitting for particle $i$. Here it is assumed that it is possible to locally apply a control field $C(t)$ via coupling to dipole moments $\mu_i$. The strength of the  Heisenberg exchange interaction between the spins is given by $\gamma_{ij}$.

\subsection{State--selective versus state--independent optimal control}\label{SSvsSI}
\begin{figure}[h]
\begin{tabular*}{8.5cm}{ccc}
		\multicolumn{3}{c}{(a) State--selective transformation}\\
		\includegraphics[width=4cm]{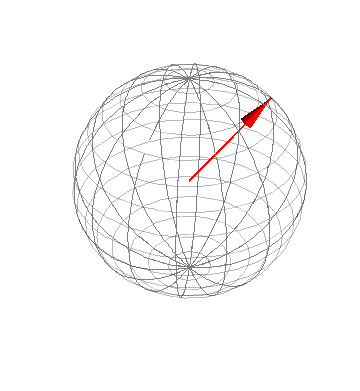} & \raisebox{2.3cm}{$\Rightarrow$}& \includegraphics[width=4cm]{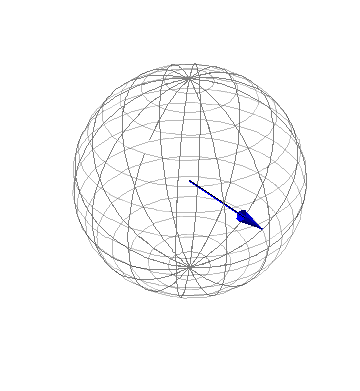} \\
		\\
		\multicolumn{3}{c}{(b) State--independent transformation}\\
		\includegraphics[width=4cm]{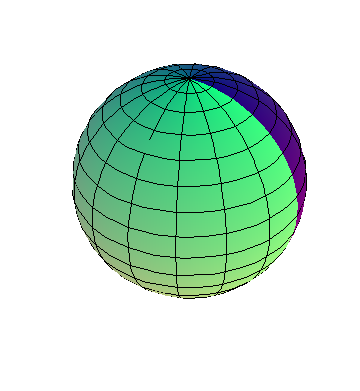} & \raisebox{2.3cm}{$\Rightarrow$}& \includegraphics[width=4cm]{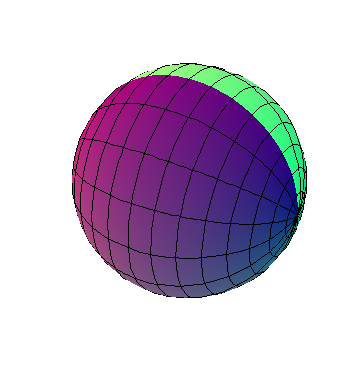} \\
	\end{tabular*}
\caption{\label{sd_si} (a) ``State--selective'' corresponds to a transformation of a predetermined initial state (denoted by a red Bloch vector) to a particular given final state (blue Bloch vector). (b) ``State--independent'' means that we want to perform a desired mapping, irrespective of the initial state of our qubit. For unitary operations, as common in the QIP context, this corresponds to a rigid rotation of the Bloch sphere.}
\end{figure}

When we apply OCT to QIP systems, we often have to change our focus from optimization of state--to--state transitions (e.g. $\ket{\psi_i} \rightarrow \ket{\psi_f}$) to an optimization of quantum dynamical mappings, see Fig.~\ref{sd_si}. This means that for quantum computation it is not sufficient to find a control field which manages to steer a quantum system starting from a particular initial state to a predetermined final state.
This form of control is usually termed ``state--dependent''. A quantum gate has to perform the corresponding desired transformation ${\mathcal O}$ regardless of the initial state. In fact, the initial state often is unknown. This form of optimization is usually termed ``state--independent''.  If we are dealing with unitary time evolution, described by a Schr\"odinger equation $\frac{d}{dt}\ket{\psi(t)}=-i/\hbar H_S(t) \ket{\psi(t)}$, state--independent optimization is conveniently implemented by optimizing the time propagator $U_S(t)$, with $\dot U_S(t)=-i/\hbar H_S(t) U_S(t)$ and $U_S(0)=\openone$. Essentially, one minimizes a cost functional of the form $J=\vectornorm{U_S(t_f)-{\mathcal O}}^2$ or similar.~\cite{Palao02,Palao03} \par
The task of state--independent optimization is especially interesting for noisy quantum gates. In general, a superoperator $\mathcal{X}(t)$ defines the time evolution of the system for the interval $[0,t]$, \textit{i.e.},
\begin{equation}\label{SUOP}
\rho_S(t)=\mathcal{X}(t)\rho_S(0), \hspace{5mm}
\mathcal{X}(0)=\mathbbm{1},
\end{equation}
where $\rho_S(t)$ denotes the subsystem's density matrix at time $t$. The most general form of such an admissible superoperator is given in terms of Kraus operators $K_{mn}(t)$,~\cite{Niel02}
\begin{eqnarray}
\rho_S(t)&=&\sum\limits_{m,n}{K_{mn}(t)\rho_S(0)K^\dag_{mn}(t)}, \nonumber \\
K_{mn}(t)&=&\left( \bra{n}\rho_B(0)\ket{n}\right)^{1/2} \bra{m}U(t)\ket{n},
\end{eqnarray}
where $U(t)$ denotes the unitary time--evolution operator of the composite system. $\rho_B$ denotes the density operator of the bath. $\left\lbrace \ket{n} \right\rbrace $ is a complete set of bath modes so that $\bra{n}\rho_B(0)\ket{n'}=\delta_{n,n'}\bra{n}\rho_B(0)\ket{n'}$.
Given a microscopic model,  it is, in general, not possible to find analytic expressions for the Kraus operators in terms of the Hamiltonian of the composite system. In fact, even an exact numerical treatment may be intractable. However, one can find equations of motion for the superoperator, which can be approximated. One possibility is to switch to a Liouville space description.~\cite{Muk95} Optimal control schemes which are based upon such a description can be found in Ref.~\onlinecite{Spoerl07} and Ref.~\onlinecite{Sklarz06}.

\section{State--selective optimal control of open quantum systems}\label{SD}

Early applications of  optimal control to quantum systems have been formulated mostly for  state--dependent  cost functionals for closed quantum systems based on pure states within the time--dependent Schr\"{o}dinger equation, with applications mostly in quantum chemistry.\cite{Rabitz,Gord97}  Then, mixed--state optimal control,  formulated within the von Neumann equation and, finally,  
within optimum control for open quantum systems has followed.  
State--dependent optimal control has applications in many aspects of quantum physics.  Originally, it was motivated for driving a 
quantum systems, such as a molecule,  from an initial state, usually the ground state, into a certain final state, for example,  a particular fragmented state of the molecule.  Later, combination with coherent control, exploiting quantum interference, has been proposed and executed in semiconductor nanostructures.\cite{Dupond95,vanDriel97,Sipe00,Smirl03,Smirl04,Poetz99,hohenester}  

In this section, we review several cases for state--dependent optimal control. Rather than specifying particular physical realizations, we keep the presentation general and distinguish between Markovian and non--Markovian quantum systems, {\it i.e.}, the nature of their dissipator.  
The motivation for this study is the use of quantum interference between competing interactions as a 
principle of operation for electronic and electro--optic nanoscale devices.  
Particularly in a solid--state environment, electronics-- and spin--based quantum interference effects are difficult 
to establish and to maintain.\cite{Breu03}  
One of the potential solutions is to steer the  quantum systems along a suitable quantum trajectory so that  
one eliminates or minimizes the system--environment interaction by destructive quantum interference.
This naturally leads to an optimization problem (inverse problem) where one  seeks optimal control fields  
which stabilize coherence of a quantum system  or maximize induced quantum interference effects.

For the remainder of the paper we shall denote the quantum subsystem ``system" and the reduced subsystem density operator 
$\rho_S$ by $\rho$, and $U_S(t)$ by $U(t)$,  for brevity.  Furthermore, the subsystem Hamiltonian Eq. \eqref{HSYS} will generally be denoted by $H(t)$, except when stated otherwise.

\subsection{Markovian kinetic equations}

The Lindblad equation Eq. \eqref{LIND} captures the dynamics of a quantum system in the Markovian regime, \textit{i.e.}, on a time--scale of the quantum system which is large compared to the memory--loss time of the environment.\cite{Breu03}  On this time--scale,  the damage to coherent dynamics of a quantum system caused by its environment partially has become irreparable, however, limited  reduction of coherence loss has been shown to be possible, particularly, if the effective rates $\gamma_\mu$ in  Eq. \eqref{LIND} feature a dependence upon the adjustable control fields.  
We shall first  consider the situation of constant $\gamma_\mu$'s.   In this case, the optimal control problem of a dissipative qubit may be solved 
analytically by direct inversion, as is shown in the following subsection.  In the second part of this section we briefly discuss the case of control--field dependent effective rates $\gamma_\mu$ which allow for greater control potential.

\subsubsection{Direct inversion}
For direct inversion of the Lindblad equation one first selects a physically allowed trajectory $\rho(t)$
which is compatible with a specified initial and final state  $\rho(0)$ and  $\rho(t_f)$, from which a suitable Hamiltonian 
$H(t)$ is extracted,~\cite{Romero,Zhao,Wenin06,Proc}
\begin{equation}\label{INV}
\rho(0)\stackrel{H(t)}\longrightarrow\rho(t_f)\Rightarrow H(t).
\end{equation}
The main problem lies in the existence and identification of such a trajectory. 
The problem is even more complicated in open
quantum systems than it is for pure coherent dynamics under unitary 
time--evolution.~\cite{Wu2}   In this case, there is a trajectory and a solution to the problem if the 
eigenvalues of $\rho(0)$ and  $\rho(t_f)$ are identical.  Unfortunately such a simple criterion cannot be formulated 
for open quantum systems.
In fact there are several open questions for open quantum systems : How does
one determine or even prove the existence of an allowed trajectory?  Which role does the kinetic
equation play on the existence? 
The procedure (\ref{INV}) may lead to non--local solutions for $H(t)$.
Hence, the question arises, what conditions have to be met in order to derive experimentally feasible solution to such inversion problems.
Finally, if a solution has been identified, is it unique or are there equivalent solutions which may be better
suited for physical realization?
Obviously, the difficulties in answering these questions increase rapidly with the dimension of the Hilbert space and
the complexity of the dissipator.
Nevertheless, some nontrivial and interesting results on
dissipative two--level systems have been found, where the system
is described by a Lindblad equation. The reader can find a
detailed discussion of this topic in Ref.~\onlinecite{Wenin06}. Here we
give a brief overview to demonstrate possibilities and
difficulties associated with the direct inversion strategy using a two--level
system.
\subsubsection{Choice of the trajectory: decoherence free subspace}\label{DSUB}
In many OCT problems the decoherence free subspace (DFS) plays a
central role. Using the kinetic equation Eq.~\eqref{MEQ}, the latter is
defined as the set of states $\rho_{DF}$ fulfilling,~\cite{Braun,Karasik,Lidar}
\begin{equation}\label{DFSdef}
D[\rho_{DF}]=0.
\end{equation}
Depending on the nature of the dissipator, Eq.~\eqref{DFSdef}
defines a subspace of  density operators, which
interesting for  OCT  because within the DFS the system dynamics  is
completely coherent. In many systems, the DFS is constructed by
dynamic decoupling processes, such as an application of control
pulses known as ``bang--bang" control.\cite{Viola,Viola1,Lidar2,Vitali} To
study such methods, the spin--boson model again is well suited,
since it explicitly displays the influence of the external control on
the system--environment interaction.\cite{Poetz2} If a state
$\rho_{DF}$ exists and is known and if the system allows complete
control, the optimal trajectory is given by,
\begin{equation}\label{DFStraj}
\rho(0)\rightarrow\rho_{DF}\rightarrow\rho(t_f),
\end{equation}
whereby the switching  into and out of $\rho_{DF}$ has to be executed rapidly. 
Using inversion formulas one can evaluate Eq.~\eqref{DFStraj} to
obtain an optimal control Hamiltonian. We remark that in such
cases it is sufficient to consider the inversion of the von
Neumann equation because one can minimize environment--induced dissipation  by ``instantaneous" 
switching in Eq.~\eqref{DFStraj} in principle.~\footnote{The inversion of the von Neumann
equation for a $N$--level system is possible, when one can
diagonalize $\rho(t)$.}
Recently we have extended the concept of the DFS from states to
evolution superoperators. This strategy 
can be applied for both
state--dependent and state--independent OCT.\cite{Wenin08c}

\subsubsection{Two--level system}\label{Inversion}

The kinetic equation for the density matrix $\rho(t)$ (quantum
trajectory) is given by Eq.~\eqref{MEQ}. Considering the inverse
problem, we begin with the selection of a quantum trajectory
$\rho(t)$ for a specified time interval $t\in [0,t_f]$. We set,
\begin{equation}\label{rhodes}
\rho(t)=\left(%
\begin{array}{cc}
  \rho_{11}(t) & a(t)+ib(t) \\
  a(t)-ib(t) & 1-\rho_{11}(t) \\
\end{array}%
\right).
\end{equation}
Here $\rho_{11}(t)$, $a(t)$, $b(t)$ are real valued functions. To
solve the inversion problem we put the dissipation part on the
left--hand side of Eq.~\eqref{MEQ} and subtract it from
$\dot{\rho}(t)$. In particular we set,
\begin{equation}
\dot{\tilde{\rho}}_{11}(t)\equiv
\dot{\rho}_{11}(t)-\frac{1}{i\hbar}(D[\rho(t)])_{11},
\end{equation}
\begin{equation}
\dot{\tilde{a}}(t)\equiv
\dot{a}(t)-\frac{1}{\hbar}\mathrm{Im}(D[\rho(t)])_{12},
\end{equation}
\begin{equation}
\dot{\tilde{b}}(t)\equiv
\dot{b}(t)+\frac{1}{\hbar}\mathrm{Re}(D[\rho(t)])_{12}.
\end{equation}
Insertion of Eq.~\eqref{rhodes} into Eq.~\eqref{MEQ} leads to,
\begin{equation}\label{z1}
\hbar\dot{\tilde{a}}=bu+w(1-2\rho_{11}),
\end{equation}
\begin{equation}\label{z2}
-\hbar\dot{\tilde{b}}=au+v(1-2\rho_{11}).
\end{equation}
Here we set $u\equiv H_{11}-H_{22}$ and
$v\equiv\mathrm{Re}(H_{12})$, $w\equiv\mathrm{Im}(H_{12})$ as our
unknowns. Note that $\rho_{11}(t)$, $a(t)$ and $b(t)$ in
Eq.~\eqref{rhodes} depend on each other via the condition,
\begin{equation}\label{Zusatz0}
\dot{\tilde{a}}a+\dot{\tilde{b}}b=\frac{\dot{\tilde{\rho}}_{11}}{2}(1-2\rho_{11}),
\end{equation}
which follows from Eq.~\eqref{MEQ} and links the variables
$\rho_{11}(t)$, $a(t)$ and $b(t)$. This relation must
hold for any allowed trajectory of the dissipative quantum system, 
independent of the structure of $D[\rho(t)]$. We
remark that for a dissipation--less system, Eq.~\eqref{Zusatz0}
represents conservation of purity, {\it i.e.},  constant length  of the Bloch vector. If the dissipation is
described by fixed rates, this relation is independent of the
control, which means that there is no Hamiltonian which
allows independent control of all matrix elements of the density
matrix. Eq.~\eqref{z1}, Eq.~\eqref{z2} and Eq.~\eqref{Zusatz0} are
the basis for the solution of the inversion problem. One can see
that Eq.~\eqref{Zusatz0} defines a condition which in simple
cases is a differential equation.  Depending on the dissipator,
more complicated integro--differential equations may occur.
\subsubsection{Example}
We consider a simple model, given by the Hamiltonian,
\begin{equation}\label{Inv}
H(t)=\varepsilon(t)\sigma_x,
\end{equation}
where the control $\varepsilon(t)$ is a real--valued function (here
$H_o$ in Eq.~\eqref{HSYS}  is chosen 
zero). For the Lindblad operators we choose,
$L_1=\sqrt{\gamma_1}|0\rangle\langle 1|$,
$L_2=\sqrt{\gamma_2}|1\rangle\langle 0|$. For simplicity we set
$\gamma_1=\gamma_2\equiv \gamma$. In this case, there is
restricted control over the system only, \textit{i.e.}, the Bloch vector can be rotated
around the $x$--axis only (see
Fig.~\ref{bloch}). We consider the case where we seek to manipulate 
the population $\rho_{11}(t)$. By inversion of the Lindblad equation
one obtains,
\begin{equation}\label{ea}
\varepsilon(t)=\hbar\frac{\dot{b}(t)+4\gamma
b(t)}{-1+2\rho_{11}(t)}.
\end{equation}
Here $b(t)$ is given by,
\begin{eqnarray}
b(t)&=&e^{-4\gamma t}\left( b(0)^2+\int_0^t\left\lbrace  e^{8 \gamma
t'}[1-2\rho_{11}(t')]\times \right.\right. \nonumber \\
&&\left. \left.[\dot{\rho}_{11}(t')+8\gamma\rho_{11}(t')-4\gamma] \vphantom{e^{8 \gamma
t'}}\right\rbrace dt'\right)^{\frac{1}{2}} ,
\end{eqnarray}
as follows from Eq.~\eqref{Zusatz0}. We require 
a real Hamiltonian, which leads to $a(t)=a(0)e^{-4\gamma t}$. So
one has the choice of $\rho_{11}(t)$ as only remaining freedom. The other
variable $b(t)$ and the control $\varepsilon(t)$ are deduced 
quantities. For a specific example, we wish to maintain Rabi oscillations in presence of dissipation and set,
\begin{equation}\label{r11}
\rho_{11}(t)=(1-2A)\cos^{2}(\Omega t)+A,
\end{equation}
where $A\geq0$ is a constant.  Induction and maintenance of Rabi oscillations in a realization of a two--level system is generally viewed as a test for its qubit potential.

\begin{figure*}
\begin{center}
\includegraphics[height=50mm, angle=0]{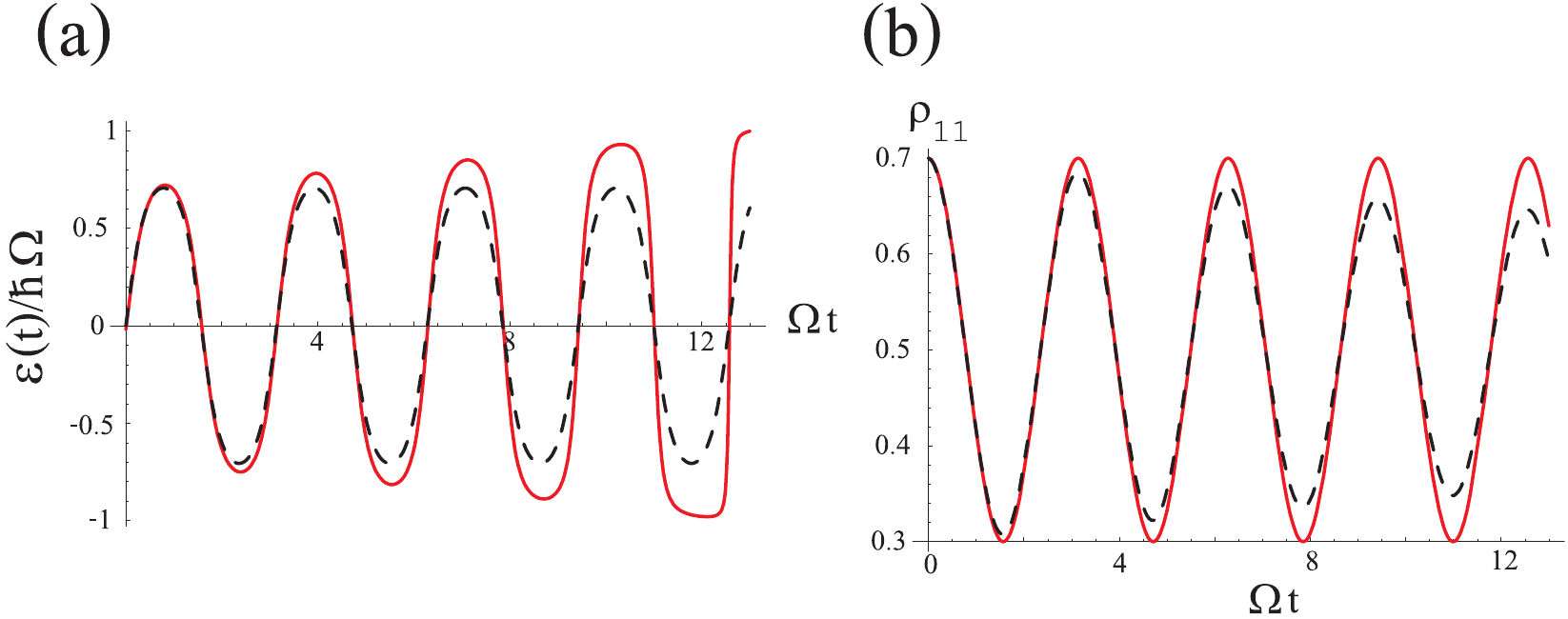}
\caption{(a) The control fields versus time. Dashed line:
solution in absence of dissipation, given by Eq.~\eqref{e0}. Red solid
line: control field which corrects for dissipation. (b) Trajectory
$\rho_{11}(t)$ versus time. Dashed line: uncorrected, damped
trajectory, obtained with the dissipation--less solution. Red,
solid line: corrected trajectory as prescribed by
Eq.~\eqref{r11}. Parameters: $\gamma/\Omega=0.0071$. $A=0.3$,
$b_0=0.2$.}\label{Grafik2}
\end{center}
\end{figure*}

The field for the special case $\gamma=0$ is,
\begin{equation}\label{e0}
\varepsilon(t)=\hbar\Omega\frac{\sqrt{2}(1-2A)\sin(2\Omega
t)}{\sqrt{8b_0^2+(1-2A)^2[1-\cos(4\Omega t)]}}.
\end{equation}
For $b(0)\equiv b_0=0$ this expression reduces to a constant
field $\varepsilon=\hbar\Omega$. The expression for  $\gamma\neq 0$ is
also available in analytical form, but is a quite lengthy expression.
Fig.~\ref{Grafik2} shows a numerical example. Part (a) shows the
field for Eq.~\eqref{e0} (dotted line) and the field with $\gamma\neq
0$ (solid line). Part (b) shows $\rho_{11}(t)$, when the
respective fields are used to drive the system. As the figure
shows, a complete restoration of the prescribed trajectory,
Eq.~\eqref{r11}, is possible. Further inspection of the field
shows, however, that this complete restoration is possible only for
a limited time;  for how long , depends on the parameters $A$,
$b_0$, and the damping constant $\gamma$.  One can compute this
time numerically using Eq.~\eqref{ea}.

Dissipative $N$--level systems have also been modeled fully numerically within the Lindblad equation accounting for population and polarization decay, based on constant decay rates.\cite{Poetz3}  The problem addressed was population transfer from a  non-degenerate stable  ground state to an unstable target state at $t_f$, using weak electric fields whenever possible. Work concentrated on a ladder--type $N$--level system for which an electric field can induce dipole transitions between adjacent energy levels, as typical for systems with inversion symmetry, and schemes as used in stimulated--Raman--adiabatic--passage (STIRAP) experiments. For all systems studied, {\it i.e.}, $\mbox{two--,}$ $\mbox{three--,}$ and four--level systems, it was found that the weak--dissipation limit which is relevant for atomic and many molecular systems, poses no serious threat to success. However, as dissipative effects increase in strength, so do the electric fields which are necessary for reaching the selected target state.  In addition, their onset is delayed closer and closer to target time. Clearly, with increasing complexity of the system, such as the number of  unstable  levels and decay channels, the degree of success decreases. Since purity decay cannot be controlled directly by the external control for constant Lindblad decay rates,  ``last--minute transfer" from a stable state remains to be the only solution for a complete transfer into an unstable target state when there is no control over decay rates.

\subsubsection{State--dependent optimization within the spin--boson model}

Dissipative two--level systems naturally are treated within the spin--boson model introduced in Sec. \ref{DISSI}.  In spite of it's simplicity, 
it allows coverage of a large spectrum of physical effects by proper mapping of more complex quantum systems and appropriate choice for the spectral function.  Unfortunately analytic solutions are not available, in general.  In Ref. \onlinecite{Poetz2}
the Bloch--Redfield approach was used to derive Markovian kinetic
equations for the spin--boson model in the
strong electron--boson coupling limit. Using the polaron
transformation, kinetic equations for the Bloch vector
with an effective coupling
within the spin system are obtained.  This coupling  is renormalized  by the spin--boson
interaction and displays a retarded control--field dependence which arises from interference between 
the system--bath and system--control field interaction. 
Several physical situations of spin flips have been investigated numerically
to demonstrate quantum--interference--based optimal
control of this model for an open quantum system.  Using simple analytical forms for the 
control fields it was shown at the example of this model how external control of quantum interference 
can be used to control the effective system--bath interaction for minimization or maximization of the coherence loss.
In the following section, we outline the generalization to non--Markovian kinetic equations.

\subsection{Non--Markovian kinetic equation}\label{NMARK}

The study of control--field dependent system--bath interactions  has recently been extended to a study of optimal control of  qubit realizations with non--Markovian dynamics  which captures and clearly demonstrates the role and potential of quantum interference effects in the control of effective system--bath interaction.\cite{Wenin08,Rol07b}  The basic idea is that, when the system--bath interaction is either treated theoretically or can be addressed experimentally  on a quantum--mechanical level, quantum interference effects can be utilized to control the effective coupling strength.\cite{Hu99}  Needless to say, this requires microscopic models for open quantum systems, on the theoretical side, and 
poses considerable challenges for experimentalists regarding precision, strength, and time--scales for the control
to be administered.   

It appears that, until recently,  classical and quantum mechanics applications of optimal control have concerned themselves with mostly, if not exclusively,  Markovian systems, \textit{i.e.}, systems for which knowledge of the state of the system at any given time is sufficient to uniquely specify their past and future.   However, when one studies the dynamics of a subsystem,  non--Markovian kinetic equations arise naturally when the degrees of freedom of the environment are integrated out.  This has become of particular importance to the study and control of quantum subsystems (``realizations of quantum systems") in the context of implementation of quantum algorithms into real systems.  If one can address and control a system in its quantum regime, quantum interference leads to new 
pathways (``control knobs") for steering the system and for controlling dissipation.  Put simply, in the classical regime 
only the diagonal matrix elements of the density matrix (in a suitable basis) can be manipulated, while in the quantum 
regime {\it all} matrix elements are available for  manipulation.\cite{Poetz99}   Simple decay rates become time-- and control--dependent  greatly enhancing control over the system.

\par
Within quantum mechanics, the emergence of non-locality with respect to time is most directly seen within the projection operator method whereby the bath (environment) degrees of freedom are projected out formally by a suitable projector  $Q$, such that, in the simplest case when using the identity representation $\one =P+Q$, one obtains two coupled sets of kinetic equations of first order in time, one each for $P\rho$ and one for $Q\rho=(\one-P)\rho$.   Formally integrating the kinetic equation for $Q\rho=(\one-P)\rho$ and inserting the result into the one for $P\rho$, the latter becomes a non--Markovian differential equation for  $P\rho$.~\cite{Breu03}  Similarly, non--Markovian equations are obtained readily within the density matrix approach when truncating the BBGKY--type hierarchy of  higher correlation functions.~\cite{Rossi02}  In many cases, non--Markovian kinetic 
equations can be approximated or even exactly be replaced by an enlarged number of Markovian equations.~\cite{Breu03,Xua04}  

We consider for the quantum subsystem a non-Markovian time--evolution of the general form,~\cite{Poetz1,Poetz06,WP}

\be
{\dot \rho}(t) = \int_0^{t} dt' K(t,t';\varepsilon, \rho), \mbox{  for } \rho(0)=\rho_o.
\label{KE}
\ee
The integral kernel $K(t,t';\varepsilon, \rho)$ is given by the following form. It  depends on two times:  the current time $t$ and a time $t'$ accounting for the past $t'< t$.   Furthermore, the kernel functionally depends  on the control field $\varepsilon(t'')$, for $t''< t$ such that causality is fulfilled, as well as on $\rho(t')$ for $t'\leq t$. 
In general, it is a non--linear super--operator which maps $\rho(t'), ~ t'\leq t$ onto a linear operator of trace zero.    The latter is needed to ensure trace preservation for $\rho(t)$.   Moreover, the memory kernel $K(t,t';\varepsilon, \rho)$ must preserve 
positivity of $\rho(t)$ at any time $0\leq  t \leq t_f$.   It is assumed that Eq. (\ref{KE}) is integrable, \textit{i.e.}, there exists  at least one solution to any  
given control field $\varepsilon(t), ~ t\in [0,t_f]$ which itself is square-integrable and bounded over the interval $[0,t_f]$.  In other  words, the constraint is holonomic.   These conditions are equivalent to the assumption of the existence of a Kraus representation for the time--evolution operator defined by Eq.~(\ref{KE}). 
\footnote{For nonlinear kinetic equations such a rigorous  existence proof may be rather difficult and we shall not concern ourselves with this mathematical issue here.}

In what follows we shall review and extend a recent generalization of optimal control problems to non--Markovian constraints.\cite{Poetz1}  For didactic reasons we present  this  approach pointing out analogies to classical mechanics.  
Referring to Sec.~\ref{OPT}  we make the following assignments.
The state vector $\boldsymbol x$ is interpreted as the density operator of the subsystem $\rho(t)$ and 
the scalar product $\left\langle \boldsymbol x, \boldsymbol y \right\rangle$ between two vectors $\boldsymbol x$ and $\boldsymbol y$ is interpreted as the Froboenius product between operators $X$ and $Y$, 
$\Tr\{ X^\dagger Y\}$.\cite{Poetz3}

The objective is formulated by means of a cost functional for which 
we choose the general form,
\bea
\label{eq:obj_1}
J(\varepsilon,t_f) & =& \mbox{Tr}\left\{\Phi_{o}(\rho(t_f)),{\dot \rho(t_f)}\right\} \nonumber \\
& + &
\int_{0}^{t_f}\mbox{Tr}\left\{\Phi(\rho(t),\varepsilon(t),{\dot \rho(t)},{\dot \varepsilon(t)},t)\right\}\,dt .
\eea
The time derivatives (``generalized velocities") ${\dot \rho(t)}$ and  ${\dot \varepsilon(t)}$ are included to preserve the analogy to the formalism of classical mechanics,  as well as to allow for conditions on the rate of change of the state of the system and a means 
for suppressing unphysically rapid variations of the control field in the cost functional explicitly.
The real--valued functionals $\Phi_{o}(\rho(t_f),{\dot \rho(t_f)})$ and $\Phi(\rho(t),\varepsilon(t),{\dot \rho(t)},{\dot \varepsilon(t)},t)$ 
are  bounded from below and 
continuously differentiable with respect to their arguments.  They account for the specific 
physical objective  at target time $t_f$ 
and intermediate times $t\in [0,t_f)$.   The target time itself, in general,  is variable in the optimization  process.\cite{Bryson75}
The dependence of $\Phi$ on $\varepsilon(t)$ and ${\dot \varepsilon(t)}$ allows the implementation of additional constraints on the control, such as 
shape, duration, rate of change, or intensity.  This may be essential to arrive at solutions which are experimentally feasible.  The (indirect) dependence of $J$ on ${\dot{\rho}}$ is included mostly for analogy to  classical mechanics.  
In principle, specification of the rate of change of $\rho$ may be useful.  For example, state trapping can be treated by
implementation of the condition ${\dot{\rho}}(t)=0,\;t\in[0,t_f]$.  However, for a quantum subsystem the state of the system is fully determined by $\rho$ and its kinetic equation is of first order in time.  In contrast, for classical mechanics the state of the system is  specified by generalized coordinates and velocities and the kinetic equations for the coordinates are of second order in time.  Hence, simultaneous specification of $\rho$ and ${\dot{\rho}}$ may over--determine the system of equations.

An optimum control field is one which minimizes the cost functional under the constraint that $\rho(t)$ obeys the kinetic equation.   
Thus the control field  $\varepsilon(t)$ represents independent variables and the density matrix elements play the role of dependent variables.  
In the following we will formulate the necessary conditions using an indirect method   
which provides the gradient of the cost functional with respect to  control field and target time.  

The total differential of the cost functional is given by,
\begin{widetext}
\bea
\label{varJ}
d J(\varepsilon,t_f) & = & \mbox{Tr}\left\{\left.\left[\frac{\delta \Phi_{o}}{\delta \rho} + 
\frac{\delta \Phi}{\delta {\dot\rho}}\right]\right|_{t_f} d\rho(t_f)+
\left.\frac{\delta \Phi_{o}}{\delta {\dot\rho}}\right|_{t_f} d{\dot\rho}(t_f)+ \left.\frac{\delta \Phi}{\delta {\dot\varepsilon}}\right|_{t_f} d\varepsilon(t_f)\right\}\\ \nonumber
&+& 
\Tr \left.\left\{ \frac{\delta \Phi_o}{\delta t}+ \Phi(\rho,\varepsilon,{\dot\rho},{\dot \varepsilon},t) - 
\frac{\delta \Phi}{\delta {\dot\rho}}{\dot\rho}-\frac{\delta \Phi}{\delta {\dot\varepsilon}}{\dot\varepsilon}\right\}\right|_{t_f}dt_f
-\left.\frac{\delta \Phi}{\delta {\dot\varepsilon}}\right|_0 d\varepsilon(0)
\\\nonumber
&+& \int_{0}^{t_f}\,dt\Tr\left\{\left[\left.\frac{\delta \Phi}{\delta \rho}\right|_t \delta \rho(t)- \frac{d}{dt}\left.\left(\frac{\delta \Phi}{\delta {\dot\rho}}\right)\right|_t\right] \delta \rho(t)+ \left[\left.\frac{\delta \Phi}{\delta \varepsilon}\right|_t 
- \left.\left(\frac{d}{dt}\frac{\delta \Phi}{\delta {\dot \varepsilon}}\right) \right|_t\right] \delta \varepsilon(t)
\right\}.
\eea
\end{widetext}
Here the variation $\delta \rho(t)$ is dependent upon the variation of $\varepsilon(t')$, for $t' < t$ via the kinetic equation Eq.~\eqref{KE}. 
${\dot \varepsilon}=\frac{d\varepsilon}{dt}$ and ${\dot \rho}=\frac{d\rho}{dt}$.

Since it is assumed that the constraint is holonomic, one may use either the general method of Lagrangean multipliers 
or Hamilton's variation principle to derive the necessary conditions for an extremum of the cost functional.\cite{Grei03} 
In the first derivation given below, the dependence of $\rho(t)$ on $\varepsilon(t)$ is incorporated by a Lagrangean multiplier which,  in this context, is termed co-state or adjoint state.   
The variation is made with respect to the generalized coordinates $\rho$ and $\varepsilon$ and the associated velocities 
${\dot\rho}$ and ${\dot\varepsilon}$.
The second, equivalent formulation offered here is based on Hamilton's variational principle which uses both $\varepsilon(t)$ and $\rho(t)$ and their  canonically conjugated variables (``canonical momenta") as  variation parameters.  In optimization theory, the latter method is known as Pontryagin's minimum principle.~\cite{Betts} 

\subsubsection{Minimality conditions via the Lagrangean multiplier technique}

The Lagrangean multiplier method is a powerful tool for incorporation of general constraints, holonomic or non--holonomic,  into an extremum problem.\cite{Grei03} In order to establish a tractable relation between the variation of the control $\varepsilon(t')$ and the 
density operator $\rho(t)$ a Lagrangean multiplier $\lambda(t)$ is introduced to implement the kinetic equations 
into the variation of the cost functional.  For each constraint a Lagrangean multiplier is introduced and an extended cost functional is constructed,
\bea
\label{Jhat}
{\hat J}&=&{\hat J}(\varepsilon,t_f)\equiv  J(\varepsilon,t_f) \\
&+&\int_{0}^{t_f}dt \Tr\{\lambda(t)\left[{
\int_0^{t} dt' K(t,t';\varepsilon, \rho)-\dot \rho(t)} \right]\}. \nonumber
\eea
$\lambda(t)$ is a linear operator in the Hilbert space of the system but it does not have the properties of a density operator.   The components of $\lambda(t)$ are chosen such that the variation with respect to the dependent variables vanishes.  Note there are exactly as many constraints 
as there are density matrix elements.    
The kinetic equation for $\rho(t)$ results from the condition of stationarity of ${\hat J}$ with respect to variation of $\lambda$. Implementing causality of the kernel, {\it i.e.}, that $K(t,t';\varepsilon,\rho)$ depends on 
$\varepsilon(t'')$ only for $t''\leq t$, using 
$\delta\rho(t_f)=d\rho(t_f)-{\dot\rho}(t_f) dt_f$ and $\delta\varepsilon(t_f)=d\varepsilon(t_f)-{\dot\varepsilon}(t_f) dt_f$, the total  differential  of ${\hat J}$ gives,  after integration by parts,
\bea
\label{varJhat}
d {\hat J}(\varepsilon,t_f) & = & d J(\varepsilon,t_f) -\Tr\left\{ \lambda(t_f) d\rho(t_f) -\lambda(t_f){\dot \rho}(t_f)dt_f\right\} \nonumber \\
& + & \int_{0}^{t_f}dt\int_{0}^{t_f}dt'\Theta(t-t')\Tr\left\{\left[\vphantom{\frac{1}{2}}
\delta(t_+ -t'){\dot \lambda}(t')\right.\right. \nonumber \\
&+&\left.\left. \lambda(t)\left.\frac{\delta K(t,t';\varepsilon,\rho)}{\delta \rho}\right|_{t'}\right]\delta\rho(t')\right. \\
&+&\left.\int_{0}^{t_f}dt''\Theta(t-t'')\lambda(t)\frac{\delta K(t,t';\varepsilon,\rho)}{\delta \varepsilon(t'')}\delta \varepsilon(t'')  \right\}.
\nonumber
\eea
Variation with respect to $\rho$,  $\varepsilon$, and $t_f$, respectively, gives the following necessary conditions,
\begin{widetext}
\begin{equation}
\label{Vlambda}
\frac{d}{dt}\left(\lambda(t')-\left.\left(\frac{\delta \Phi}{\delta {\dot\rho}}\right)\right|_{t'}\right) = -\int_{t'}^{t_f}dt \lambda(t)\left.\frac{\delta K(t,t',\varepsilon,\rho)}{\delta \rho}\right|_{t'} - \left.\frac{\delta \Phi}{\delta \rho}\right|_{t'},\; \lambda(t_f)-\left.\frac{\delta \Phi}{\delta {\dot\rho}(t)}\right|_{t_f}=\left.\frac{\delta \Phi_o}{\delta \rho(t)}\right|_{t_f}.
\end{equation}
\be
\label{Ve}
\frac{\delta J(\varepsilon,t_f)}{\delta \varepsilon(t'')}= 
\int_{0}^{t_f}dt\int_{0}^{t_f}dt'\Theta(t-t')\Theta(t-t'')\Tr\left\{\lambda(t)
\frac{\delta K(t,t';\varepsilon,\rho)}{\delta \varepsilon(t'')}\right\} + \Tr\left\{\left.\frac{\delta \Phi}{\delta \varepsilon}\right|_{t''}-\frac{d}{dt}
\left.\frac{\delta \Phi}{\delta {\dot\varepsilon}}\right|_{t''}\right\}=0,
\ee
\end{widetext}
\be
\label{Vt}
\left.\frac{\delta J(\varepsilon,t_f)}{\delta t}\right|_{t_f}= 
\Tr\left\{ \left(\frac{\delta \Phi_o}{\delta t} + 
\Phi(\varepsilon,{\dot\varepsilon},\rho,t)+\left.\frac{\delta \Phi_o}{\delta{\rho}}{\dot\rho}\right)\right|_{t_f}\right\},
\ee
\be
\left.\frac{\delta \Phi}{\delta{\dot\varepsilon}}\right|_{t_f}=\left.\frac{\delta \Phi}{\delta{\dot\varepsilon}}\right|_{t=0}=0,
\ee
and
\be
\label{rdot}
\frac{\delta J(\varepsilon,t_f)}{\delta {\dot\rho}}= 
\Tr\left\{ \left.\frac{\delta \Phi_o}{\delta {\dot\rho}}\right|_{t_f}\right\}.
\ee

We  conclude this formulation with a few comments: 

(i) The optimality problem of dynamic control of a quantum system has been 
reformulated as a coupled set of two initial--value problems embedded in an iterative scheme: For given control and target time $t_f$ 
the initial--value problem of finding $\rho(t)$ for given $\rho(0)$ is solved.   Then the co--state is determined from the initial--value problem Eq.  \eqref{Vlambda}
starting at time $t_f$ and going back to time zero.  Finally, the gradients of the cost functional  $J$ with respect to control and target time are computed
to aid the approach towards  a minimum of $J$ using a suitable numerical procedure.  

(ii) Here we have used a continuum notation.   Numerical implementations, however, will use a time grid and care must be taken  to use a 
consistent grid for density operator, co--state, and the gradients to ensure optimal numerical efficiency.  

(iii) In the  relations above we have used step 
functions $\Theta(t)$ which ensure causality for clarity,only.  In fact, physical kernels will ensure causality on their own. 

(iv) When the cost functional is chosen to be independent of the ``velocities" ${\dot \rho}(t)$ and
${\dot \varepsilon}(t)$,  the Euler--Lagrange equations reduce to  $\frac{\partial L(q,t)}{\partial q_i}=0$, where $q_i$ stands for the components of 
control and density operator.    
 
\subsubsection{Minimality conditions via Hamilton's variation principle (Pontryagin's principle)}

Hamilton's variation principle applied to the action, whereby the integrand is interpreted as the Legendre--transformed 
Lagrange function, yields the canonical equations of motion of a classical mechanical system.\cite{Grei03}  
The expression for 
${\hat J}$, Eq.(\ref{Jhat}), lends itself to this procedure if one rewrites it as,
\begin{widetext}
\be
\label{Jhat1}
{\tilde J}(\rho,\varepsilon,\lambda,p,t_f)={\hat J}+\int_{0}^{t_f}dt p(t)\left( \frac{d\varepsilon}{dt}-{\dot \varepsilon}(t)\right) = \mbox{Tr}\left\{\Phi_{o}(\rho,{\dot\rho},t_f)\right\} + 
\int_{0}^{t_f} dt\Tr\{H(\varepsilon,\rho,\lambda,t)-\lambda(t){\dot \rho}(t) -p(t){\dot \varepsilon}(t)\},
\ee
where,
$$
H(\varepsilon,\rho,\lambda,t)\equiv \Tr\left\{\Phi(\rho(t),{\dot\rho}(t),\varepsilon(t),{\dot\varepsilon}(t),t) + p(t)\frac{d \varepsilon}{dt} + \lambda(t) \int_0^{t} dt' K(t,t';\varepsilon, \rho)\right\}.
$$
\end{widetext}
Written in this form, $\lambda(t)$ and $p(t)$ play the role of the canonical momenta, respectively,  associated with the variables $\rho(t)$
and $\varepsilon(t)$.
Variation is carried out independently with respect to the ``generalized coordinates" $\rho(t)$ and $\varepsilon(t)$ and 
``canonical momenta" $\lambda(t)$ and $p(t)$.    After integration by parts,  one obtains the following necessary conditions for an extremum of ${\hat J}$:

\begin{widetext}
\be
\label{Hro}
{\dot \rho}(t)= \frac{\partial H(\varepsilon,\rho,\lambda,t)}{\partial \lambda(t)}, \mbox{  with } 
\rho(0)=\rho_o,\;  {\dot \varepsilon}(t)=\frac{\partial H(\varepsilon,\rho,\lambda,t)}{\partial p(t)}=\frac{d\varepsilon}{dt}, 
\ee

\be
\label{Hlambda}
{\dot \lambda}(t)= -\int_{0}^{t_f}dt' \frac{\partial H(\varepsilon,\rho,\lambda,t')}{\partial \rho(t)}\Theta(t'-t), \mbox{  with } \lambda(t_f)=\left.\frac{\delta \Phi_o}{\delta \rho(t)}\right|_{t_f},
\ee
\bea
\label{He}
\frac{\partial {\tilde J}}{\partial \varepsilon(t')}=\frac{\partial  J}{\partial \varepsilon(t')}={\dot p}(t') + \int_{0}^{t_f}dt\Theta(t-t')\frac{\partial H(\varepsilon,\rho,\lambda,t)}{\partial \varepsilon(t')}=0,\; \mbox{ with } p(t_f)=p(0)=0, 
\eea
\be
\left.\frac{\delta \Phi}{\delta{\dot\varepsilon}}\right|_{t_f}=\left.\frac{\delta \Phi}{\delta{\dot\varepsilon}}\right|_{t=0}=0.
\ee
\end{widetext}

These equations may be interpreted as ``canonical equations of motion"  (Hamilton's equations of motion) 
generalized to systems with a non--Markovian dependence on ``generalized coordinates".  
 Eqs. (\ref{Hro}) and (\ref{Hlambda}), respectively, give the kinetic equations for the density operator and co--state, the latter playing the role of the canonical momentum of $\rho$.   Eq. (\ref{He}) gives an implicit relation for the optimal control field in terms of 
the solutions $\rho(t)$ and $\lambda(t)$, as well as the gradient of the cost functional $J$ with respect to variation of $\varepsilon(t)$.  Inserting the definition of $H$ into Eqs.~(\ref{Hro}), (\ref{Hlambda}), and (\ref{He}), respectively, gives a system of optimality conditions which is equivalent to Eq.~(\ref{KE}), Eq.~(\ref{Vlambda}), Eq.~(\ref{Vt}), and Eq.~(\ref{Ve}).  

Kernels which are linear in the density operator and fixed target time represent a specially important case,
$$
K(t,t'\varepsilon,\rho)=k(t,t',\varepsilon)\rho(t')+ d(t,t',\varepsilon).
$$
If, for example, we consider driving of the subsystem along a desired trajectory 
$\rho_o(t)$, with $\rho_o(0)=\rho(0)=\rho_o$ and $\rho_o(t_f) = \rho_f$ for fixed $t_f$, the following cost functional,
\bea\label{CNM}
J(\varepsilon)&  = & \frac{w_1}{2}\Vert\rho(t_f)-\rho_o\Vert^2 + \frac{w_2}{2t_f}\int_0^{t_f}\,dt\,\Vert\rho(t)-\rho_o(t)\Vert^2\   \nonumber \\
& + & \frac{1}{2}
\int_{0}^{T}\alpha(t)\left|\varepsilon\right|^2(t)\,dt,
\eea
is useful.
Here $\Vert A\Vert\equiv \Tr\{A A^\dagger\}$  is the Froboenius norm and $\alpha(t)$, $w_1$, and $w_2$, with $w_1+w_2=1$ are real-valued weight factors to specify driving ($w_1=1$) and trapping ($w_2=1$). $\alpha(t)$ 
is real-valued and can be used to taylor the control pulse shape by penalizing high intensity.  In case of certain linear control problems the third term is necessary to make the problem regular.\cite{Krotov96}

For this case the optimality conditions are,
\begin{widetext}
\be
\label{Vlambda-lin}
{\dot \lambda}(t') = -\int_{t'}^{t_f}dt \lambda(t) k(t,t',\varepsilon)  - w_2(\rho(t')- \rho_o(t')), \; \lambda(t_f)=w_1(\rho(t_f)-\rho_o),
\ee
\be
\label{Ve-lin}
\frac{\delta J(\varepsilon,t_f)}{\delta \varepsilon(t'')}= 
\int_{0}^{t_f}dt\int_{0}^{t_f}dt'\Theta(t-t')\Theta(t'-t'')\Tr\left\{\lambda(t)\left[
\frac{\delta k(t,t';\varepsilon)}{\delta \varepsilon(t'')}\rho(t')+ 
\frac{\delta d(t,t';\varepsilon)}{\delta \varepsilon(t'')}\right]
\right\} + \alpha(t'')\varepsilon(t'').
\ee
\end{widetext}

\subsubsection{Application}

\begin{figure}
\includegraphics[width=8cm]{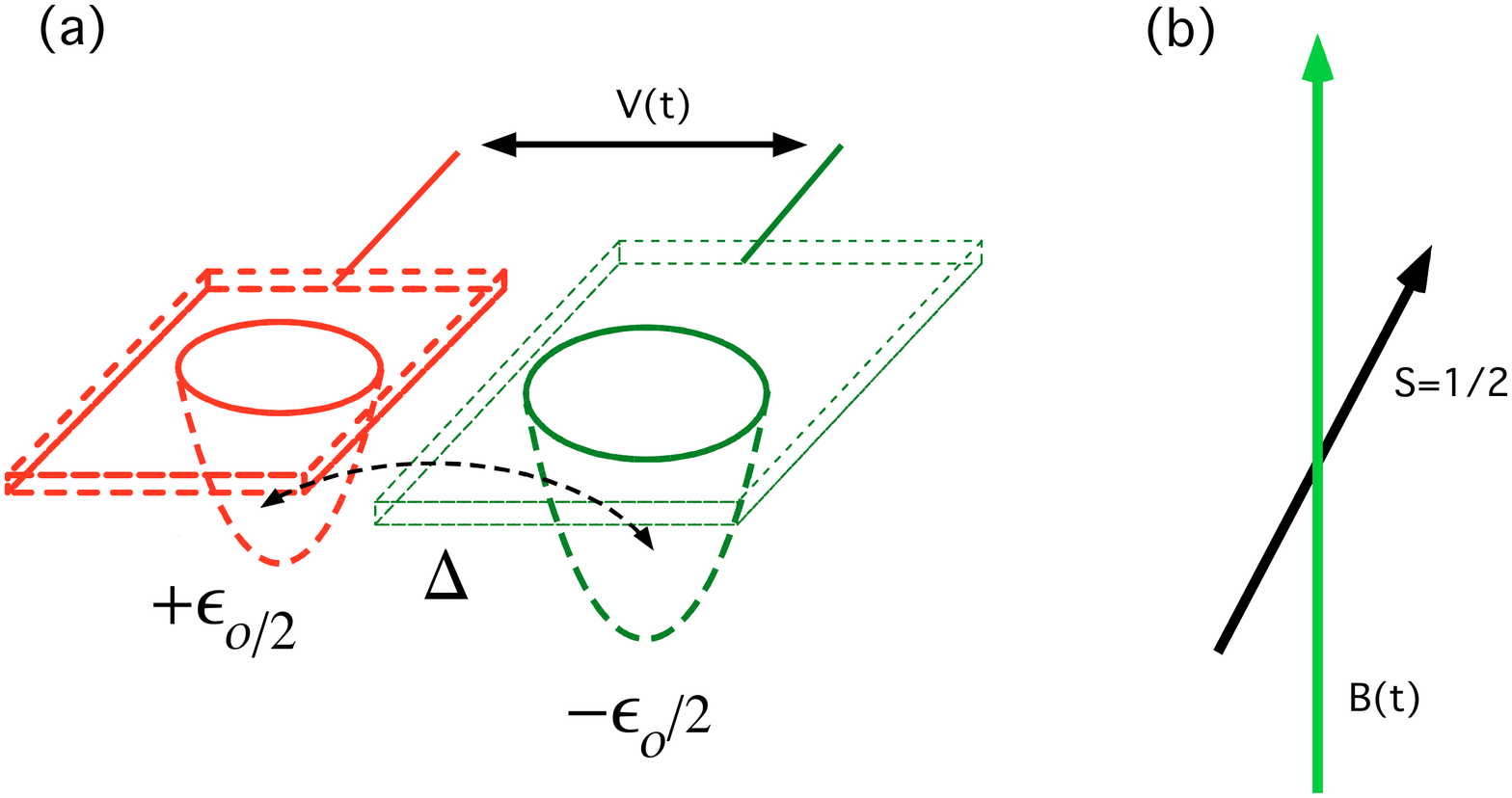}
\caption{\label{D-DOT} Two realizations of a dissipative qubit: (a) bias--controlled location of an electron in a semiconductor double dot;
(b) spin 1/2 orientation controlled by a magnetic field.}
\end{figure}

In an effort to demonstrate optimal control  by quantum interference for a non--Makovian 
quantum system, the two--level spin--boson model with $\sigma_z$ coupling to both bosons (phonons) and control field and 
constant $\sigma_x$ coupling between the two levels was employed.\cite{Poetz06,WP}  Examples for physical realizations of this model are shown in Fig. \ref{D-DOT}.

The polaron--transformed 
Hamiltonian may be written,~\cite{Leg87}  
\be
\label{eq:spinbosonH_Trans}
H'_{\rm tot}=H_S(t)+ H_B+ H_{\rm int}\,.
\ee
$H_S(t)=-\frac{\hbar}{2}\left(\varepsilon_0+\varepsilon(t)\right)\sigma_z$ is the new Hamilton operator of the driven qubit, and 
\be
H_{\rm int}=-\frac{1}{2}\hbar\Delta\left(\sigma_+ e^{-i\Omega}+\sigma_- e^{i\Omega}\right)\, ,
\ee
gives the new interaction which is now proportional to $\Delta$, renormalized by the electron--phonon interaction. Here, 
$\sigma_+=(\sigma_x+i\sigma_y)/2$ and $\sigma_-=(\sigma_x-i\sigma_y)/2$ and  $\Omega =\sum_i\Omega_i,\quad
\Omega_i= 
\left(q_0c_i/\hbar m_i\omega_i^2\right)p_i$. 

\begin{figure}
\includegraphics[width=6cm,angle=-90]{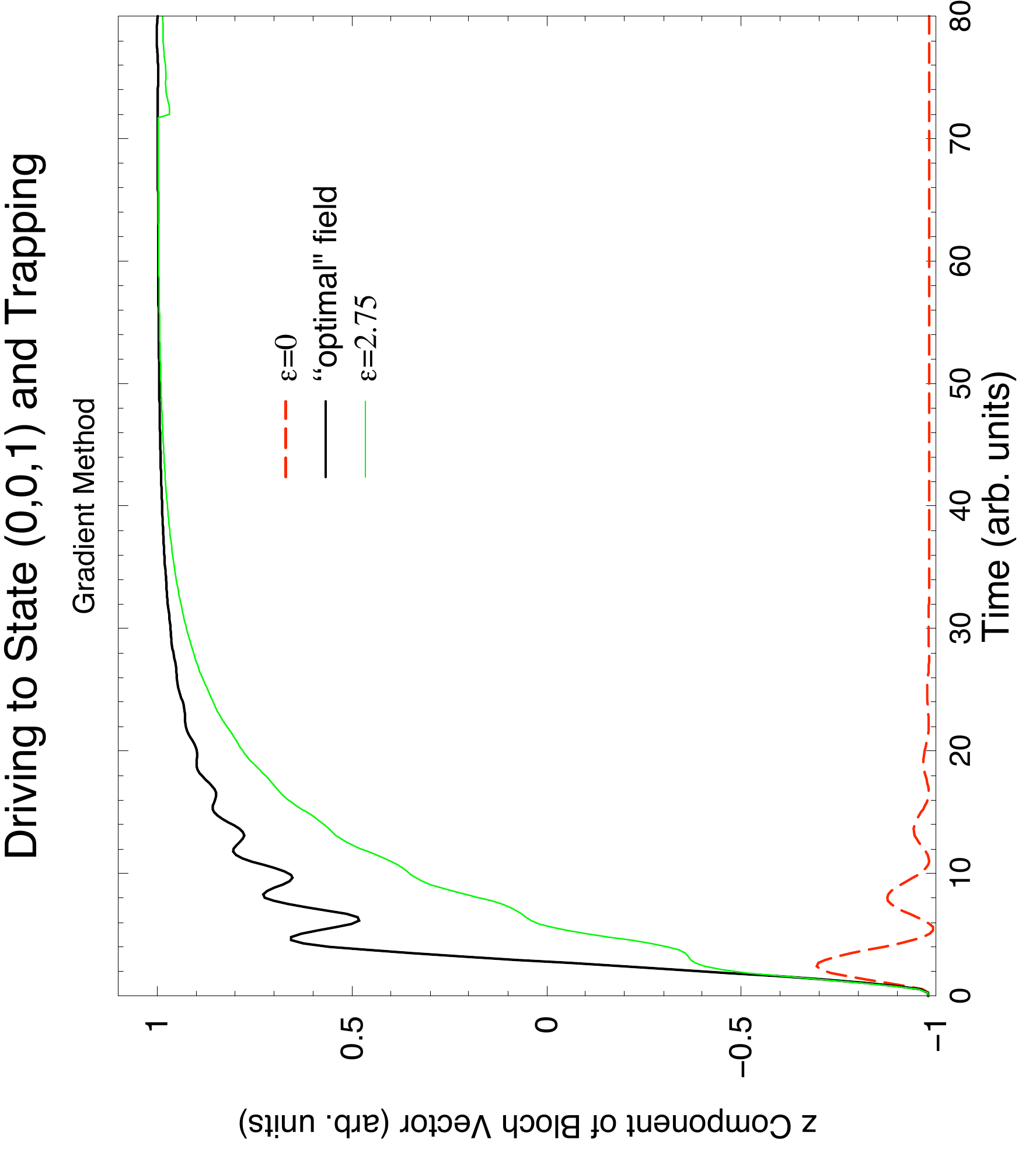}
\caption{\label{popz-easy} Driving a qubit from its ground state into the ``up" state and trapping:  red dashed line: ground state (control field=0); solid green line: step pulse (optimized constant field); black solid line: indirect method.
Parameters: $\varepsilon=-1$,$\Delta=0.75$, $\omega_c=4$, $T=0.2$, $\alpha=0.216$. From Ref.~\onlinecite{WP}.}
\end{figure}

\begin{figure}
\includegraphics[width=6cm,angle=-90]{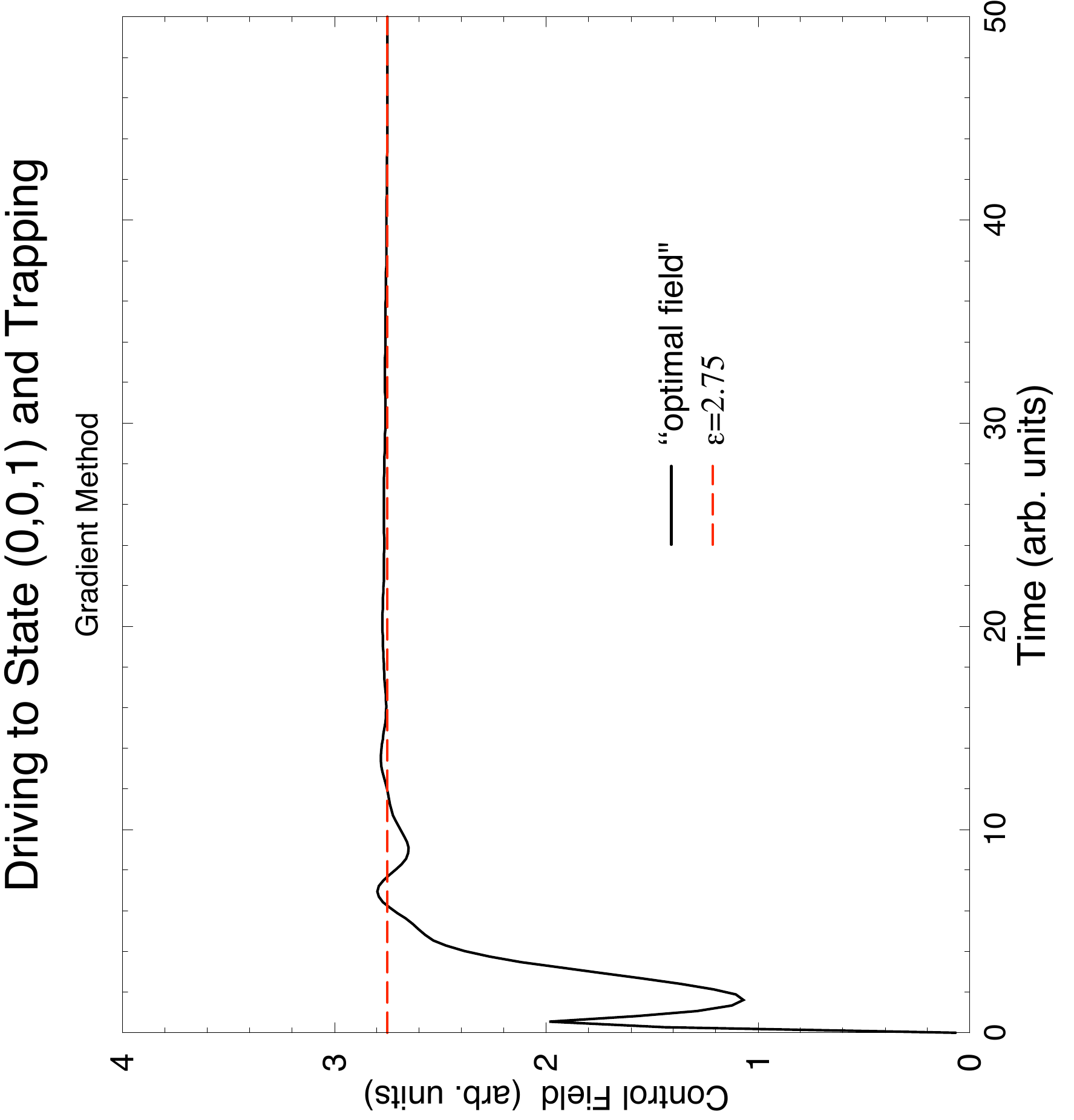}
\caption{\label{E-trapz-easy}Control field for driving a qubit from its ground state into the ``up" state and trapping:  solid black line: indirect method; dotted red line: step pulse (optimized constant field). From Ref.~\onlinecite{WP}.}
\end{figure}

A non-Markovian kinetic equation is readily obtained within the  Nakajima--Zwanzig projection--operator method.~\cite{Breu03} Applied to the present model, the kinetic equations for the Bloch vector ${\bf R}$ up to second order in $\Delta$ and arbitrary spin--phonon coupling strength take the form,\cite{Poetz06} 
\be
{\dot {\bf  R}}(t)=  M_\varepsilon(t){\bf  R} + \int_{0}^{t}dt' K(t,t'){\bf  R}(t')+ {\bf \Gamma}(t),
\label{eq:rdyn}
\ee
where,
\bea
M_\varepsilon(t)=\left( \begin{array}{ccc} 0 & (\varepsilon_o+\varepsilon(t)) & 0  \\ -(\varepsilon_o+\varepsilon(t)) & 0 &  0\\ 0 & 0& 0\end{array}\right),
\eea
with the kernel,
\bea
K(t,t') & = &  \Delta^2 e^{-Q_2(t-t')}\cos Q_1(t-t')\nonumber \\
& \times & \left(\begin{array}{ccc} 0 & 0 & 0 \\ 0 & -1 & 0 \\ 0 & 0 & -\cos(f(t,t'))\end{array}\right),
\eea
and
${\bf \Gamma}(t)=( 0,0,- \Gamma_{o}(t))$, with,
\be\label{Gamma0}
\Gamma_{o}(t)=\Delta^2\int_0^t dt' e^{-Q_2(t-t')}\sin (f(t,t'))\sin (Q_1(t-t')),
\ee
and 
$$
f(t,t')=\varepsilon_0(t-t') + \int_{t'}^{t}dt''\varepsilon(t''). 
$$
An Ohmic bath with phonon cut--off frequency $\omega_c$ is chosen.~\cite{Leg87}  
Using the cost functional Eq. \eqref{CNM} containing two real positive weight factors $w_i$, several objectives were posed:~\cite{Poetz06,WP}  
``Instantaneous" population transfer and subsequent trapping ($w_1=w_2=1/2$) is illustrated in Fig.~\ref{popz-easy}
with corresponding control fields given in Fig.~\ref{E-trapz-easy}.  Model parameters are given in the caption. 
Oscillations seen for the control--free case (red line in Fig.~\ref{popz-easy}) are a signature of non--Markovian behavior: the system is released in its ground state (in thermal equilibrium with its bath) but ``does not know it ".  Only after probing its environment it settles  into this state.


\begin{figure}
\includegraphics[width=6cm,angle=-90]{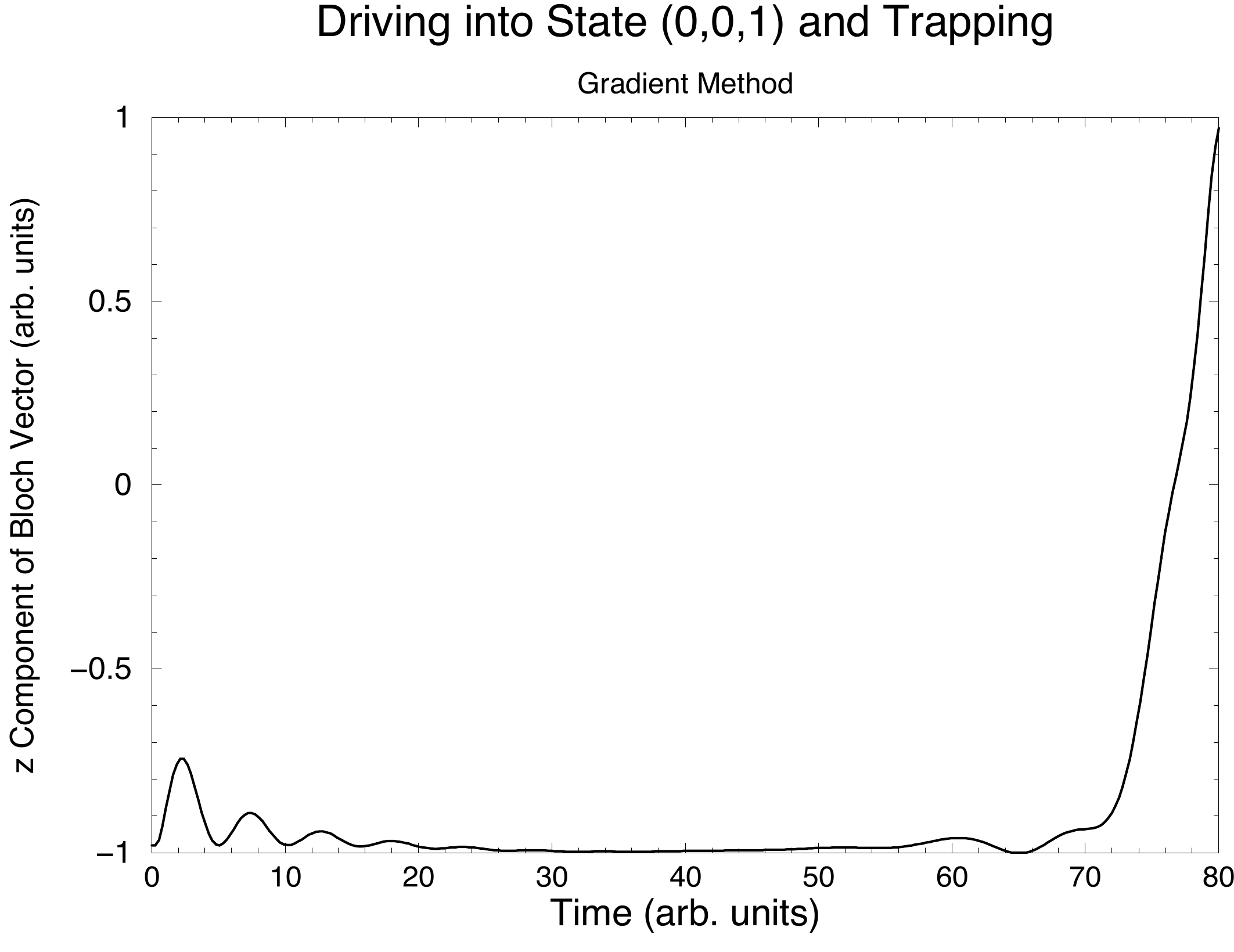}
\caption{\label{pop-drive}
Optimized control for driving a qubit from its ground state into the ``up" state at given target time. From Ref.~\onlinecite{WP}.}
\end{figure}

\begin{figure}
\includegraphics[width=6cm,angle=-90]{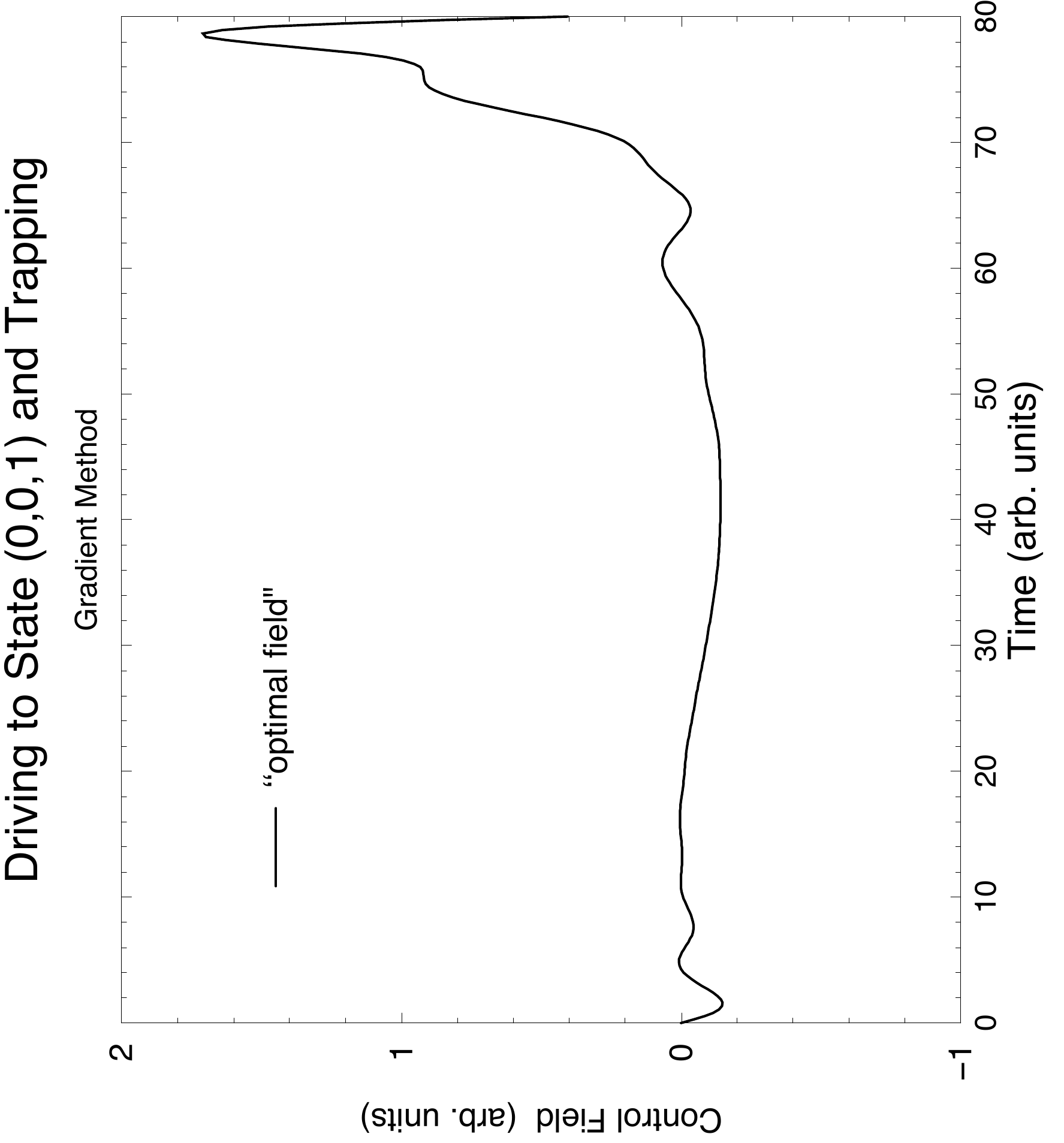}
\caption{\label{ES-drive}Driving a qubit from its ground state into the ``up" state at given target time $t_f=80$.
(Intensity minimization). $\varepsilon=-1$, $\Delta=0.75$, $\omega_c=4$, $T=0.2$, $\alpha=0.216$. From Ref.~\onlinecite{WP}.}
\end{figure}

The example of population transfer at target time $t_f$  ($w_1=1, w_2=0$) is illustrated in Fig.~\ref{pop-drive}
with corresponding control fields given in Fig.~\ref{ES-drive}.  Again, model parameters are given in the caption. The ``last--minute switching" strategy is clearly evident.

%
\begin{figure}
\includegraphics[width=6cm,angle=-90]{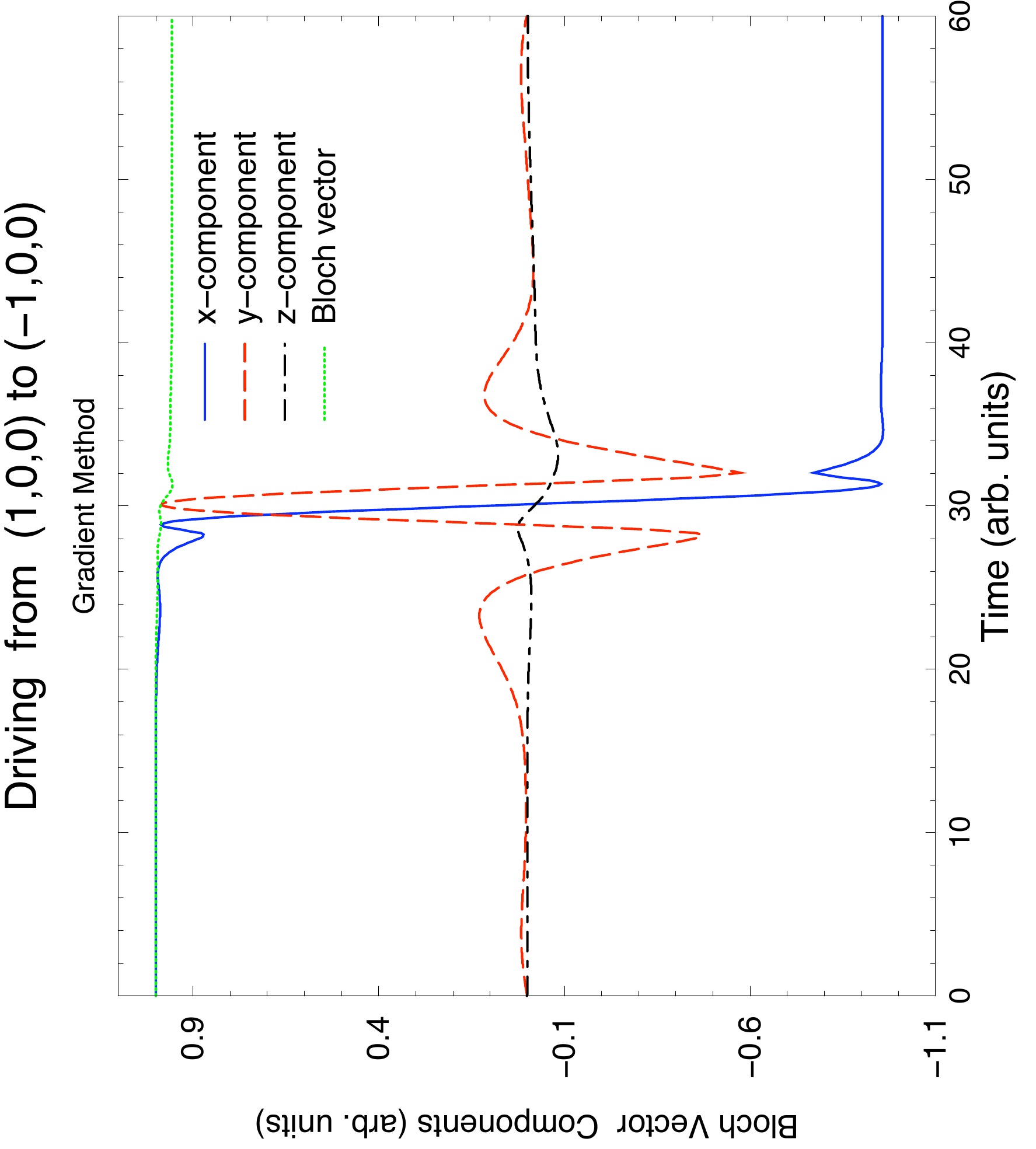}
\caption{\label{popxyz}Flipping the Bloch vector from state $(1,0,0)$ into $(-1,0,0)$.
$\varepsilon=-1$,$\Delta=0.25$, $\omega_c=0.5$, $T=0.5$, $\alpha=0.25$. From Ref.~\onlinecite{WP}.}
\end{figure}

\begin{figure}
\includegraphics[width=6cm,angle=-90]{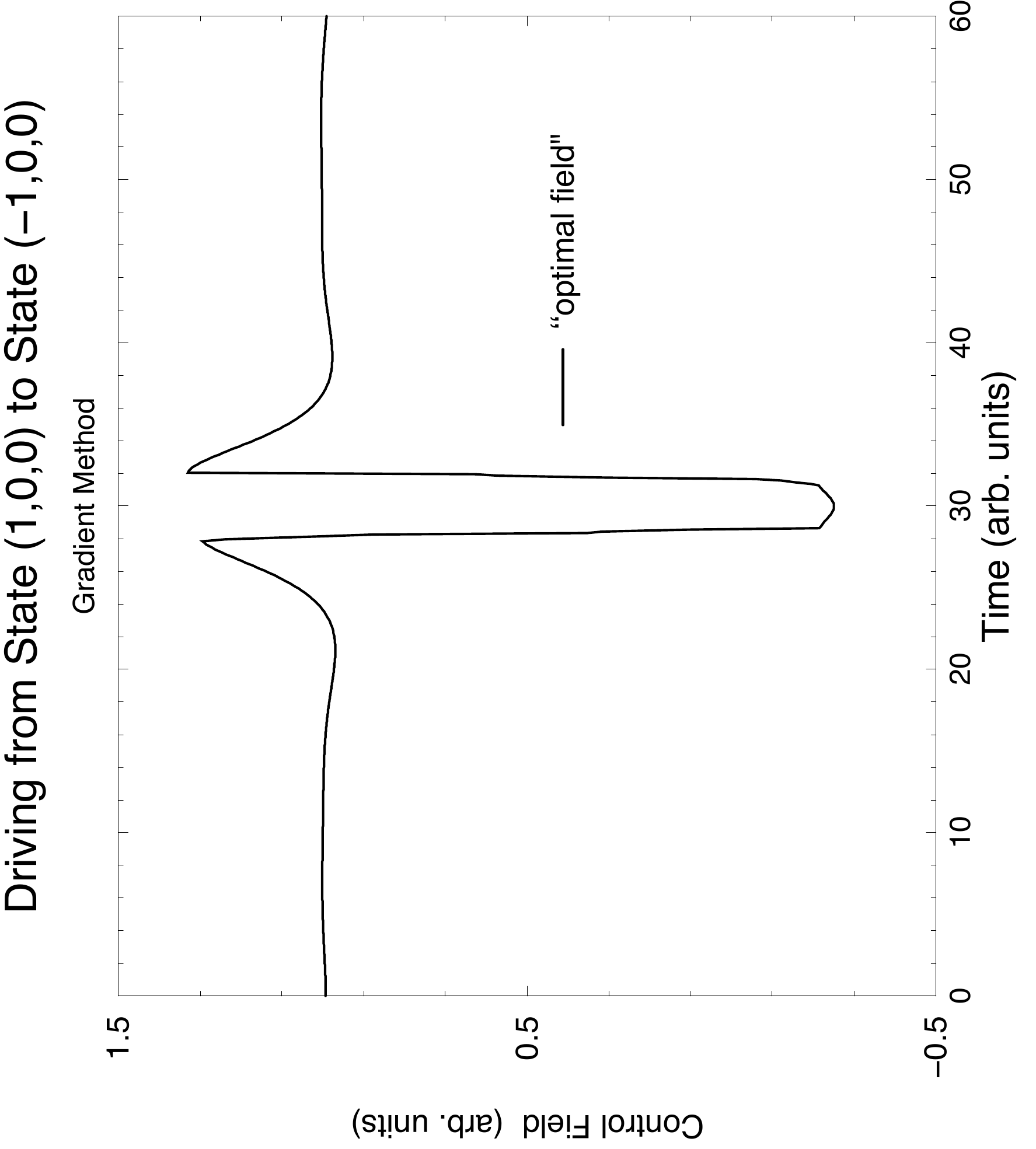}
\caption{\label{ESxyz}Flipping the Bloch vector from state $(1,0,0)$ into $(-1,0,0)$.
$\varepsilon=-1$,$\Delta=0.25$, $\omega_c=0.5$, $T=0.5$, $\alpha=0.25$. From Ref.~\onlinecite{WP}.}
\end{figure}

%
The third task is a flipping of the Bloch vector from a stable state into another.  Results are  illustrated in Fig.~\ref{popxyz} and Fig.~\ref{ESxyz}. Detailed analysis of this case shows that the control decisively adjusts the effective coupling to minimize dissipative losses.  In some cases multiple (rather than single) switching has been found to lead to best results.~\cite{Poetz06}

\begin{figure}
\centering
\includegraphics[width=7.5cm,angle=-90]{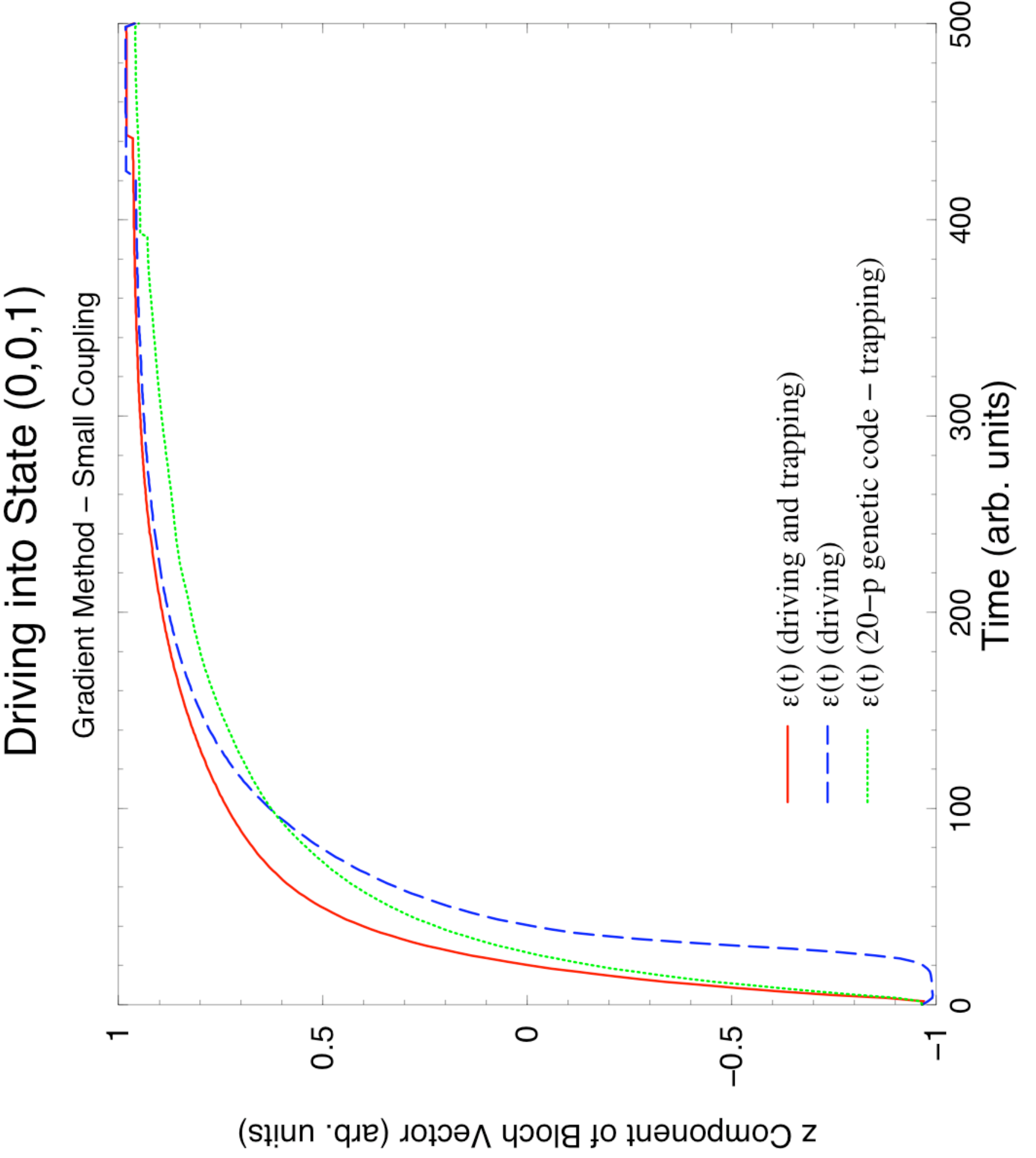}%
\caption{\label{popz-smD} Driving of the spin--boson system from thermal equilibrium at $z_i=-0.96$  to $z=-1$ (target state) and subsequent trapping. $\varepsilon_o=-2$, $\Delta=0.25$, $\eta=0.45$, temperature $T=\beta^{-1}=0.5$, and $\omega_c=2$.}
\end{figure}

\begin{figure}
\centering
\includegraphics[width=7.5cm,angle=-90]{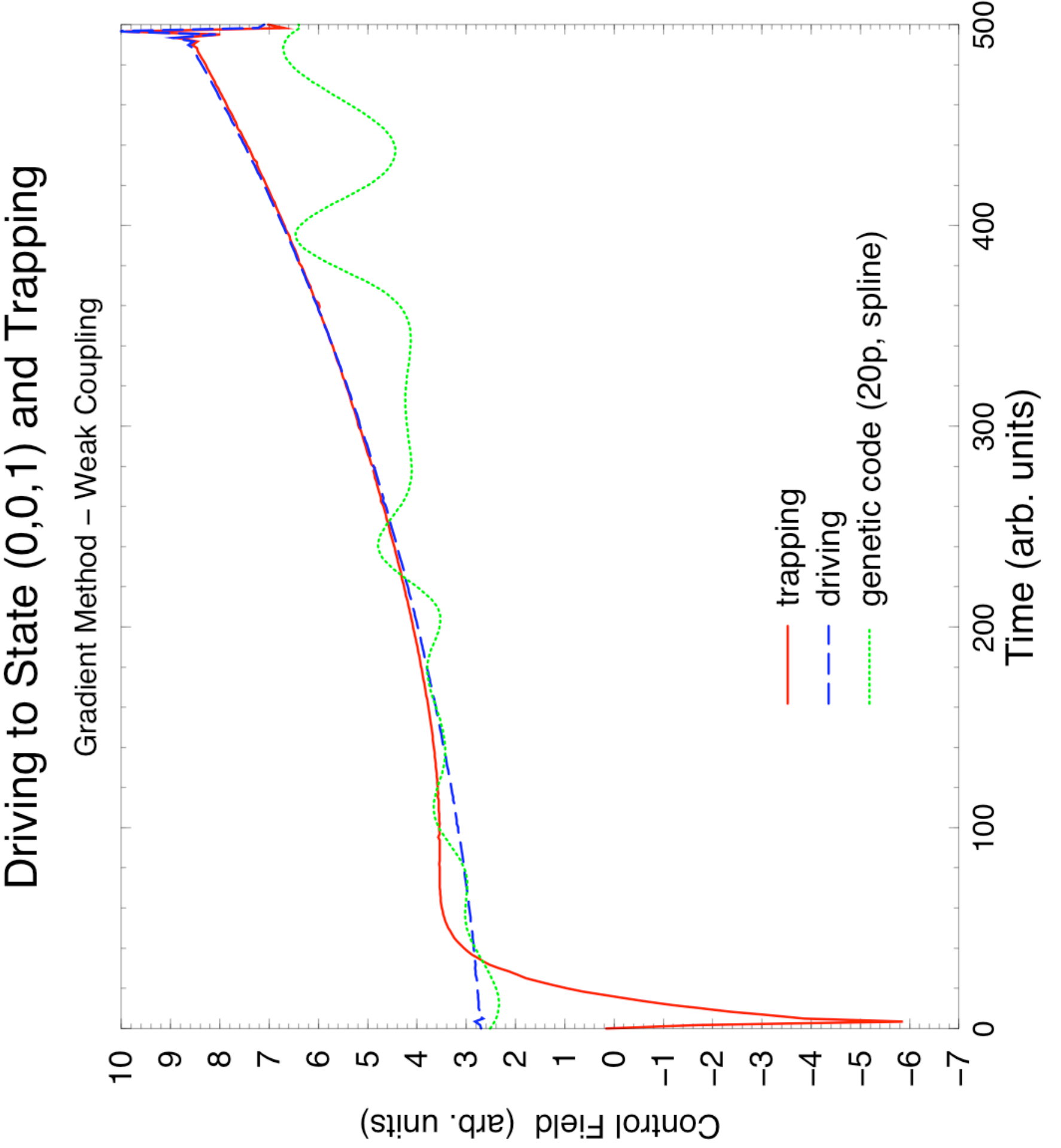}%
\caption{\label{ES-weak} Selected optimal control fields $\varepsilon$ for driving the spin--boson system from 
thermal equilibrium at $z_i=-0.96$ to $z=-1$ (target state) and subsequent trapping.
$\varepsilon_o=-2$, $\Delta=0.25$, $\eta=0.45$, temperature $T=\beta^{-1}=0.5$, and $\omega_c=2$.}
\end{figure}

\begin{figure}
\centering
\includegraphics[width=7.5cm,angle=-90]{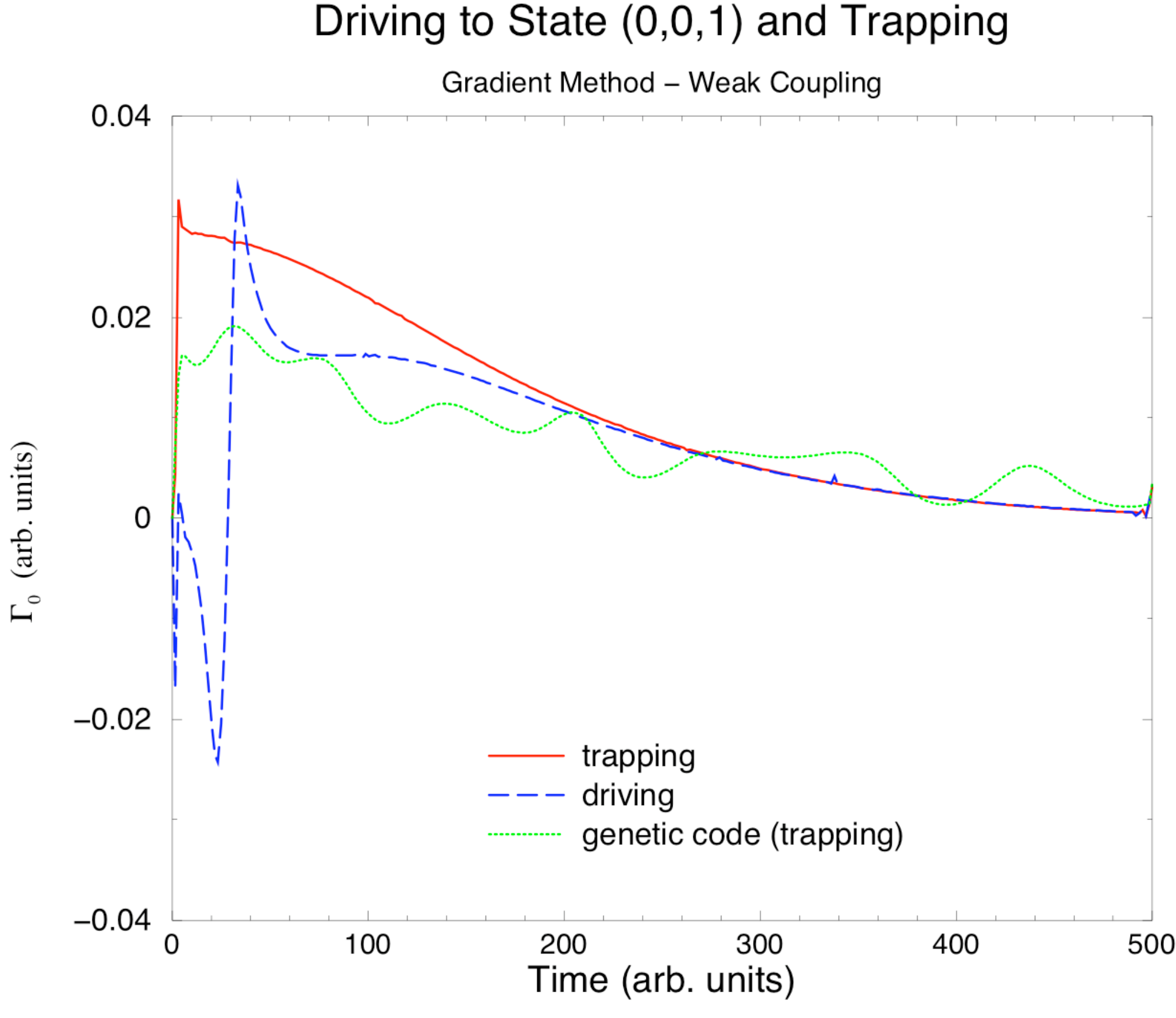}%
\caption{\label{rat0} Control of the inhomogeneous part of the kinetic equations Eq. \eqref{eq:rdyn} via optimal control fields $\varepsilon$ for driving of spin--boson system from 
thermal equilibrium at $z_i=-0.96$  to $z=-1$ (target state) and subsequent trapping.}
\end{figure}
 
Finally, we consider a weak coupling situation with $\varepsilon_o=-2$, $\Delta=0.25$,  $\eta=0.45$, temperature $T=\beta^{-1}=0.5$, and $\omega_c=2$.
In thermal equilibrium the Bloch vector is  ${\bf R} =(0,0,-0.96)$. The task is  to flip the spin into state ${\bf R} =(0,0,1)$ and to trap it there.   
In this case, the shortest possible flipping time is of the same order as $t_f$.
We consider three cases: driving and trapping $w_1=w_2=1/2$ in Eq.~\eqref{CNM}, pure driving $w_1=1, w_2=0$ using the conjugate gradient method, and a genetic code for driving and trapping $w_1=w_2=1/2$, whereby the control is represented by a 20--parameter spline function over the interval $[0,t_f]$. 
Results for the $z$-component of the Bloch vector are shown in Fig.~\ref{popz-smD}, while corresponding control fields are given in Fig.~\ref{ES-weak}. It is seen that quite different control field solutions provide similar results.  Fig.~\ref{ES-weak} shows $\Gamma_{o}(t)$ for the three different solutions, clearly demonstrating the influence of the control on the effective system--bath coupling.


\subsection{Application to electron spin in quantum dots}
Among the proposed physical implementations of a qubit the most prominent are based on superconducting devices,~\cite{Shnir97,Mak01} quantum dots,~\cite{DiVi98,Burk99} ion traps,~\cite{Cir95} nuclear spins in molecules~\cite{Ger97} and optical systems.~\cite{Knill01} Because of their potential scalability, solid state architectures seem promising candidates to build quantum information processing devices.\par
Here we review optimal control of the spin dynamics of an excess electron in a semiconductor quantum dot within the spin--boson model,  see Sec.~\ref{DISSI} and Ref.~\onlinecite{Rol07}.

The electron, as spin--$\frac{1}{2}$ particle, provides a natural two--level system. Trapping single electrons by means of semiconductor quantum dots enables one to use this quantum two level system as qubit.  The spin directions \textit{up} ($\ket{1}$) and \textit{down} ($\ket{0}$) with respect to an external magnetic field represent the computational basis states of the qubit. This implementation has first been proposed by Loss and DiVincenzo.~\cite{DiVi98} Quantum dots can be realized, for example, by means of surface gates on top of a GaAs/AlGaAs heterostructure which holds a two dimensional electron gas.  Controlling and monitoring the number of conducting electrons in each dot is possible by means of well--established experimental methods.~\cite{Elz} In a system consisting of two neighboring quantum dots, each populated by one excess electron, exchange interaction results in a Heisenberg--like Hamiltonian, $H_S^{(12)}=J(t)\vec S_1 \cdot \vec S_2$, if the dots are coupled via a tunable tunneling barrier.~\cite{Burk99,DiVi98} Two--qubit arrays of quantum dots are realized by extending heterostructures as described above with additional gate electrodes, thus defining an appropriate electric potential to trap several electrons at different sites.~\cite{Elz}\par
In this section, we first define the Hamilton operators for the double quantum dot, the environment, and the interactions between spin and bath. To describe the dynamics of electron spins in the double dot, a Markovian quantum master equation approach is used.~\cite{Car02,Breu03,Scully}\par
We examine two conduction electrons, each sitting in its own quantum dot, where each is described by a Hamiltonian of the form,
\begin{eqnarray}
H_S^{(i)}(t)&=&-g^{*}\frac{e}{2m_e} S_z B^{(i)}_z(t)=-\frac{g^{*}}{2}\mu_B B^{(i)}_z(t) \sigma_z^{(i)} \nonumber \\
&\equiv&-\hbar \tilde B^{(i)}_z(t) \sigma_z^{(i)},\quad i=1,2 \label{H_S} \\
\vec S^{(i)}&=& \frac{\hbar}{2} \vec \sigma^{(i)}.
\end{eqnarray}
$g^{*}$ denotes the gyromagnetic ratio which depends, as well as the effective mass $m_e^*$, on the types of semiconductors used to fabricate the double--dot system.
Eq.~\eqref{H_S} describes the interaction of the $i$th electron spin with an external magnetic field applied in the $z$ direction. Beside $B^{(i)}_z(t)$, which is a control field that can be used to adjust the Zeeman splitting associated with the electron spin in quantum dot $i$, one can, in principle, apply nonzero $x$ and $y$ components to perform rotations of the spin around other axes.\par
In spin--quantum--dot systems, the main cause for dephasing arises from charge fluctuations in the vicinity of the quantum dots, phonons, and interaction with nuclear spins.~\cite{Tay06,Gol04,Moz02,Yu02,Hu06}
For double--dot systems, modeling the environment can be achieved by coupling uncorrelated baths of harmonic oscillators to each of the spins,
\begin{equation}
{H_B}^{(i)} = \sum\limits_k{\hbar \omega_k^{(i)} b_k^{(i)\dag}b_k^{(i)}}, \; i=1,2, \label{H_R}
\end{equation}
where $b_k^{(i)\left[ \dag\right] }$ is the bosonic annihilation [creation] operator for the mode with frequency $\omega_k^{(i)}$.
We describe the interaction of the spins with the baths according to Sec.~\ref{DISSI},
\begin{eqnarray}
& &H_{SB}^{(i)}=\hbar \sigma_z^{(i)}\Gamma^{(i)}, \label{H_SR}\\
& &\Gamma^{(i)}=\sum\limits_k{g_k^{(i)}\left( b_k^{(i)}+b_k^{(i)\dag}\right). \label{gamma_def}}
\end{eqnarray}
The Heisenberg--type interaction, $H_S^{(12)}(t) = J(t){\vec\sigma^{(1)}\cdot\vec\sigma^{(2)}}$, is needed to produce entanglement and conditional operations.~\cite{DiVi98,Burk99} ($\vec\sigma$ denotes a vector containing the $x,y$ and $z$ Pauli matrices.)
In the interaction picture with respect to $H_S+H_B$ (operators denoted by a tilde),
the master equation in Born-Markov approximation for the present system-bath interaction is of the form,~\cite{Car02}
\begin{eqnarray}
& &\frac{d}{dt} \tilde \rho_S(t)= \label{master_born_markov} \\
& &-\frac{1}{\hbar^2} \int_0^t{dt'\operatorname{tr_R}\left\lbrace \left[\tilde H_{SR}(t),\left[\tilde H_{SR}(t'),\tilde \rho_S(t) \otimes \tilde \rho_R(0) \right] \right] \right\rbrace  }. \nonumber
\end{eqnarray}
When evaluating the master equation, one encounters correlation functions of the form,
\begin{equation}
\left\langle {\tilde \Gamma^{(i)}(t) \tilde \Gamma^{(i)}(t') } \right\rangle _R=\operatorname{tr_R}\left\lbrace \tilde \Gamma^{(i)}(t) \tilde \Gamma^{(i)}(t') \tilde \rho_R(0) \right\rbrace. \label{corrf0}
\end{equation}
By choosing an Ohmic spectral density,~\cite{Leg87,Weiss99}
\begin{eqnarray}
\sum_k {\left\lbrace g_k^2 ...\right\rbrace} &\rightarrow& \int\limits_0^\infty{d\omega \left\lbrace J\left( \omega \right) ...\right\rbrace}, \nonumber \\
J(\omega) &\rightarrow & \eta \omega e^{-\frac{\omega}{\omega_c}}, \label{ohmic}
\end{eqnarray}
the correlation functions can be calculated analytically. $\omega_c$ is a cutoff frequency, which depends on the physical properties of the dephasing process, and $\eta$ is a parameter which describes the effective coupling strength of the bosons to the qubit.
For a bath in thermal equilibrium we get,
\begin{eqnarray}
& &\left\langle {\tilde \Gamma^{(i)}(t) \tilde \Gamma^{(i)}(t') } \right\rangle _R=\frac{2\eta}{\hbar^2 \pi \beta^2}\left\lbrace \psi' \left( 1+ \frac{1-i \omega_c\left( t-t'\right) }{\hbar \omega_c \beta}\right) \right.\nonumber \\
& &+\left.\psi' \left( \frac{1+i \omega_c\left( t-t'\right) }{\hbar \omega_c \beta}\right)\right\rbrace, \label{corrf1}
\end{eqnarray}
where $\psi'$ is the derivative of the digamma function. We choose $\eta=0.8\times 10^{-13}\einheit{meVs}$, $\beta=1/\left( k_B\;50\einheit{mK}\right) $ and $\omega_c=5\einheit{meV}$.\par
As an example, we try to steer the double--spin system in the maximally entangled Bell state $|\psi^+\rangle$, starting from the initial state $\ket{\psi_I}$,
\begin{equation}
|\psi_I\rangle \langle\psi_I| \doteq \begin{bmatrix}
0 & 0 & 0 & 0 \\
0 & 1 & 0 & 0 \\
0 & 0 & 0 & 0 \\
0 & 0 & 0 & 0 \\
\end{bmatrix},
\end{equation}
where,
\begin{equation}
|\psi^\pm\rangle = \frac{1}{\sqrt{2}} (|1\rangle \otimes |0\rangle \pm |0\rangle \otimes |1\rangle) \doteq \frac{1}{\sqrt{2}}\left( 0,1,\pm 1,0\right)^T.
\end{equation}
A proper choice for the cost functional, reflecting the objective given above, is,
\begin{eqnarray}
\mathcal{J}\left[ \varepsilon \right] &=& \vectornorm {\rho_S(t_f)-\ket{\psi^+}\bra{\psi^+}}^2, \label{J_trapping} \\
\mbox{with}&& \quad \vectornorm {A}^2 \equiv \operatorname{tr}\left\lbrace A A^\dag \right\rbrace, 
\end{eqnarray}
where $\mathcal{J}$ denotes the cost functional [to be distinguished from the Heisenberg--coupling $J(t)$] and $\varepsilon$ the control field.
\begin{figure*}
\begin{tabular*}{17.4cm}{lll}
(a) & (b) & (c)	\\
\includegraphics[width=5.8cm]{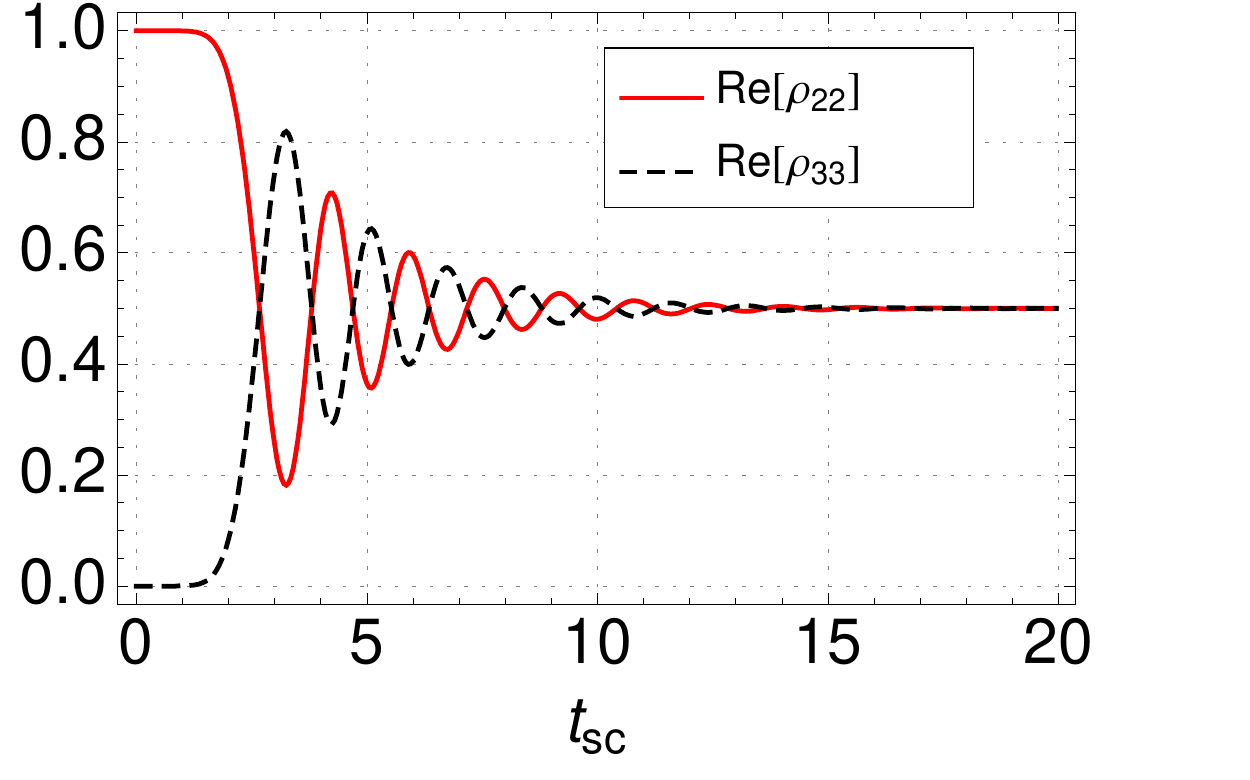}&\includegraphics[width=5.8cm]{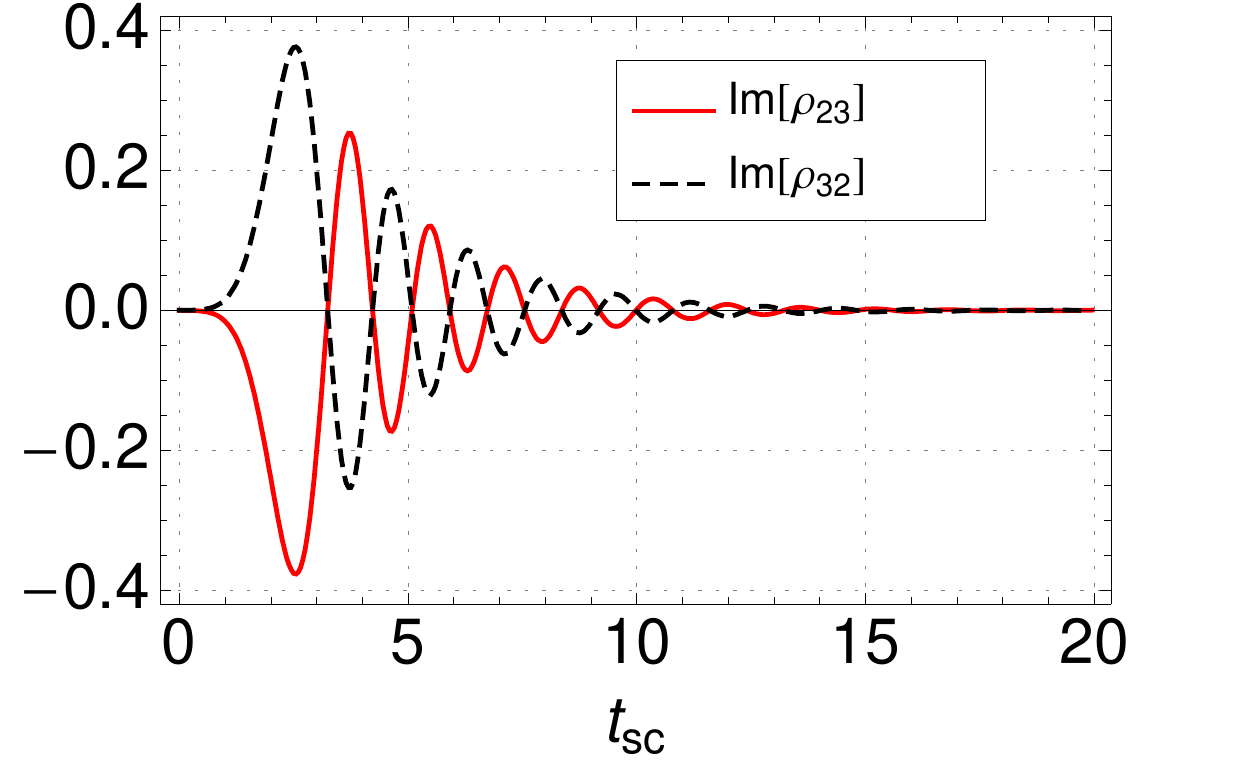} &\includegraphics[width=5.8cm]{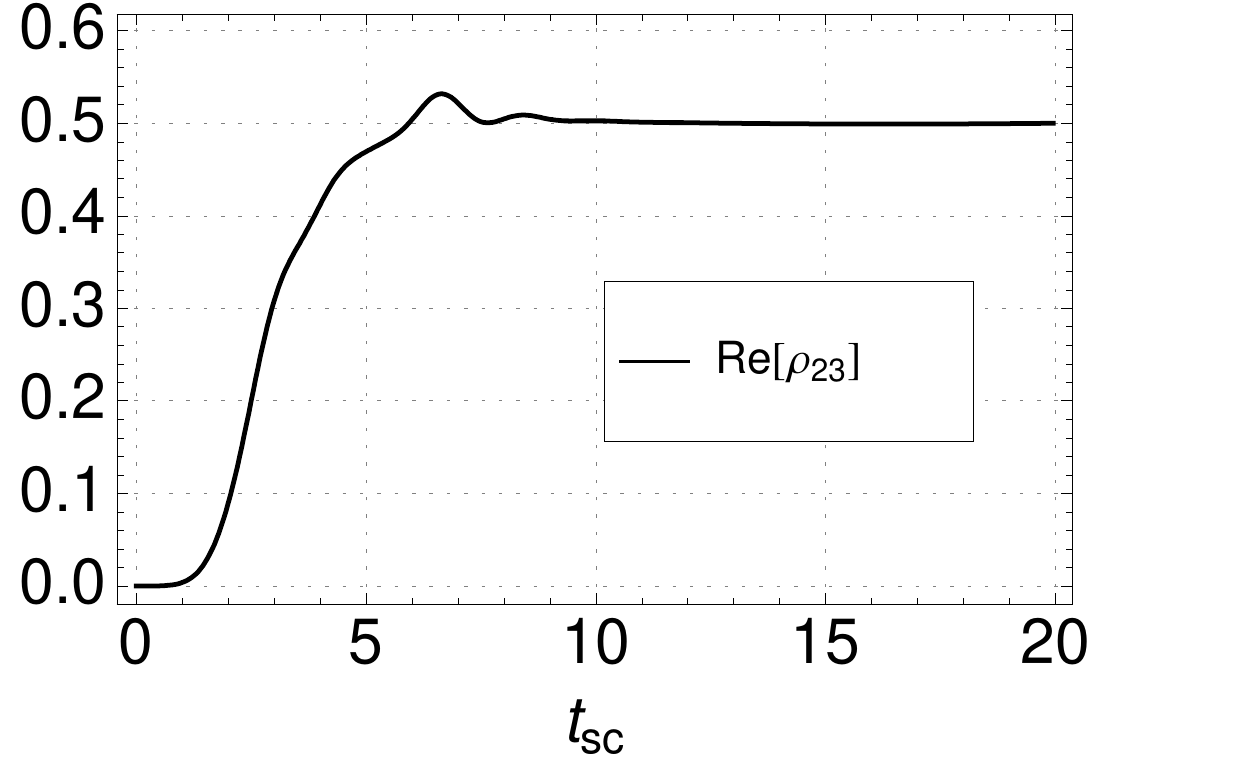} \\
(d) & (e) & 	\\
\includegraphics[width=5.8cm]{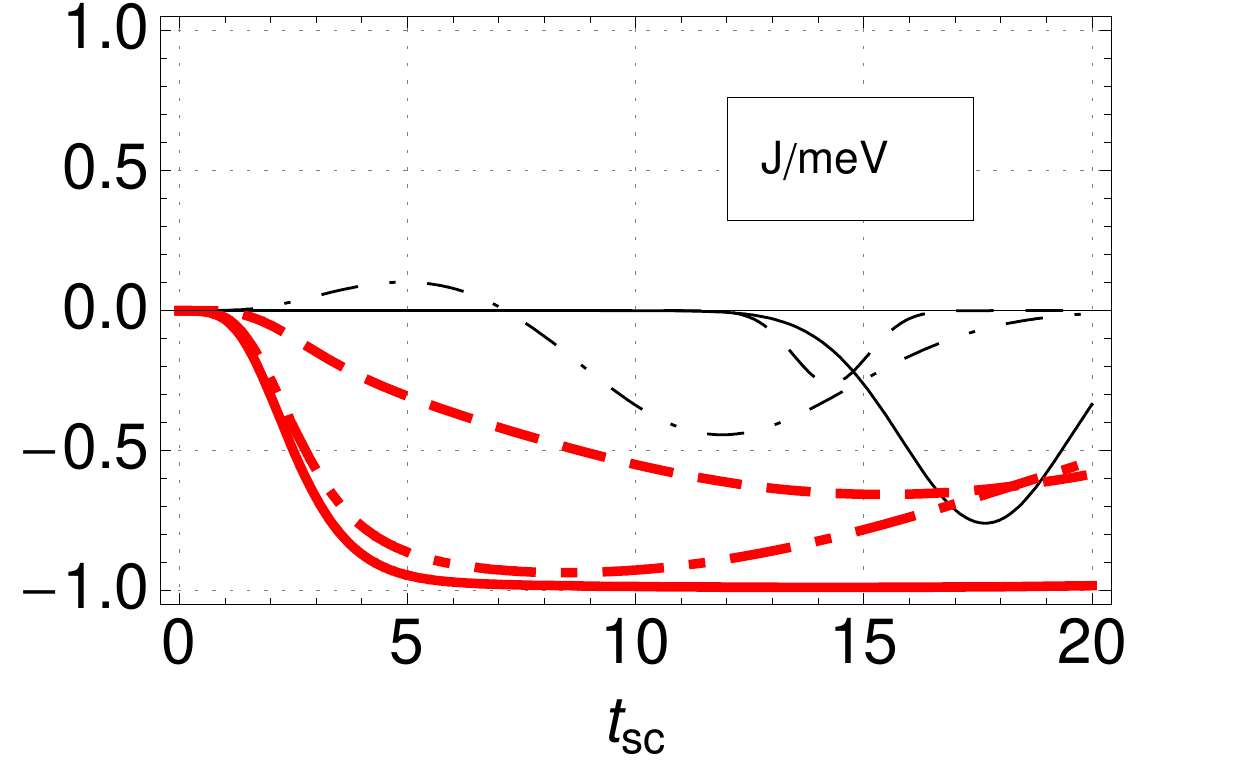} & \includegraphics[width=5.3cm]{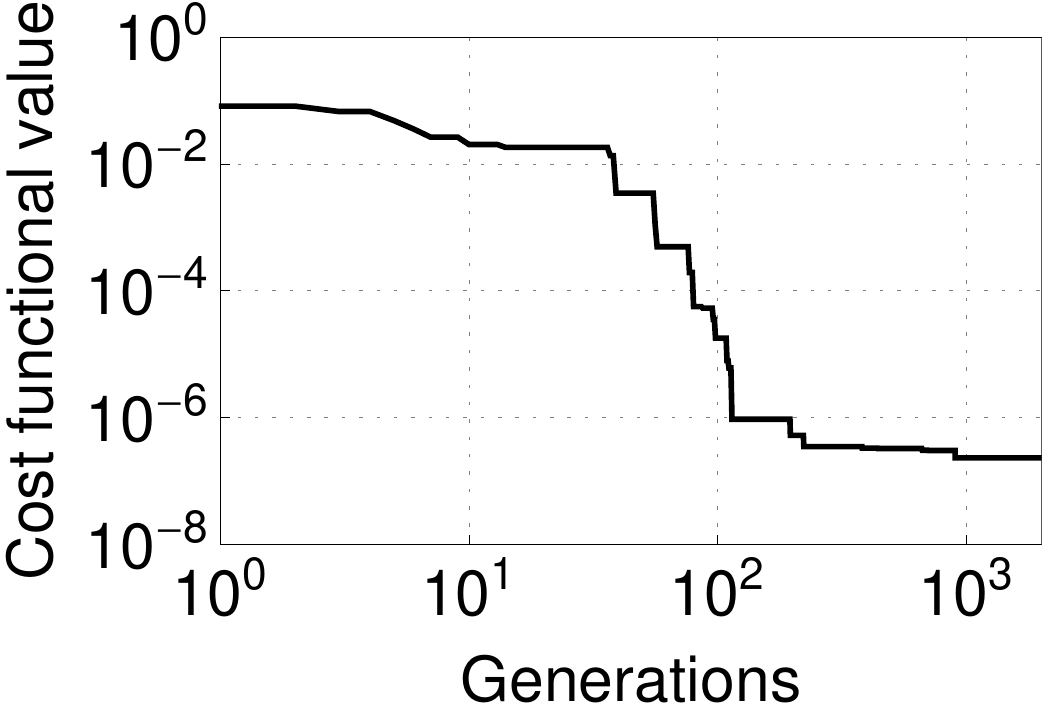}
\end{tabular*}
\caption{\label{qdot2}(a,b,c) Density matrix elements $\left\lbrace \operatorname{Re}\left[ \rho_{22}\right],\;\operatorname{Re}\left[ \rho_{33}\right]\right\rbrace ,\;\left\lbrace \operatorname{Im}\left[ \rho_{23}\right],\;\operatorname{Im}\left[ \rho_{32}\right]\right\rbrace ,\;\operatorname{Re}\left[ \rho_{23}\right]$ for the system subjected to the optimal control field. $t_{sc}=\omega_{sc}t$, where $\hbar \omega_{sc}=1\einheit{meV}$. (d) Best solutions after $n$ generations. Black~(thin)~dashed:~$n=2$, black~(thin)~dotdashed:~$n=13$, black~(thin)~solid:~$n=30$, red~(thick)~dashed:~$n=70$, red~(thick)~dotdashed:~$n=500$, red~(thick)~solid:~$n=2000$. (e) Value of the cost functional for the best solution of each generation.}
\end{figure*}
We parametrize the control field by $\varepsilon(t)\equiv J(t)=g(t) E_0 \sin{\left(  \omega t +\phi_0\right)} e^{-\gamma \left( t -t_0\right)^2}$, with $\left\lbrace E_0, \omega, \phi_0, \gamma, t_0 \right\rbrace $ being free parameters to be optimized. The function $g(t)$ provides a vanishing control field for $t=0$, as well as smooth increase of $\varepsilon(t)$ for $t>0$. As optimization procedure we choose a constrained, parallelized differential evolution algorithm with 230 individuals and 2000 generations.~\cite{DEA} The results are given in Fig.~\ref{qdot2} and Tab.~\ref{tab1}.

The desired final entangled state $\ket{\psi^+}$ can be realized with high accuracy ($\mathcal{J}[\varepsilon^*]\approx 10^{-7}$). As can be seen in Fig.~\ref{qdot2}(d), the differential evolution algorithm converges to a control field with large negative constant value ($J\approx -1\einheit{meV}$), which corresponds to the maximal allowed magnitude of the Heisenberg qubit--qubit coupling for the present double--quantum--dot system. Qubit--qubit couplings of several $100\einheit{$\mu$ eV}$ have been reported in Ref.~\onlinecite{Elz}. A large negative value of $J(t)$ makes the desired entangled state $\ket{\psi^+}$ to the approximate non--degenerate ground--state of the system. States with an accumulated relative phase with respect to $\ket{\psi^+}$, \textit{i.e.}, $\ket{\psi}=1/\sqrt{2}\left(  \ket{10}+e^{i\phi}\ket{01}\right) $, become energetically well separated from $\ket{\psi^+}$ and, hence, transitions to these are suppressed. From a different point of view, decoherence helps to relax the system from $\ket{\psi_I}$ into the new ground state. The stronger the environmental coupling the faster the system is able to relax. By choosing appropriate coupling constants $\eta$, one can, in analogy to classical mechanics, encounter underdamped and critically damped regimes, see Ref.~\onlinecite{Rol07}.
\begin{table}[h]
\begin{center}
\begin{tabular}{p{1.5cm}|p{1.5cm}|p{1.5cm}|p{1.5cm}|p{1.5cm}}
\hline
\hline
$E_0 / (\hbar \omega_{sc})$		& 	$\omega / \omega_{sc}$		&	$\phi_0$	&	$\gamma / \omega_{sc}^2$	&	$t_0 \omega_{sc}$		\\
\hline
-0.99	&	0.013	 		&	1.34		&	0.0			&	10.71	\\
\hline
\hline
\end{tabular}
\caption{\label{tab1} Optimal control--field parameters obtained by the differential evolution algorithm, where $\hbar \omega_{sc}=1\einheit{meV}$.}
\end{center}
\end{table}

\subsection{Control strategies - summary}

Basically all control strategies which have emerged from recent work  may be classified in the following way:
 
(i) $\pi$-flip control: In cases where pure dephasing occurs on a shorter time--scale, such as in SQUID--based qubits or spin quantum dots in presence of nuclear magnetic moments, $\pi$--flip--inducing control fields can successfully reverse dephasing 
in spin--echo--type fashion.\cite{slichter}  This works up to time scales at which population decay becomes 
important.     

(ii)  Storage in or transfer into a dissipation--free subspace:   This strategy utilizes the presence a decoherence--free subspace, see Sec.~\ref{DSUB}. 

(iii) Dynamical generation of a dissipative sub--space (high frequency ``bang--bang" control):  Here an intense high--frequency 
external perturbation is used to stabilize the system, as discussed in Sec.~\ref{DSUB}.

(iv) Quantum--interference (low--frequency/intensity control with state--specific optimization): 
This case was discussed above in Sec.~\ref{NMARK}.

(v) ``Last minute switching": The transfer from a state within the decoherence--free subspace (frequently thermal equilibrium) into another state at specified target time is accomplished  by an intense pulsed control
applied just prior to target time $t_f$.  If the target state is also decoherence free, this pulse may be administered at any time 
within the allotted  time interval.   This type of control is usually of little relevance, since frequently the system--control interaction strength is small, so that switching times are comparable to decoherence times or the target time. 

(v) Custom design of the control field:  This is not really an independent strategy but it allows, via  control theory, an identification of equivalent optimal control fields.  Among these, one selects the one which can most easily be realized in experiment, for example, regarding intensity and temporal behavior.  Ideal quantum gates should be perfectly shielded from the environment yet, upon demand, 
couple strongly to the control.   This clearly constitutes contradicting requirements which can be alleviated 
by a compromise developed within an optimal control scheme.

\section{State-independent optimal control}\label{SID}

As outlined in the introductory part of this review, state--independent optimal control
plays an important tool for the identification of the most efficient quantum gate realizations.
Indeed, the control of quantum subsystems using external forces is the
 basis for many recent experiments on Bose condensates, qubits and
 quantum gates, molecules, and nanostructures.~\cite{Brumer,Haroche,Blatt,Chu,Dupond95,NMR,Krausz,Blatt2,Nowack,Yanwen}
 For most applications seeking the observation or utilization of quantum
 interference effects, a minimization of the interaction between
 the quantum system and its environment is required.  For
 most efficient cooling of a quantum system, however, maximizing the latter
is desirable.~\cite{Bartana}
In this section, we shall briefly review three new approaches to state--independent optimal control
for open quantum systems and applications to physical qubit and quantum gate implementations.
    
\subsection{Perturbative approach}

Here we discuss a general approach for state--independent
optimal control for weakly dissipative quantum systems. The task posed 
 is the execution of a unitary operation $\mathcal{O}$ within the 
 dissipative quantum system as perfectly as possible in spite of the presence of
an environment and independent of the (unknown) initial state of the system.  

 This approach differs from earlier work in
 several major and nontrivial aspects:  The cost
 functional  for state--independent optimized control of inherently
 {\it dissipative} quantum systems  avoids the need for 
 repeated solution of the kinetic equation of the quantum subsystem's
 density operator during the optimization process and the 
 need for a co--state altogether, in contrast to mainstream
 optimal control for quantum systems discussed in previous chapters  which is based on cost functionals
 constrained by the system's kinetic equations.~\cite{Krotov96,Bryson75}. 

 Let us consider the general kinetic equation Eq. \eqref{MEQ} for the density operator 
 $\rho(t)$  of a quantum subsystem
  with a contribution from the time evolution under the Hamiltonian
 $H(t)$ containing the external control fields  and the dissipator
 $D[\rho]$. The latter  may be quite
 general but is assumed to be small in the sense of time--dependent perturbation theory. In particular, it may be of Markovian (Lindblad) or
 non-Markovian form, see Eqs.~\eqref{LIND} and \eqref{KE}.~\cite{Lindblad76,Poetz2,Poetz06,WP} 
The objective is to find an optimal control to realize a specified unitary operation $\mathcal{O}$  which is to be executed within 
a prescribed time interval $[0,t_f]$. 
The basic idea here is to seek a control which minimizes the effect of the dissipator while, at the same time, 
executing $\mathcal{O}$ as perfectly as possible.  
In order to motivate the cost functional selected below, Eq.~\eqref{MEQ} is written in the interaction picture,
$$
\tilde{\rho}(t)=U^{\dagger}(t)\rho(t)U(t),
$$
where $U(t)$ is the time--evolution operator for $H(t)$.
Eq.~\eqref{MEQ} takes the form,
\begin{equation}\label{INTP}
i\hbar\dot{\tilde{\rho}}=U^{\dagger}(t)D[U(t)\tilde{\rho}(t)U^{\dagger}(t)]U(t),
\end{equation}
and describes the action of the dissipator for given
 control Hamiltonian and initial state. In the absence of the
 dissipator, $\tilde{\rho}(t)=\rho(0)$ and the time--evolution is
 perfectly unitary. Confining ourselves
 to the case of weak dissipation  Eq.~\eqref{INTP} is solved iteratively.  Replacing, on the r.h.s. of Eq. \eqref{INTP}, 
$\tilde{\rho}(t')$ by $\rho(0)$  gives the solution within first--order 
perturbation theory,
\begin{equation}\label{integraleq}
\tilde{\rho}(t)-\rho_0\approx \frac{1}{i\hbar}\int_{0}^{t}U^{\dagger}(t')D[U(t')\rho(0)U^{\dagger}(t')]U(t')dt'.
 \end{equation}

The optimization approach uses two different cost functionals: an auxiliary cost functional $J$ to determine the optimal solution, and a second one, $J_Z$,  is used to test the quality of the solution for a set of $Z$ randomly selected initial states.   The auxiliary cost functional $J$ for identification of optimized control fields consists of three contributions,
\begin{equation}\label{Jtot}
J=J_O + J_D + J_\varepsilon.
\end{equation}
The contribution $J_O$ enforces completion of the  
 desired operation $\mathcal{O}$ (up to a phase) at time $t_f$.   
Possible choices are,~\cite{Grace07}
\begin{equation}\label{J}
 J_O=N^2-|\mathrm{Tr}\{\mathcal{O}^\dagger U(t_f,0)\}|^2,
 \end{equation}
 where $N$ is the dimension of the Hilbert space of the quantum
 system,  
 \begin{equation}\label{JP}
 J_O'= {\left|\mathrm{Tr}\{\mathcal{O}^\dagger U(t_f,0)-\openone \}\right|}^2,
 \end{equation}
$\left(\operatorname{Im}\mathrm{Tr}\{\mathcal{O}^\dagger U(t_f,0)\}\right)^2$, or $\left(\operatorname{Re}\mathrm{Tr}\{\mathcal{O}^\dagger U(t_f,0)-\openone\}\right)^2$.  
The second contribution, $J_D$, seeks to minimize the
 undesirable action of the dissipator at final time $t_f$ and is written,
 \begin{equation}\label{JD}
 J_D=s_D
 \langle\langle\mathrm{Tr}[(\tilde{\rho}(t_f)-\rho(0))^2]\rangle\rangle,
 \end{equation}
 where $\mathrm{Tr}$ and $\langle\langle ...\rangle\rangle$, respectively, denote
 the trace and an average over all (pure or mixed) possible initial states. $s_D$
 is a real--valued weight factor to guide convergence. 
The distribution of initial states is chosen according to the specific physical system. 
 
The third contribution controls shape and intensity of the control,
\begin{equation}
 J_\varepsilon = \int_{0}^{t_f} s(t)  \left|{\bf\varepsilon}(t) \right|^2 dt,
\end{equation}
 where ${\bf \varepsilon}(t)$ is the vector containing the external control fields.  
 $s(t)$ is a weight factor, as used before.   
Note that this cost functional $J$ avoids the use of  the dependent variable $\rho(t)$, thus avoiding a co--state.
The control may, in general, enter  $J_\varepsilon$ in the propagator $U(t,0)$ and the dissipator explicitly. Therefore, there is no need to evaluate the full kinetic
equation Eq.~\eqref{MEQ}. Evaluation of $J$ during the optimization process requires calculation of $U(t,0)$, $t\in (0, t_f]$.  
This is done best by solving the Schroedinger equation for $H(t)$ as an initial value problem with $U(0,0)=\openone$, or by using a discretized version of the formal time--ordered solution Eq. \eqref{PROP}. 
 
An arbitrary minimization algorithm can be employed.   For algorithms requiring 
cost functional and gradients as an input, 
 variations of $J$ with respect to the control ${\bf \varepsilon}(t)$  can be computed directly by using,
\be
 \frac{\delta U(t,0)}{\delta \varepsilon(t_k)}=\lim_{M\rightarrow
 \infty}U(t,t_{k+1})\left(-\frac{i\Delta t}{\hbar} \frac{\delta
 H(t_k)}{\delta \varepsilon(t_k)}\right) U(t_{k-1},0).
 \label{DU} \ee
This works also for Hamiltonians in which the field enters nonlinearly,
as long as its dependence remains local in time.
For complicated dissipators with retarded dependence
upon the control fields this provides a significant reduction in
computational effort.\cite{Wenin08}  
For microscopic models of dissipation, the computation of the retarded dissipation kernel is by far the most time--consuming 
part in the optimization loop.    
It should also be emphasized that the present approach, apart from being designed  for weak dissipation in a perturbative sense, 
per se does not provide a novel physical mechanism  for optimization nor does it introduce a bias regarding the selected optimization solutions.  As will be shown in the simple examples below, there is a multitude of practically equivalent solutions whose selection is mainly determined by the initial guess and the form of $J_\varepsilon$.  By the nature of the problem, all these optimal solutions are also good solutions for vanishing dissipator.

Verification of the quality of a solution is performed 
by computation of a test functional $J_Z$ 
 which is averaged over a set of Z initial states $\rho_j$,
 \be\label{JZ}
 J_Z=\frac{1}{Z}\sum_{j=1}^{Z}\mathrm{Tr}[(\rho_j(t_f)-\mathcal{O}\rho_{
 j}\mathcal{O}^\dagger)^{2}]. \ee
 The initial states $\rho_j$ are
 distributed randomly according to their likelihood of occurrence, consistent with the state average in $J_D$. $\rho_j(t_f)$ are
 the density operators obtained from the kinetic equation
 Eq.~\eqref{MEQ} using the optimized control fields.  Note that merely this performance test
requires repeated evaluation of the full kinetic equations.  Moreover, it does not resort to a perturbative account of dissipation.

This approach was originally applied to a dissipative qubit, with basis states $\mid 0\rangle$  and $\mid 1\rangle$,  treated within the Lindblad equation.~\cite{Wenin08b}
Numerical examples will be published elsewhere.\cite{Wenin08b}  A general result, however, should be mentioned here also.  We find that for the weak dissipation limit and where (local) optimal  minima exist, there is a large number of equivalent solutions. This is similar to the case of unitary systems.\cite{Rabitz04}   As a consequence there is considerable freedom regarding the shape of the control fields which thus can 
be used to aid experimental implementation.    

This approach has recently been applied to a study of the upper limit for the fidelity of single--qubit gate fidelities within currently available technology 
for Josephson charge qubit realizations.  Josephson--junction--based qubits, such as flux and charge qubits, have become systems of significant attention  due to their  potential for scalable qubit and quantum gate realizations.~\cite{Mak01,Nak99,Yam03,Clarke,Nakahara,Mooij,Spoerl07}
Typical for qubit realizations they suffer from two  shortcomings:  they are quantum two--level systems merely within approximation and their dynamics is influenced, apart from the externally applied control, by an undesirable coupling to the environment. This leads to state leakage from the two--dimensional computational Hilbert space of the ideal qubit, as well as undesirable decoherence.~\cite{Monta07} Both effects need to be suppressed before scaling to larger numbers of qubits becomes meaningful.   Tackled with the perturbative approach outlined above, 
among the infinitely  many equivalent control fields which execute the operation for the ideal qubit exactly one has to select 
and optimize these which minimize state leakage and dissipation.

\begin{figure}[H]
\includegraphics[width=8.5cm]{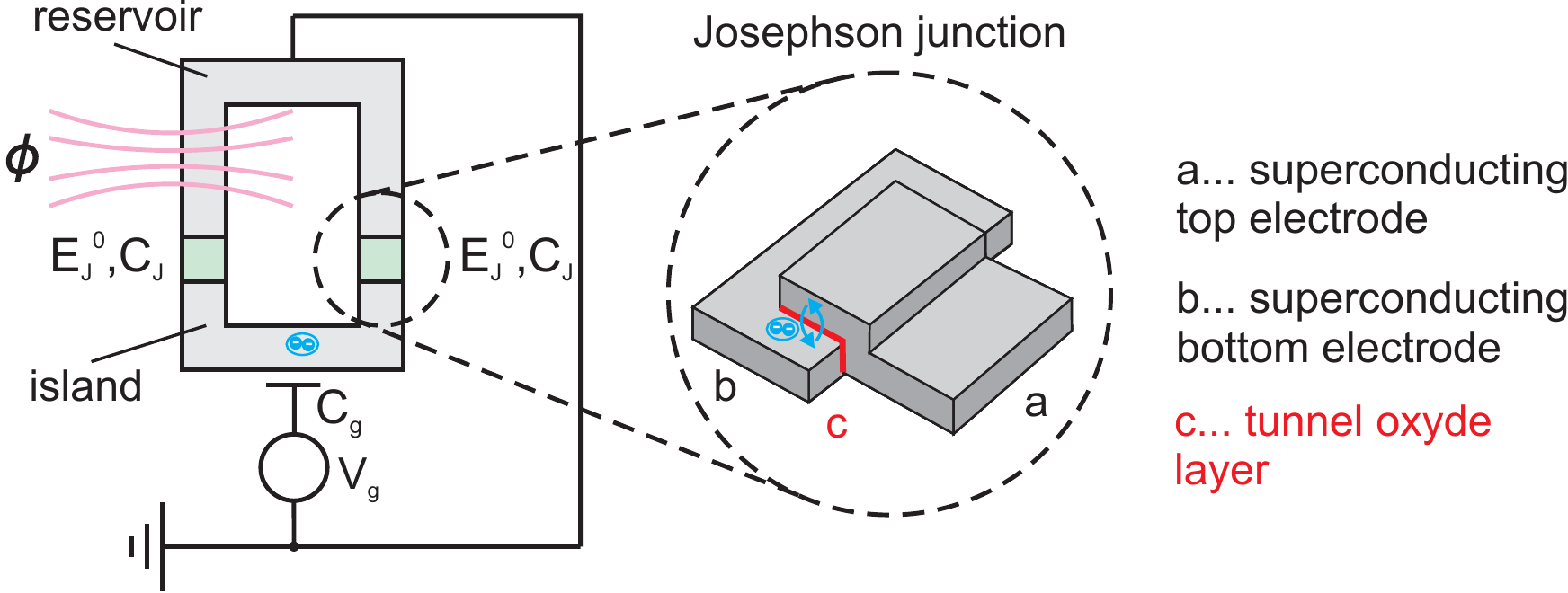}
\caption{\label{charge_qubit} A superconducting ring is divided by an oxide layer (c) into a superconducting reservoir and an island. By proper tuning of the controls $\phi$ and $V_g$, one can control the tunneling of cooper pairs through the Josephson junctions, which are characterized by the Josephson coupling energy $E_J^0$ and the capacitance $C_J$. For a detailed description see Ref.~\onlinecite{Mak01}}
\end{figure}

In a Josephson charge qubit, (see Fig.~\ref{charge_qubit}), the qubit is encoded in the number of additional Cooper pairs on a superconducting island (none $\ket{0}$ and one $\ket{1}$).~\cite{Mak01}

The two control fields ultimately consist of a gate voltage $V_g$ which determines the equilibrium charge state of the island and a magnetic flux $\phi$ which controls the Josephson energy. 
Leakage from the computational basis $\{\ket{0}, \ket{1}\}$ is accounted for by the adoption of the two charge states $\ket{-1}$ and $\ket{2}$ leading to a four--dimensional Hilbert space for the ``leaky" qubit. The effective Hamiltonian of the superconducting qubit including leakage states and proper system bias reads,~\cite{Rol08} 
\begin{eqnarray}
H_o(n_g,\Phi)=&&  \begin{bmatrix}
8 E_C & -\frac{1}{2}E_J & 0 & 0 \\
-\frac{1}{2}E_J &  0 & -\frac{1}{2}E_J & 0 \\
0 & -\frac{1}{2}E_J & 0 & -\frac{1}{2}E_J \\
0 & 0 & -\frac{1}{2}E_J & 8 E_C
\end{bmatrix}+ \nonumber \\&&4E_C n_g\begin{bmatrix}
3 & 0 & 0 & 0 \\
0 &  1 & 0 & 0 \\
0 & 0 & -1 & 0 \\
0 & 0 & 0 & -3
\end{bmatrix} \nonumber,
\end{eqnarray}
where $E_J=2 E_J^0 \cos{\left( \pi \frac{\Phi}{\Phi_0}\right) }$. $E_C$ denotes the single electron charging energy of the island and $E_J=E_J(\phi)$ is the Josephson coupling energy, which depends on the externally applied magnetic flux $\phi$; $n_g=\frac{C_g V_g}{2e}$, with $C_g$  denoting  the gate capacitance.
It is believed that the dominant dephasing mechanism is due to background charge fluctuations in the vicinity of the gate electrodes.~\cite{Asta04,Schriefl06}  Two Lindblad operators $i=x,z$ capturing dephasing and population decay were
identified as,
$$
L_i= \sqrt{\gamma_i(t})H_i,
$$
with,
$$
\gamma_i(t)\equiv \frac{2}{\hbar^2} \int_0^{t}\,dt'c_i(t,t'),
$$
and,
\begin{equation}
H_z=
\begin{bmatrix}
3 & 0 & 0 & 0 \\
0 &  1 & 0 & 0 \\
0 & 0 & -1 & 0 \\
0 & 0 & 0 & -3
\end{bmatrix}, H_x=
 \begin{bmatrix}
0 & 1 & 0 & 0 \\
1 &  0 & 1 & 0 \\
0 & 1 & 0 & 1 \\
0 & 0 & 1 & 0
\end{bmatrix}.
\end{equation}
$c_i(t,t')$ is a bath correlation function. 
Thus, fluctuations in the gate voltage  ($\propto L_z$) lead  to dephasing and fluctuations in the magnetic flux  and/or $E_j^0$  ($\propto L_x$) lead  to combined population decay and dephasing.  The former typically  occur on a significantly shorter time--scale than the latter.

In addition to the optimization scheme outlined above, the superoperator--based scheme introduced in the following chapter was used to 
optimize Hadamard gate operation.
This study has revealed that leakage from the computational subspace can be largely suppressed, but dissipative effects, predominantly due to dephasing due to fluctuations in the 
charging energy, lead to  noticeable reduction in fidelity below 100 percent.  
While the latter is readily achievable in the sole presence of state leakage, for switching times  of about 0.5 ns and typical system parameters taken from experiment, the predicted optimal fidelity is about 98 percent in presence of both state leakage and dephasing.   Effects from fluctuations in the Josephson energy are found to be negligible.  These results are in qualitative agreement with a 
superoperator optimization scheme based on a spin--boson model for noise in the leaky qubit.\cite{Rol08}

\subsection{Superoperator formulation of state--independent control}

The time evolution of the density operator is expressed in terms
of a time evolution of a super--operator which acts as a linear
map on the initial state. As a consequence the initial state is
detached from the time evolution. We give an overview of the
theory, using a simple example, where dissipation is described by
a Lindblad dissipator. A more elaborate discussion is given in
Ref.\onlinecite{Wenin08c} where we consider a microscopic
dissipator based on the spin--boson model and various control
Hamiltonians.
\subsubsection{Kinetic equations}
We start with the kinetic equation for the reduced density matrix
of the open quantum system, $\rho(t)$, Eq. \eqref{MEQ}. The
Hamiltonian of the system, $H(t)$, contains the unperturbed system
Hamiltonian $H_0$ and the external control $H_c(t)$, see Eq.
\eqref{HSYS}. At this point the dissipator $D[\rho]$ can be quite
general.\cite{Thorwart,Storcz,Mucciolo,Poetz06,Poetz2,Wenin06,Uhrig,Palao03,Rabitz}
We rewrite Eq.~\eqref{MEQ} using superoperators,
\begin{equation}\label{MEQ2}
i\hbar\dot{\rho}(t)=[\mathcal{L}(t)+\mathcal{D}(t)]\rho(t).
\end{equation}
Here $\mathcal{L}(t)$ is the usual Liouville--superoperator,
whereas $\mathcal{D}(t)$ represents the dissipator.
For the superoperator $\mathcal{X}(t)$, defined in Eq.
\eqref{SUOP}, we have the equation of motion,
\begin{equation}\label{chid}
i\hbar\dot{\mathcal{X}}(t)=[\mathcal{L}(t)+\mathcal{D}(t)]\mathcal{X}(t).
\end{equation}
Eq.~\eqref{chid} represent the central differential equation for
the evolution superoperator to describe dissipative quantum
systems.
\subsubsection{Representation of the superoperators and the Lindblad dissipator}
We represent the various superoperators in a basis and work with
the components, using always Einstein summation convention. The
Liouville--superoperator $\mathcal{L}(t)$ depends on the
time--dependent system Hamiltonian $H(t)$. The elements are given
by,
\begin{equation}
\mathcal{L}_{ijmn}(t)=H_{im}(t)\delta_{nj}-H_{nj}(t)\delta_{im}.
\end{equation}
For a dissipator in Lindblad form, $\mathcal{D}^{L}$, we have the
expression,

\ea \label{diss} \mathcal{D}^{L}_{ijmn}& = &
i\hbar\sum_{\mu}\Big\{(L_\mu)_{im}(L_{\mu}^{\dagger})_{nj}
\nonumber{}\\
&- &  \frac{1}{2}(L_{\mu}^{\dagger}L_{\mu})_{im}
\delta_{jn}-\frac{1}{2}(L_{\mu}^{\dagger}L_{\mu})_{nj}
\delta_{im}\Big\}. \eea
Here $L_\mu$ are Lindblad--operators, describing the structure of
the dissipator. For completeness we give also Eq.~\eqref{MEQ2} in
components,
\begin{equation}\label{GlA}
i\hbar\dot{\rho}_{ij}(t)=[\mathcal{L}_{ijrs}(t)+\mathcal{D}^{L}_{ijrs}]\rho_{rs}(t).
\end{equation}
For the evolution superoperator $\mathcal{X}(t)$ we have the
equations,
\begin{equation}\label{Xt}
i\hbar\dot{\mathcal{X}}_{ijrs}(t)=[\mathcal{L}_{ijmn}(t)+\mathcal{D}^{L}_{ijmn}]\mathcal{X}_{mnrs}(t),
\end{equation}
with $\mathcal{X}_{ijrs}(0)=\delta_{ir}\delta_{js}$ and the
state--evolution of an initial state $\rho(0)$ is given by,
\begin{equation}
\rho_{ij}(t)=\mathcal{X}_{ijrs}(t)\rho_{rs}(0).
\end{equation}
When we compare  Eq.~\eqref{GlA} with  Eq.~\eqref{Xt} we can see
that the only, but in practice relevant difference of both is the
number of differential equations. In Eq.~\eqref{GlA} this number
grows as $N^2$, whereas in Eq.~\eqref{Xt} it grows as $N^4$,where
$N$ is the dimension of the Hilbert space. This makes the OCT
problem for state independent control including
dissipation numerically more expensive.\\
For dissipation--less systems, $\mathcal{D}^{L}_{ijmn}\equiv 0$ in
Eq.~\eqref{Xt}, one can express the evolution superoperator
components $\mathcal{X}_{ijrs}(t)$, using the usual
time--evolution operator $U(t)$,
\begin{equation}\label{Xdf}
\mathcal{X}_{ijrs}(t)=U_{ir}(t)U^{\dagger}_{sj}(t).
\end{equation}

\subsubsection{Cost functional}
We formulate in this section the cost functional and the
optimization problem. Let $\mathcal{O}$ the target operation,
unique defined up to a physical irrelevant global phase. We
consider first the transformation of a state, represented by a
initial density operator $\rho(0)$. In the ideal case the final
state is,
\begin{equation}
\rho(t_f)=\mathcal{O}\rho(0)\mathcal{O}^\dagger,
\end{equation}
for any $\rho(0)$. Considering a pure coherent dynamic we obtain
the target superoperator $\mathcal{X}_T$,
\begin{equation}
\mathcal{X}_{ijrs}(t_f)=U_{ir}(t_f)U^{\dagger}_{sj}(t_f)\rightarrow\mathcal{O}_{ir}\mathcal{O}^{\dagger}_{sj}
=(\mathcal{X}_T)_{ijrs}.
\end{equation}
A natural and simple choice for the cost functional is therefore,
\begin{equation}\label{Jn}
J=||\mathcal{X}(t_f)-\mathcal{X}_T||^2=\sum_{p,q,m,j}|\mathcal{X}_{pqmj}(t_f)-\mathcal{O}_{pm}\mathcal{O}^{+}_{jq}|^{2}.
\end{equation}
Minimization of $J$ by variation of the control, $H_c(t)$ in
Eq.~\eqref{HSYS}, which enters in Eq.~\eqref{Xt}, defines the task
of state independent optimal control of dissipative quantum
systems. One can solve this mathematical
problem using different methods.\cite{Bryson75}
We remark that Eq.~\eqref{Jn} applied  to a dissipation--less
system takes the form Eq. \eqref{J}.
\subsubsection{An example: The CNOT gate}
We consider a simplified model for a Josephson two--qubit system
and apply the theory presented before to the CNOT
gate.\cite{Niel02,Niskanen2} The example is chosen in order to
demonstrate the strategy. To describe the system we use the
product states of the computational basis,
$|0\rangle|0\rangle\equiv|1\rangle$,
$|0\rangle|1\rangle\equiv|2\rangle$,
$|1\rangle|0\rangle\equiv|3\rangle$,
$|1\rangle|1\rangle\equiv|4\rangle$. The target is the unitary
operator,
\begin{equation}\label{UCnot}
\mathcal{O}=\left(%
\begin{array}{cccc}
  1 & 0 & 0 & 0 \\
  0 & 1 & 0 & 0 \\
  0 & 0 & 0 & 1 \\
  0 & 0 & 1 & 0 \\
\end{array}%
\right).
\end{equation}
\subsubsection{Hamiltonian of the system}
Josephson charge qubits are optimal controllable quantum systems.
Control of the wave function can be achieve by tuning gate
voltages and magnetic fluxes.\cite{Mak01} The single qubit
Hamiltonians ($i=1,2$),
\begin{equation}
H^{(i)}(t)=\varepsilon^{(i)}_x(t)\sigma_x+\varepsilon^{(i)}_z(t)\sigma_z.
\end{equation}
offer full control over both independent qubits. The control
fields are $\varepsilon^{(i)}_{x,z}(t)$. For the qubit--qubit
interaction $H^{12}(t)$ we set (with $C$ a constant),
\begin{equation}
H^{12}(t)=C\varepsilon^{1}_x(t)\varepsilon^{2}_x(t)\sigma_y\otimes\sigma_y.
\end{equation}
We note that this structure of the qubit--qubit interaction
recently was studied for dissipation--less systems, where a
minimization of a cost functional of the type Eq.~\eqref{J} has
been carried out.\cite{Niskanen2}
\subsubsection{Dissipation and Lindblad--operators}
We model the effect of the environment in a simplified version
also, where two parameters are enough to describe the total
effects of the system--environment interactions.
To proceed, we need to specify the Lindblad operators $L_\mu$. As
we consider two identical qubits, which without coupling,
$H^{12}(t)=0$, should evolve independently, the Lindblad operators
must have a tensor product structure. For each qubit we use two
Lindblad operators to describe transitions between the two basis
states $|0\rangle$ and $|1\rangle$. In particular we set for the
four Lindblad operators,
\begin{equation}
L_1=\sqrt{\gamma_1}|0\rangle\langle 1|\otimes
\mathbbm{1}=\sqrt{\gamma_1}(|1\rangle\langle 3|+|2\rangle\langle
4|),
\end{equation}
\begin{equation}
L_2=\sqrt{\gamma_2}|1\rangle\langle 0|\otimes
\mathbbm{1}=\sqrt{\gamma_2}(|3\rangle\langle 1|+|4\rangle\langle
2|),
\end{equation}
\begin{equation}
L_3=\sqrt{\gamma_1}\mathbbm{1}\otimes|0\rangle\langle
1|=\sqrt{\gamma_1}(|1\rangle\langle 2|+|3\rangle\langle 4|),
\end{equation}
\begin{equation}
L_4=\sqrt{\gamma_2}\mathbbm{1}\otimes|1\rangle\langle
0|=\sqrt{\gamma_2}(|2\rangle\langle 1|+|4\rangle\langle 3|).
\end{equation}
For simplicity we use the same rates $\gamma_{1,2}$ for both
qubits. Dephasing and relaxation in the one--qubit system occurs
during the times,
\begin{equation}
T_2=\frac{2}{\gamma_1+\gamma_2}~,\hspace{5mm}T_1=\frac{1}{\gamma_1+\gamma_2}.
\end{equation}
In our model the unbiased single qubit, $\varepsilon_x(t)=0$,
relaxes to the equilibrium state,
\begin{equation}\label{req}
\rho^{(1)}_{\mathrm{eq}}=\left(%
\begin{array}{cc}
  \frac{\gamma_1}{\gamma_1+\gamma_2} & 0 \\
  0 & \frac{\gamma_2}{\gamma_1+\gamma_2} \\
\end{array}%
\right).
\end{equation}
We note that only for $\gamma_1=0$ or $\gamma_2=0$ this is a pure
state, in general hoverer, Eq.~\eqref{req} describes a mixed
state. The equilibrium state of the two--qubit--system is the
product state,
$\rho_{\mathrm{eq}}=\rho^{(1)}_{\mathrm{eq}}\otimes\rho^{(2)}_{\mathrm{eq}}$.
\subsubsection{Numerical study}
\begin{figure*}
\begin{center}
\includegraphics[height=90mm, angle=0]{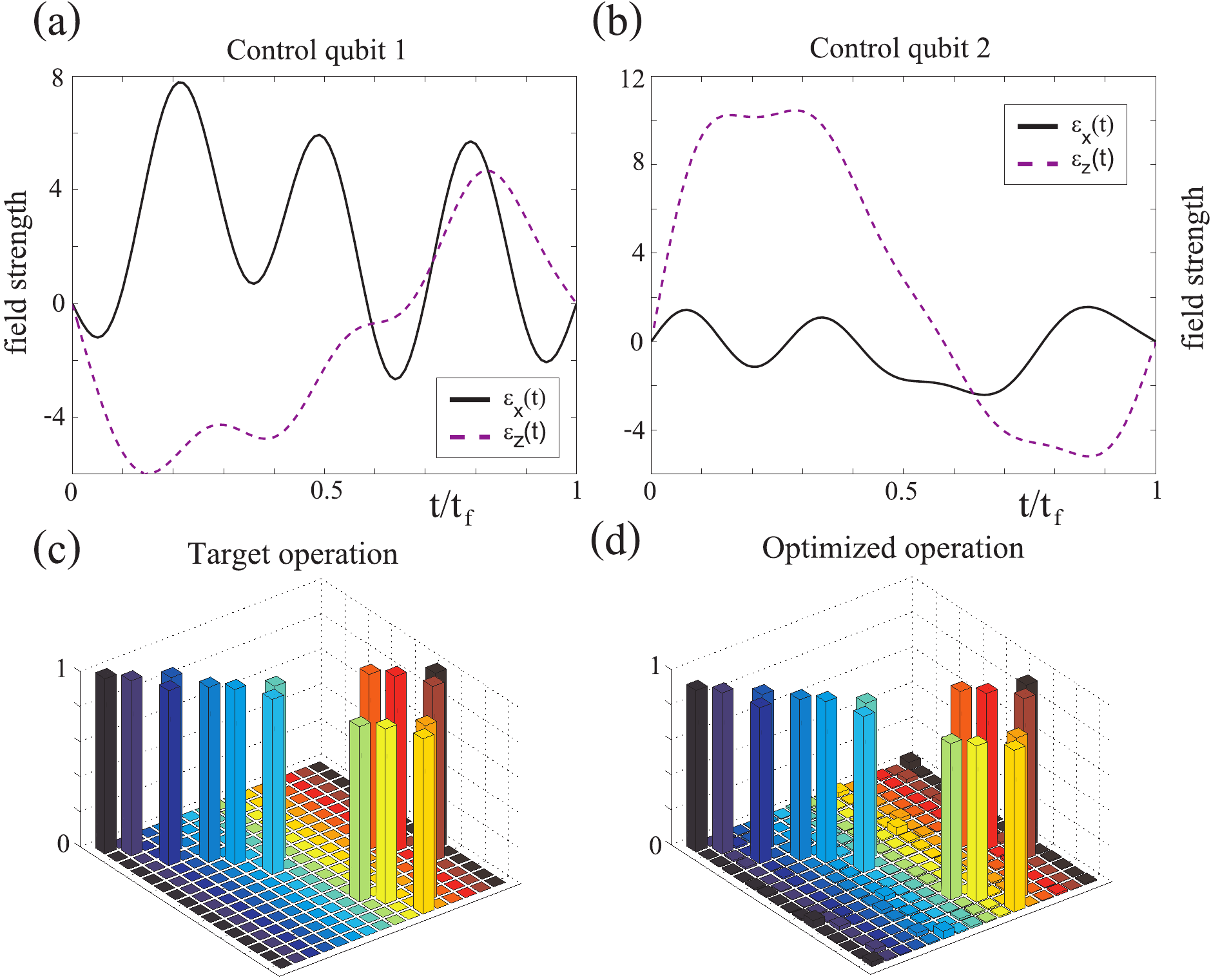}
\caption{CNOT gate: (a), (b) shows the control fields of both
qubits respectively in units of $\hbar/t_f$. In (c) are shown the
$16\times 16$ elements of the target superoperator, which must be
compared to the result of numerical optimization (d).
}\label{Grafik}
\end{center}
\end{figure*}

In order to obtain further results we have to compute the kinetic
equation Eq.~\eqref{Xt} numerically. We assume given rates
$\gamma_{1,2}$. To solve the optimization problem, we set up the
fields by a Fourier series,
\begin{equation}\label{Fourier}
\varepsilon(t)=\sum_{k=1}^{F}a_k\sin(\omega_k
t)~,\hspace{5mm}\omega_k=\frac{k\pi}{t_f}.
\end{equation}
All four control fields are expressed by such a decomposition,
with a priori unknown coefficients. Whit this Ansatz we have for
$F\rightarrow\infty$ a complete function system and we have
incorporated also boundary conditions, in our example,
$\varepsilon(0)=\varepsilon(t_f)=0.$
When the coefficients are known, we can compute the evolution
superoperator using the kinetic equation, Eq.~\eqref{Xt} and the
cost functional Eq.~\eqref{Jn}. We start with a set of guess
coefficients and minimize Eq.~\eqref{Jn} using standard line
search routines. We set $F=8$ coefficients for each control field.
Decay rates are chosen as $\gamma_1 t_f=\gamma_2 t_f=0.1$, the
constant $C=t_f/\hbar$. The cost functional is in the unitary case
of the order $J\approx10^{-7}$ and with dissipation is $J=0.1678$.
Results are shown in Fig.~\ref{Grafik}. The optimal control fields
are plotted in (a) and (b). The lower part of the figure shows the
absolute values of the superoperator elements. The bars in (c)
represent the ideal values and in (d) we plot the values obtained
by using the fields shown in (a) and (b). The norm of $\chi$
decreases from $\|\chi(0)\|^{2}=16$ to $\|\chi(t_f)\|^{2}=13.15$.
In order to demonstrate the action of the dissipator we choose
very strong decay rates. After one gate operation the norm of
$\chi$ lose about 20\% of its initial value.


\subsection{Optimal control within the Kraus operator representation}

Recently, a theoretical study of the Josephson charge qubit was performed to identify its performance limits within current experimental means.\cite{Rol08}  It accounts for state leakage, dissipation and decoherence and is based on a Kraus representation of the time--evolution superoperator, see Sec.~\ref{SSvsSI}. The Hilbert space of the system is divided into the computational subspace $\mathcal{C}$ over which the desired quantum operation is defined, and the remaining space $\mathcal{L}$ is spanned by potential ``leakage'' states. The full Hilbert space (without environment) is the direct sum, $\mathcal{C} \oplus \mathcal{L}$.

State--independent optimization is performed  as follows. One considers a map, $\rho(0) \mapsto \rho(t)=\mathcal{E}_t\left\lbrace  \rho(0)\right\rbrace$, for which  the superoperator $\mathcal{E}_t$ is functionally dependent upon the control field $\varepsilon(t)$,  \textit{i.e.}, $\mathcal{E}_t=\mathcal{E}_t[\varepsilon]$.   Since positivity of the density operator $\rho(t)$ must be conserved, the map $\mathcal{E}$ has to be \textit{completely positive} and thus can be represented by Kraus operators $K_{m}$,\cite{Niel02,Havel03}
\begin{equation}
\rho(t)=\sum\limits_{m}{K_{m}[\varepsilon](t)\rho(0)K^\dag_{m}[\varepsilon](t)}. \label{kraus}
\end{equation}
The optimal  control field $\varepsilon^*(t)$ is selected such that, at some final time $t_f$, $\mathcal{E}_{t_f}[\varepsilon^*]$ approaches the desired mapping
$\mathcal{E}^{\mathcal{O}}$ as closely as possible.  For given quantum gate operations, 
\begin{equation}
\mathcal{E}^\mathcal{O}\left\lbrace .\right\rbrace = \mathcal{O}\left( .\right) \mathcal{O}^\dag, \label{opt_unitary}
\end{equation}
where $\mathcal{O}$ denotes the desired unitary operation. For quantum information theory it is useful to formulate the cost functional
within the language of process tomography (see \textit{e.g.} Refs.~\onlinecite{Alt03,Niel02b}).  The mapping $\mathcal{E}$ can be expressed by expanding the Kraus operators, $K_m(t)=\sum_n{\alpha_{mn} \tilde K_n}$, with $\alpha_{mn}\in\mathbb{C}$ and  $\tilde K_{n} \in \mathcal{A}$, where $\mathcal{A}$ denotes a complete basis set of $M \times M$ matrices.~\cite{Alt03}  $M=M_C+M_L$, where $M_C$ and $M_L$ correspond to the number of (orthonormal) computational and leakage basis states, respectively.
The state of the quantum system at $t_f$ and starting out in $\rho(0)$ now reads, $\rho(t_f)=\mathcal{E}_{t_f}\left\lbrace \rho(0) \right\rbrace = \sum_{m,n}{\tilde K_{m} \rho(0) \tilde K^\dag_{n} \chi_{mn}(t_f)}$, with $\chi_{mn}=\sum_k{\alpha_{km}\alpha_{kn}^*}$.
For evaluation of the process tomography matrix $\chi$, one chooses a fixed set of operators $\left\lbrace \sigma_j \right\rbrace=\mathcal{B}$ (for simplicity we choose $\mathcal{B}=\mathcal{A}$) and determine the time--evolution of these operators with respect to the mapping $\mathcal{E}$, $\sigma_j(t_f)\equiv\mathcal{E}_{t_f}\left\lbrace \sigma_j \right\rbrace=\sum_k{c_{jk}\sigma_k}$.
Eq.~\eqref{kraus} is a linear mapping and one can obtain $\chi$ by computing the time evolution of the set $\mathcal{B}$ to fully characterize the quantum operation performed. However, calculating the time evolution using Eq.~\eqref{kraus} is often intractable. Hence, one may want to approximate the dynamics of the system by employing e.g. perturbative methods. A short description of appropriate techniques for the spin--boson model is given in Sec.~\ref{DISSI}.

For convenience, we define operators $\hat \chi=\sum_{m,n}{\left( {\tilde K_{n}}^*\otimes \tilde K_{m}\right) \chi_{mn}}$,~\cite{Havel03} and formulate a simple cost functional,\cite{Rol08}
\begin{equation}
J\equiv \vectornorm{P \hat \chi - \hat \chi^\mathcal{O}}^2=\operatorname{tr}\left\lbrace \left[ P \hat \chi - \hat \chi^\mathcal{O}\right] \left[ P \hat \chi - \hat \chi^\mathcal{O}\right]^\dag \right\rbrace, \label{cost}
\end{equation}
where $P$ denotes the projector onto the $M_C$-dimensional computational Hilbert space $\mathcal{C}$ and $0\leq J\leq J_{max}=2 M_C^2$. $J$ measures the norm distance between the target operation $\hat\chi^\mathcal{O}$, corresponding to the mapping $\mathcal{E}^\mathcal{O}$, and the actual operation $\hat\chi$ executed at time $t_f$ for control field $\varepsilon$. 

\section{Conclusion and outlook}\label{SUM}

Over the last two decades optimal control of quantum dynamics has become a mature field with applications 
in many areas of physics  and chemistry.  In its beginning it dealt with state--dependent control of unitary systems within the Schr\"{o}dinger equation.  Nowadays, one deals with open quantum systems obeying more or less complex master equations to study dissipative effects in quantum systems, such as quantum gate realizations  or Bose condensates.

In this article we have presented  recent progress in optimal control theory applied to the dynamics 
of open dissipative quantum systems.  While we have exclusively presented our own results, we have tried to 
reference relevant work and alternative approaches which are available in the literature.  Special focus was given to approaches which allow external control of the effective system--bath interaction.  The motivation for investigating this aspect in particular lies in the recognition that most effective control of the system--bath interaction has to be administered at the quantum level, \textit{i.e.}, on a time scale for which the system--bath interaction reveals its quantum nature.  

The presented work covers two main tasks: state-dependent control and state--independent control.
The first task is important when one wishes to prepare a quantum system in a specific state, starting from a predetermined initial state, usually its ground state.  For this case, we reviewed standard optimal control theory and an extension to non--Markovian systems.  The latter approach has been applied to a model system to demonstrate quantum interference effects in the control of the effective system--bath interaction (dissipation).  Furthermore, we reviewed analytic solutions for dissipative two--level systems which solve the optimization problem by direct inversion.  As an example, optimal control fields for inducing Rabi oscillations in a dissipative two--level system have been derived.  While direct inversion  is limited to simple models, it allows the derivation of absolute bounds for controllability.  In general, numerical methods provide local minima only. 
This holds particularly for conjugate--gradient methods.  Global minima can be found, in principle, by evolutionary algorithms, however, at the expense of a large number of evaluations of the cost functional.

Three recently developed approaches to state--independent optimal control were presented.  A first--order perturbative approach of the system--bath interaction was formulated which determines optimal control fields using an auxiliary cost functional consisting of three parts which, respectively, enforce execution of the quantum operation, minimization (maximization) of the system--bath interaction, and the desired shape of the control.  Application to 
SQUID--based qubit realization was reviewed.  A completely general approach based on the time--superoperator representation 
was presented.  It provides the most general possible representation of state--independent optimal control for open quantum systems.  A third related approach based on the Kraus operator representation of the time--evolution superoperator
and a formulation of the cost functional within process tomography we specially developed for application to quantum--information processing.     

An attempt was made to render the presentation widely applicable throughout quantum physics.  Our physical examples, however, were 
restricted to solid--state realizations, in particular quantum dots and SQUIDS.  

The need for a microscopic description to fully exploit the potential of control of quantum systems naturally leads to 
numerical complexity limiting such models to elementary quantum systems at present. 
Moreover, quantum information processing will require studies regarding scaling behavior of dissipative effects with 
quantum gate array size.  
It is therefore important to reduce 
complicated many--body systems to simple effective models, whenever possible. Several such models were reviewed: the Lindblad equation, the spin--boson model and the spin--bath model.    

While this article has discussed optimal control from a theoretical point of view, it can also be executed in conjunction with experiment.  For example, using feedback control, optimal control fields can be determined in a learning cycle.  
Pulse shaping in ultra--fast laser spectroscopy is a representative example for this strategy.  Clearly, interplay between theory and experiment will be beneficial and is inevitable for modeling experimental setups.  

\section{Acknowledgment}
This work was supported by FWF, project number P18829.


\begin{thebibliography}{122}
\expandafter\ifx\csname natexlab\endcsname\relax\def\natexlab#1{#1}\fi
\expandafter\ifx\csname bibnamefont\endcsname\relax
  \def\bibnamefont#1{#1}\fi
\expandafter\ifx\csname bibfnamefont\endcsname\relax
  \def\bibfnamefont#1{#1}\fi
\expandafter\ifx\csname citenamefont\endcsname\relax
  \def\citenamefont#1{#1}\fi
\expandafter\ifx\csname url\endcsname\relax
  \def\url#1{\texttt{#1}}\fi
\expandafter\ifx\csname urlprefix\endcsname\relax\def\urlprefix{URL }\fi
\providecommand{\bibinfo}[2]{#2}
\providecommand{\eprint}[2][]{\url{#2}}

\bibitem[{\citenamefont{Loss and DiVincenzo}(1998)}]{DiVi98}
\bibinfo{author}{\bibfnamefont{D.}~\bibnamefont{Loss}} \bibnamefont{and}
  \bibinfo{author}{\bibfnamefont{D.}~\bibnamefont{DiVincenzo}},
  \bibinfo{journal}{Phys. Rev. A} \textbf{\bibinfo{volume}{57}},
  \bibinfo{pages}{120} (\bibinfo{year}{1998}).

\bibitem[{\citenamefont{Burkard et~al.}(1999)\citenamefont{Burkard, Loss, and
  DiVincenzo}}]{Burk99}
\bibinfo{author}{\bibfnamefont{G.}~\bibnamefont{Burkard}},
  \bibinfo{author}{\bibfnamefont{D.}~\bibnamefont{Loss}}, \bibnamefont{and}
  \bibinfo{author}{\bibfnamefont{D.~P.} \bibnamefont{DiVincenzo}},
  \bibinfo{journal}{Phys. Rev. B} \textbf{\bibinfo{volume}{59}},
  \bibinfo{pages}{2070} (\bibinfo{year}{1999}).

\bibitem[{\citenamefont{Makhlin et~al.}(2001)\citenamefont{Makhlin, Sch\"{o}n,
  and Shnirman}}]{Mak01}
\bibinfo{author}{\bibfnamefont{Y.}~\bibnamefont{Makhlin}},
  \bibinfo{author}{\bibfnamefont{G.}~\bibnamefont{Sch\"{o}n}},
  \bibnamefont{and} \bibinfo{author}{\bibfnamefont{A.}~\bibnamefont{Shnirman}},
  \bibinfo{journal}{Rev. Mod. Phys.} \textbf{\bibinfo{volume}{73}},
  \bibinfo{pages}{357} (\bibinfo{year}{2001}).

\bibitem[{\citenamefont{Nakamura et~al.}(1999)\citenamefont{Nakamura, Pashkin,
  and Tsai}}]{Nak99}
\bibinfo{author}{\bibfnamefont{Y.}~\bibnamefont{Nakamura}},
  \bibinfo{author}{\bibfnamefont{Y.~A.} \bibnamefont{Pashkin}},
  \bibnamefont{and} \bibinfo{author}{\bibfnamefont{J.~S.} \bibnamefont{Tsai}},
  \bibinfo{journal}{Nature} \textbf{\bibinfo{volume}{398}},
  \bibinfo{pages}{786} (\bibinfo{year}{1999}).

\bibitem[{\citenamefont{Yamamoto et~al.}(2003)\citenamefont{Yamamoto, Pashkin,
  Astafiev, Nakamura, and Tsai}}]{Yam03}
\bibinfo{author}{\bibfnamefont{T.}~\bibnamefont{Yamamoto}},
  \bibinfo{author}{\bibfnamefont{Y.~A.} \bibnamefont{Pashkin}},
  \bibinfo{author}{\bibfnamefont{O.}~\bibnamefont{Astafiev}},
  \bibinfo{author}{\bibfnamefont{Y.}~\bibnamefont{Nakamura}}, \bibnamefont{and}
  \bibinfo{author}{\bibfnamefont{J.~S.} \bibnamefont{Tsai}},
  \bibinfo{journal}{Nature} \textbf{\bibinfo{volume}{425}},
  \bibinfo{pages}{941} (\bibinfo{year}{2003}).

\bibitem[{\citenamefont{Clarke and Wilhelm}(2008)}]{Clarke}
\bibinfo{author}{\bibfnamefont{J.}~\bibnamefont{Clarke}} \bibnamefont{and}
  \bibinfo{author}{\bibfnamefont{F.~K.} \bibnamefont{Wilhelm}},
  \bibinfo{journal}{Nature} \textbf{\bibinfo{volume}{453}},
  \bibinfo{pages}{1031} (\bibinfo{year}{2008}).

\bibitem[{\citenamefont{Vartiainen et~al.}(2004)\citenamefont{Vartiainen,
  Niskanen, Nakahara, and Salomaa}}]{Nakahara}
\bibinfo{author}{\bibfnamefont{J.~J.} \bibnamefont{Vartiainen}},
  \bibinfo{author}{\bibfnamefont{A.~O.} \bibnamefont{Niskanen}},
  \bibinfo{author}{\bibfnamefont{M.}~\bibnamefont{Nakahara}}, \bibnamefont{and}
  \bibinfo{author}{\bibfnamefont{M.~M.} \bibnamefont{Salomaa}},
  \bibinfo{journal}{Phys. Ref. A} \textbf{\bibinfo{volume}{70}},
  \bibinfo{pages}{012319} (\bibinfo{year}{2004}).

\bibitem[{\citenamefont{Mooij et~al.}(1999)\citenamefont{Mooij, Orlando,
  Levitov, Tian, van~der Wal, and Lloyd}}]{Mooij}
\bibinfo{author}{\bibfnamefont{J.~E.} \bibnamefont{Mooij}},
  \bibinfo{author}{\bibfnamefont{T.~P.} \bibnamefont{Orlando}},
  \bibinfo{author}{\bibfnamefont{L.}~\bibnamefont{Levitov}},
  \bibinfo{author}{\bibfnamefont{L.}~\bibnamefont{Tian}},
  \bibinfo{author}{\bibfnamefont{C.~H.} \bibnamefont{van~der Wal}},
  \bibnamefont{and} \bibinfo{author}{\bibfnamefont{S.}~\bibnamefont{Lloyd}},
  \bibinfo{journal}{Science} \textbf{\bibinfo{volume}{285}},
  \bibinfo{pages}{1036} (\bibinfo{year}{1999}).

\bibitem[{\citenamefont{Sp\"{o}rl et~al.}(2007)\citenamefont{Sp\"{o}rl,
  Schulte-Herbr\"{u}ggen, Glaser, Bergholm, Storcz, Ferber, and
  Wilhelm}}]{Spoerl07}
\bibinfo{author}{\bibfnamefont{A.}~\bibnamefont{Sp\"{o}rl}},
  \bibinfo{author}{\bibfnamefont{T.}~\bibnamefont{Schulte-Herbr\"{u}ggen}},
  \bibinfo{author}{\bibfnamefont{S.~J.} \bibnamefont{Glaser}},
  \bibinfo{author}{\bibfnamefont{V.}~\bibnamefont{Bergholm}},
  \bibinfo{author}{\bibfnamefont{M.~J.} \bibnamefont{Storcz}},
  \bibinfo{author}{\bibfnamefont{J.}~\bibnamefont{Ferber}}, \bibnamefont{and}
  \bibinfo{author}{\bibfnamefont{F.~K.} \bibnamefont{Wilhelm}},
  \bibinfo{journal}{Phys. Rev. A} \textbf{\bibinfo{volume}{75}},
  \bibinfo{pages}{012302} (\bibinfo{year}{2007}).

\bibitem[{\citenamefont{Krotov}(1996)}]{Krotov96}
\bibinfo{author}{\bibfnamefont{V.}~\bibnamefont{Krotov}},
  \emph{\bibinfo{title}{Global Methods in Optimal Control}}
  (\bibinfo{publisher}{Marcel Dekker}, \bibinfo{year}{1996}).

\bibitem[{\citenamefont{Sussmann and Willems}(1997)}]{Suss97}
\bibinfo{author}{\bibfnamefont{H.}~\bibnamefont{Sussmann}} \bibnamefont{and}
  \bibinfo{author}{\bibfnamefont{J.}~\bibnamefont{Willems}},
  \bibinfo{journal}{IEEE Control Systems} \textbf{\bibinfo{volume}{17}},
  \bibinfo{pages}{32} (\bibinfo{year}{1997}).

\bibitem[{\citenamefont{Bryson and Ho}(1975)}]{Bryson75}
\bibinfo{author}{\bibfnamefont{A.~E.} \bibnamefont{Bryson}} \bibnamefont{and}
  \bibinfo{author}{\bibfnamefont{Y.~C.} \bibnamefont{Ho}},
  \emph{\bibinfo{title}{Applied optimal control: Optimization, estimation, and
  control}} (\bibinfo{publisher}{Hemisphere Publishing}, \bibinfo{year}{1975}).

\bibitem[{\citenamefont{Hanson}(2007)}]{Hanson07b}
\bibinfo{author}{\bibfnamefont{F.~B.} \bibnamefont{Hanson}},
  \emph{\bibinfo{title}{Applied Stochastic Processes and Control for
  Jump-Diffusions:Modeling, Analysis, and Computation}}
  (\bibinfo{publisher}{SIAM Books, Philadelphia}, \bibinfo{year}{2007}).

\bibitem[{\citenamefont{Stengel}(1994)}]{Stengel94}
\bibinfo{author}{\bibfnamefont{R.~F.} \bibnamefont{Stengel}},
  \emph{\bibinfo{title}{Optimal Control and Estiamtion}}
  (\bibinfo{publisher}{Dover Publications}, \bibinfo{year}{1994}).

\bibitem[{\citenamefont{Greiner}(2003)}]{Grei03}
\bibinfo{author}{\bibfnamefont{W.}~\bibnamefont{Greiner}},
  \emph{\bibinfo{title}{Classical Mechanics}} (\bibinfo{publisher}{Springer,
  New York}, \bibinfo{year}{2003}).

\bibitem[{\citenamefont{Peirce et~al.}(1988)\citenamefont{Peirce, Dahleh, and
  Rabitz}}]{Peirce88}
\bibinfo{author}{\bibfnamefont{A.~P.} \bibnamefont{Peirce}},
  \bibinfo{author}{\bibfnamefont{M.~A.} \bibnamefont{Dahleh}},
  \bibnamefont{and} \bibinfo{author}{\bibfnamefont{H.}~\bibnamefont{Rabitz}},
  \bibinfo{journal}{Phys. Rev. A} \textbf{\bibinfo{volume}{37}},
  \bibinfo{pages}{4950} (\bibinfo{year}{1988}).

\bibitem[{\citenamefont{Tannor and Rice}(1985)}]{Tann85}
\bibinfo{author}{\bibfnamefont{D.~J.} \bibnamefont{Tannor}} \bibnamefont{and}
  \bibinfo{author}{\bibfnamefont{S.~A.} \bibnamefont{Rice}},
  \bibinfo{journal}{J. Chem. Phys.} \textbf{\bibinfo{volume}{83}},
  \bibinfo{pages}{5013} (\bibinfo{year}{1985}).

\bibitem[{\citenamefont{Sola et~al.}(1998)\citenamefont{Sola, Santamaria, and
  Tannor}}]{Sola98}
\bibinfo{author}{\bibfnamefont{I.~R.} \bibnamefont{Sola}},
  \bibinfo{author}{\bibfnamefont{J.}~\bibnamefont{Santamaria}},
  \bibnamefont{and} \bibinfo{author}{\bibfnamefont{D.~J.}
  \bibnamefont{Tannor}}, \bibinfo{journal}{J. Phys. Chem. A}
  \textbf{\bibinfo{volume}{102}}, \bibinfo{pages}{4301} (\bibinfo{year}{1998}).

\bibitem[{\citenamefont{Palao and Kosloff}(2003)}]{Palao03}
\bibinfo{author}{\bibfnamefont{J.~P.} \bibnamefont{Palao}} \bibnamefont{and}
  \bibinfo{author}{\bibfnamefont{R.}~\bibnamefont{Kosloff}},
  \bibinfo{journal}{Phys. Rev. A} \textbf{\bibinfo{volume}{68}},
  \bibinfo{pages}{062308} (\bibinfo{year}{2003}).

\bibitem[{\citenamefont{Palao and Kosloff}(2002)}]{Palao02}
\bibinfo{author}{\bibfnamefont{J.~P.} \bibnamefont{Palao}} \bibnamefont{and}
  \bibinfo{author}{\bibfnamefont{R.}~\bibnamefont{Kosloff}},
  \bibinfo{journal}{Phys. Rev. Lett.} \textbf{\bibinfo{volume}{89}},
  \bibinfo{pages}{188301} (\bibinfo{year}{2002}).

\bibitem[{\citenamefont{Grace et~al.}(2007)\citenamefont{Grace, Brif, Rabitz,
  Walmsley, Kosut, and Lidar}}]{Grace07}
\bibinfo{author}{\bibfnamefont{M.}~\bibnamefont{Grace}},
  \bibinfo{author}{\bibfnamefont{C.}~\bibnamefont{Brif}},
  \bibinfo{author}{\bibfnamefont{H.}~\bibnamefont{Rabitz}},
  \bibinfo{author}{\bibfnamefont{I.~A.} \bibnamefont{Walmsley}},
  \bibinfo{author}{\bibfnamefont{R.~L.} \bibnamefont{Kosut}}, \bibnamefont{and}
  \bibinfo{author}{\bibfnamefont{D.~A.} \bibnamefont{Lidar}},
  \bibinfo{journal}{J. Phys. B: At. Mol. Opt. Phys.}
  \textbf{\bibinfo{volume}{40}}, \bibinfo{pages}{103} (\bibinfo{year}{2007}).

\bibitem[{\citenamefont{Borzi et~al.}(2002)\citenamefont{Borzi, Stadler, and
  Hohenester}}]{Borzi02}
\bibinfo{author}{\bibfnamefont{A.}~\bibnamefont{Borzi}},
  \bibinfo{author}{\bibfnamefont{G.}~\bibnamefont{Stadler}}, \bibnamefont{and}
  \bibinfo{author}{\bibfnamefont{U.}~\bibnamefont{Hohenester}},
  \bibinfo{journal}{Phys. Rev. A} \textbf{\bibinfo{volume}{66}},
  \bibinfo{pages}{053811} (\bibinfo{year}{2002}).

\bibitem[{\citenamefont{Hohenester et~al.}(2007)\citenamefont{Hohenester,
  Rekdal, Borzi, and Schmiedmayer}}]{Hohen07}
\bibinfo{author}{\bibfnamefont{U.}~\bibnamefont{Hohenester}},
  \bibinfo{author}{\bibfnamefont{P.~K.} \bibnamefont{Rekdal}},
  \bibinfo{author}{\bibfnamefont{A.}~\bibnamefont{Borzi}}, \bibnamefont{and}
  \bibinfo{author}{\bibfnamefont{J.}~\bibnamefont{Schmiedmayer}},
  \bibinfo{journal}{Phys. Rev. A} \textbf{\bibinfo{volume}{75}},
  \bibinfo{pages}{023602} (\bibinfo{year}{2007}).

\bibitem[{\citenamefont{Jirari and P\"{o}tz}(2006)}]{Poetz2}
\bibinfo{author}{\bibfnamefont{H.}~\bibnamefont{Jirari}} \bibnamefont{and}
  \bibinfo{author}{\bibfnamefont{W.}~\bibnamefont{P\"{o}tz}},
  \bibinfo{journal}{Phys. Rev. A} \textbf{\bibinfo{volume}{74}},
  \bibinfo{pages}{022306} (\bibinfo{year}{2006}).

\bibitem[{\citenamefont{Jirari and P\"{o}tz}(2005)}]{Poetz3}
\bibinfo{author}{\bibfnamefont{H.}~\bibnamefont{Jirari}} \bibnamefont{and}
  \bibinfo{author}{\bibfnamefont{W.}~\bibnamefont{P\"{o}tz}},
  \bibinfo{journal}{Phys. Rev. A} \textbf{\bibinfo{volume}{72}},
  \bibinfo{pages}{013409} (\bibinfo{year}{2005}).

\bibitem[{\citenamefont{Tesch and deVivie Riedle}(2002)}]{Tesch02}
\bibinfo{author}{\bibfnamefont{C.~M.} \bibnamefont{Tesch}} \bibnamefont{and}
  \bibinfo{author}{\bibfnamefont{R.}~\bibnamefont{deVivie Riedle}},
  \bibinfo{journal}{Phys. Rev. Lett.} \textbf{\bibinfo{volume}{89}},
  \bibinfo{pages}{157901} (\bibinfo{year}{2002}).

\bibitem[{\citenamefont{Roloff and P\"{o}tz}(2007)}]{Rol07}
\bibinfo{author}{\bibfnamefont{R.}~\bibnamefont{Roloff}} \bibnamefont{and}
  \bibinfo{author}{\bibfnamefont{W.}~\bibnamefont{P\"{o}tz}},
  \bibinfo{journal}{Phys. Rev. B} \textbf{\bibinfo{volume}{76}},
  \bibinfo{pages}{075333} (\bibinfo{year}{2007}).

\bibitem[{\citenamefont{Wenin and P\"{o}tz}(2006)}]{Wenin06}
\bibinfo{author}{\bibfnamefont{M.}~\bibnamefont{Wenin}} \bibnamefont{and}
  \bibinfo{author}{\bibfnamefont{W.}~\bibnamefont{P\"{o}tz}},
  \bibinfo{journal}{Phys. Rev. A} \textbf{\bibinfo{volume}{74}},
  \bibinfo{pages}{022319} (\bibinfo{year}{2006}).

\bibitem[{\citenamefont{Montangero et~al.}(2007)\citenamefont{Montangero,
  Calarco, and Fazio}}]{Monta07}
\bibinfo{author}{\bibfnamefont{S.}~\bibnamefont{Montangero}},
  \bibinfo{author}{\bibfnamefont{T.}~\bibnamefont{Calarco}}, \bibnamefont{and}
  \bibinfo{author}{\bibfnamefont{R.}~\bibnamefont{Fazio}},
  \bibinfo{journal}{Phys. Rev. Lett.} \textbf{\bibinfo{volume}{99}},
  \bibinfo{pages}{170501} (\bibinfo{year}{2007}).

\bibitem[{\citenamefont{Wenin and P\"otz}(2008)}]{Wenin08c}
\bibinfo{author}{\bibfnamefont{M.}~\bibnamefont{Wenin}} \bibnamefont{and}
  \bibinfo{author}{\bibfnamefont{W.}~\bibnamefont{P\"otz}},
  \bibinfo{journal}{Phys. Rev. B {\bf 78}, 165118}  (\bibinfo{year}{2008}).

\bibitem[{\citenamefont{Roloff and P\"{o}tz}(2009)}]{Rol08}
\bibinfo{author}{\bibfnamefont{R.}~\bibnamefont{Roloff}} \bibnamefont{and}
  \bibinfo{author}{\bibfnamefont{W.}~\bibnamefont{P\"{o}tz}},
  \bibinfo{journal}{Phys. Rev. B} \textbf{\bibinfo{volume}{79}},
  \bibinfo{pages}{224516} (\bibinfo{year}{2009}).

\bibitem[{\citenamefont{Schulte-Herbr\"{u}ggen
  et~al.}(2006)\citenamefont{Schulte-Herbr\"{u}ggen, Sp\"{o}rl, Khaneja, and
  Glaser}}]{Schulte06}
\bibinfo{author}{\bibfnamefont{T.}~\bibnamefont{Schulte-Herbr\"{u}ggen}},
  \bibinfo{author}{\bibfnamefont{A.}~\bibnamefont{Sp\"{o}rl}},
  \bibinfo{author}{\bibfnamefont{N.}~\bibnamefont{Khaneja}}, \bibnamefont{and}
  \bibinfo{author}{\bibfnamefont{S.}~\bibnamefont{Glaser}},
  \bibinfo{journal}{arXiv:quant-ph/0609037 v1}  (\bibinfo{year}{2006}).

\bibitem[{\citenamefont{Press et~al.}(1992)\citenamefont{Press, Teukolsky,
  Vetterling, and Flannery}}]{Press92}
\bibinfo{author}{\bibfnamefont{W.~H.} \bibnamefont{Press}},
  \bibinfo{author}{\bibfnamefont{S.}~\bibnamefont{Teukolsky}},
  \bibinfo{author}{\bibfnamefont{W.~T.} \bibnamefont{Vetterling}},
  \bibnamefont{and} \bibinfo{author}{\bibfnamefont{B.~P.}
  \bibnamefont{Flannery}}, \emph{\bibinfo{title}{Numerical Recipes}}
  (\bibinfo{publisher}{Cambrige University Press, Cambridge},
  \bibinfo{year}{1992}).

\bibitem[{\citenamefont{Schmitt}(2001)}]{Schmitt01}
\bibinfo{author}{\bibfnamefont{L.}~\bibnamefont{Schmitt}},
  \bibinfo{journal}{Theoretical Computer Science}
  \textbf{\bibinfo{volume}{259}}, \bibinfo{pages}{1} (\bibinfo{year}{2001}).

\bibitem[{\citenamefont{Storn and Price}(1997)}]{Storn97}
\bibinfo{author}{\bibfnamefont{R.}~\bibnamefont{Storn}} \bibnamefont{and}
  \bibinfo{author}{\bibfnamefont{K.}~\bibnamefont{Price}}, \bibinfo{journal}{J.
  Global Optimization} \textbf{\bibinfo{volume}{11}}, \bibinfo{pages}{341}
  (\bibinfo{year}{1997}).

\bibitem[{\citenamefont{Kirkpatrick et~al.}(1983)\citenamefont{Kirkpatrick,
  Gelatt, and Vecchi}}]{Kirk83}
\bibinfo{author}{\bibfnamefont{S.}~\bibnamefont{Kirkpatrick}},
  \bibinfo{author}{\bibfnamefont{C.~D.} \bibnamefont{Gelatt}},
  \bibnamefont{and} \bibinfo{author}{\bibfnamefont{M.~P.}
  \bibnamefont{Vecchi}}, \bibinfo{journal}{Science}
  \textbf{\bibinfo{volume}{220}}, \bibinfo{pages}{671} (\bibinfo{year}{1983}).

\bibitem[{\citenamefont{Amstrup et~al.}(1993)\citenamefont{Amstrup, Doll,
  Sauerbrey, Szabo, and Lorincz}}]{Amstrup93}
\bibinfo{author}{\bibfnamefont{B.}~\bibnamefont{Amstrup}},
  \bibinfo{author}{\bibfnamefont{J.~D.} \bibnamefont{Doll}},
  \bibinfo{author}{\bibfnamefont{R.~A.} \bibnamefont{Sauerbrey}},
  \bibinfo{author}{\bibfnamefont{G.}~\bibnamefont{Szabo}}, \bibnamefont{and}
  \bibinfo{author}{\bibfnamefont{A.}~\bibnamefont{Lorincz}},
  \bibinfo{journal}{Phys. Rev. A} \textbf{\bibinfo{volume}{48}},
  \bibinfo{pages}{3820} (\bibinfo{year}{1993}).

\bibitem[{\citenamefont{P\"{o}tz}(2006)}]{Poetz06}
\bibinfo{author}{\bibfnamefont{W.}~\bibnamefont{P\"{o}tz}},
  \bibinfo{journal}{Appl. Phys. Lett.} \textbf{\bibinfo{volume}{89}},
  \bibinfo{pages}{254102} (\bibinfo{year}{2006}).

\bibitem[{\citenamefont{Feynman}(1982)}]{Feyn82}
\bibinfo{author}{\bibfnamefont{R.~P.} \bibnamefont{Feynman}},
  \bibinfo{journal}{J. Theor. Phys.} \textbf{\bibinfo{volume}{21}},
  \bibinfo{pages}{467} (\bibinfo{year}{1982}).

\bibitem[{\citenamefont{Fick and Sauermann}(1990)}]{Fick90}
\bibinfo{author}{\bibfnamefont{E.}~\bibnamefont{Fick}} \bibnamefont{and}
  \bibinfo{author}{\bibfnamefont{G.}~\bibnamefont{Sauermann}},
  \emph{\bibinfo{title}{The Quantum Statistics of Dynamic Processes}}
  (\bibinfo{publisher}{Springer, Berlin}, \bibinfo{year}{1990}).

\bibitem[{\citenamefont{Nielsen and Chuang}(2002)}]{Niel02}
\bibinfo{author}{\bibfnamefont{M.}~\bibnamefont{Nielsen}} \bibnamefont{and}
  \bibinfo{author}{\bibfnamefont{I.}~\bibnamefont{Chuang}},
  \emph{\bibinfo{title}{Quantum computation and Quantum Information}}
  (\bibinfo{publisher}{Cambridge University Press}, \bibinfo{year}{2002}).

\bibitem[{\citenamefont{Bouwmeester et~al.}(2000)\citenamefont{Bouwmeester,
  Ekert, and Zeilinger}}]{Zeil00}
\bibinfo{editor}{\bibfnamefont{D.}~\bibnamefont{Bouwmeester}},
  \bibinfo{editor}{\bibfnamefont{A.}~\bibnamefont{Ekert}}, \bibnamefont{and}
  \bibinfo{editor}{\bibfnamefont{A.}~\bibnamefont{Zeilinger}}, eds.,
  \emph{\bibinfo{title}{The Physics of Quantum Information: Quantum
  Cryptography, Quantum Teleportation, Quantum Computation}}
  (\bibinfo{publisher}{Springer, Berlin Heidelberg New York},
  \bibinfo{year}{2000}).

\bibitem[{\citenamefont{Barenco et~al.}(1995)\citenamefont{Barenco, Bennett,
  Cleve, DiVincenzo, Margolus, Shor, Sleator, Smolin, , and
  Weinfurter}}]{Bar95}
\bibinfo{author}{\bibfnamefont{A.}~\bibnamefont{Barenco}},
  \bibinfo{author}{\bibfnamefont{C.~H.} \bibnamefont{Bennett}},
  \bibinfo{author}{\bibfnamefont{R.}~\bibnamefont{Cleve}},
  \bibinfo{author}{\bibfnamefont{D.~P.} \bibnamefont{DiVincenzo}},
  \bibinfo{author}{\bibfnamefont{N.}~\bibnamefont{Margolus}},
  \bibinfo{author}{\bibfnamefont{P.}~\bibnamefont{Shor}},
  \bibinfo{author}{\bibfnamefont{T.}~\bibnamefont{Sleator}},
  \bibinfo{author}{\bibfnamefont{J.~A.} \bibnamefont{Smolin}}, ,
  \bibnamefont{and}
  \bibinfo{author}{\bibfnamefont{H.}~\bibnamefont{Weinfurter}},
  \bibinfo{journal}{Phys. Rev. A} \textbf{\bibinfo{volume}{52}},
  \bibinfo{pages}{3457 } (\bibinfo{year}{1995}).

\bibitem[{\citenamefont{Deutsch et~al.}(1995)\citenamefont{Deutsch, Barenco,
  and Ekert}}]{Deu95}
\bibinfo{author}{\bibfnamefont{D.}~\bibnamefont{Deutsch}},
  \bibinfo{author}{\bibfnamefont{A.}~\bibnamefont{Barenco}}, \bibnamefont{and}
  \bibinfo{author}{\bibfnamefont{A.}~\bibnamefont{Ekert}},
  \bibinfo{journal}{Proc. R. Soc. Lond. A} \textbf{\bibinfo{volume}{449}},
  \bibinfo{pages}{669} (\bibinfo{year}{1995}).

\bibitem[{\citenamefont{Gardiner and Zoller}(2000)}]{Gard00}
\bibinfo{author}{\bibfnamefont{C.}~\bibnamefont{Gardiner}} \bibnamefont{and}
  \bibinfo{author}{\bibfnamefont{P.}~\bibnamefont{Zoller}},
  \emph{\bibinfo{title}{Quantum Noise: A Handbook of Markovian and
  Non-Markovian Quantum Stochastic Methods with Applications to Quantum
  Optics}} (\bibinfo{publisher}{Springer}, \bibinfo{year}{2000}).

\bibitem[{\citenamefont{Breuer and Petruccione}(2003)}]{Breu03}
\bibinfo{author}{\bibfnamefont{H.}~\bibnamefont{Breuer}} \bibnamefont{and}
  \bibinfo{author}{\bibfnamefont{F.}~\bibnamefont{Petruccione}},
  \emph{\bibinfo{title}{The Theory of Open Quantum Systems}}
  (\bibinfo{publisher}{Oxford}, \bibinfo{year}{2003}).

\bibitem[{\citenamefont{Wenin and P\"{o}tz}(2008{\natexlab{a}})}]{Wenin08}
\bibinfo{author}{\bibfnamefont{M.}~\bibnamefont{Wenin}} \bibnamefont{and}
  \bibinfo{author}{\bibfnamefont{W.}~\bibnamefont{P\"{o}tz}},
  \bibinfo{journal}{Appl. Phys. Lett.} \textbf{\bibinfo{volume}{92}},
  \bibinfo{pages}{103509} (\bibinfo{year}{2008}{\natexlab{a}}).

\bibitem[{\citenamefont{Wenin and P\"{o}tz}(2008{\natexlab{b}})}]{Wenin08b}
\bibinfo{author}{\bibfnamefont{M.}~\bibnamefont{Wenin}} \bibnamefont{and}
  \bibinfo{author}{\bibfnamefont{W.}~\bibnamefont{P\"{o}tz}},
  \bibinfo{journal}{Phys. Rev. A} \textbf{\bibinfo{volume}{78}},
  \bibinfo{pages}{012358} (\bibinfo{year}{2008}{\natexlab{b}}).

\bibitem[{\citenamefont{Leggett et~al.}(1987)\citenamefont{Leggett,
  Chakravarty, Dorsey, Fisher, Garg, and Zwerger}}]{Leg87}
\bibinfo{author}{\bibfnamefont{A.}~\bibnamefont{Leggett}},
  \bibinfo{author}{\bibfnamefont{S.}~\bibnamefont{Chakravarty}},
  \bibinfo{author}{\bibfnamefont{A.~T.} \bibnamefont{Dorsey}},
  \bibinfo{author}{\bibfnamefont{M.~P.~A.} \bibnamefont{Fisher}},
  \bibinfo{author}{\bibfnamefont{A.}~\bibnamefont{Garg}}, \bibnamefont{and}
  \bibinfo{author}{\bibfnamefont{W.}~\bibnamefont{Zwerger}},
  \bibinfo{journal}{Rev. Mod. Phys.} \textbf{\bibinfo{volume}{59}},
  \bibinfo{pages}{1} (\bibinfo{year}{1987}).

\bibitem[{\citenamefont{Weiss}(1999)}]{Weiss99}
\bibinfo{author}{\bibfnamefont{U.}~\bibnamefont{Weiss}},
  \emph{\bibinfo{title}{Quantum dissipative systems}}
  (\bibinfo{publisher}{World Scientific}, \bibinfo{year}{1999}).

\bibitem[{\citenamefont{Carmichael}(2002)}]{Car02}
\bibinfo{author}{\bibfnamefont{H.}~\bibnamefont{Carmichael}},
  \emph{\bibinfo{title}{Statistical Methods in Quantum Optics 1: Master
  Equations and Fokker--Planck Equations}} (\bibinfo{publisher}{Springer},
  \bibinfo{year}{2002}).

\bibitem[{\citenamefont{Mahan}(2000)}]{Mahan}
\bibinfo{author}{\bibfnamefont{G.}~\bibnamefont{Mahan}},
  \emph{\bibinfo{title}{Many--Particle Physics}}
  (\bibinfo{publisher}{Many--Particle Physics}, \bibinfo{year}{2000}).

\bibitem[{\citenamefont{Nesi et~al.}(2007)\citenamefont{Nesi, Paladino,
  Thorwart, and Grifoni}}]{Nesi07}
\bibinfo{author}{\bibfnamefont{F.}~\bibnamefont{Nesi}},
  \bibinfo{author}{\bibfnamefont{E.}~\bibnamefont{Paladino}},
  \bibinfo{author}{\bibfnamefont{M.}~\bibnamefont{Thorwart}}, \bibnamefont{and}
  \bibinfo{author}{\bibfnamefont{M.}~\bibnamefont{Grifoni}},
  \bibinfo{journal}{EPL} \textbf{\bibinfo{volume}{80}}, \bibinfo{pages}{40005}
  (\bibinfo{year}{2007}).

\bibitem[{\citenamefont{Hartmann et~al.}(2000)\citenamefont{Hartmann, Goychuk,
  Grifoni, and H\"{a}nggi}}]{Hart00}
\bibinfo{author}{\bibfnamefont{L.}~\bibnamefont{Hartmann}},
  \bibinfo{author}{\bibfnamefont{I.}~\bibnamefont{Goychuk}},
  \bibinfo{author}{\bibfnamefont{M.}~\bibnamefont{Grifoni}}, \bibnamefont{and}
  \bibinfo{author}{\bibfnamefont{P.}~\bibnamefont{H\"{a}nggi}},
  \bibinfo{journal}{Phys. Rev. E} \textbf{\bibinfo{volume}{61}},
  \bibinfo{pages}{4687} (\bibinfo{year}{2000}).

\bibitem[{\citenamefont{Grifoni et~al.}(1996)\citenamefont{Grifoni, Sassetti,
  and Weiss}}]{Grif96}
\bibinfo{author}{\bibfnamefont{M.}~\bibnamefont{Grifoni}},
  \bibinfo{author}{\bibfnamefont{M.}~\bibnamefont{Sassetti}}, \bibnamefont{and}
  \bibinfo{author}{\bibfnamefont{U.}~\bibnamefont{Weiss}},
  \bibinfo{journal}{Phys. Rev. E} \textbf{\bibinfo{volume}{53}},
  \bibinfo{pages}{2033} (\bibinfo{year}{1996}).

\bibitem[{\citenamefont{Aslangul et~al.}(1986)\citenamefont{Aslangul, Pottier,
  and Saint-James}}]{Aslan86}
\bibinfo{author}{\bibfnamefont{C.}~\bibnamefont{Aslangul}},
  \bibinfo{author}{\bibfnamefont{N.}~\bibnamefont{Pottier}}, \bibnamefont{and}
  \bibinfo{author}{\bibfnamefont{D.}~\bibnamefont{Saint-James}},
  \bibinfo{journal}{J. Physique} \textbf{\bibinfo{volume}{47}},
  \bibinfo{pages}{1657} (\bibinfo{year}{1986}).

\bibitem[{\citenamefont{Reina et~al.}(2002)\citenamefont{Reina, Quiroga, and
  Johnson}}]{Rei02}
\bibinfo{author}{\bibfnamefont{J.}~\bibnamefont{Reina}},
  \bibinfo{author}{\bibfnamefont{L.}~\bibnamefont{Quiroga}}, \bibnamefont{and}
  \bibinfo{author}{\bibfnamefont{N.}~\bibnamefont{Johnson}},
  \bibinfo{journal}{Phys. Rev. A} \textbf{\bibinfo{volume}{65}},
  \bibinfo{pages}{032326} (\bibinfo{year}{2002}).

\bibitem[{\citenamefont{Prokof'ev and Stamp}(2000)}]{Prokof00}
\bibinfo{author}{\bibfnamefont{N.~V.} \bibnamefont{Prokof'ev}}
  \bibnamefont{and} \bibinfo{author}{\bibfnamefont{P.~C.~E.}
  \bibnamefont{Stamp}}, \bibinfo{journal}{Rep. Prog. Phys.}
  \textbf{\bibinfo{volume}{63}}, \bibinfo{pages}{669} (\bibinfo{year}{2000}).

\bibitem[{\citenamefont{Mukamel}(1995)}]{Muk95}
\bibinfo{author}{\bibfnamefont{S.}~\bibnamefont{Mukamel}},
  \emph{\bibinfo{title}{Principles of nonlinear optical spectroscopy}}
  (\bibinfo{publisher}{Oxford University Press}, \bibinfo{year}{1995}).

\bibitem[{\citenamefont{Sklarz and Tannor}(2006)}]{Sklarz06}
\bibinfo{author}{\bibfnamefont{S.~E.} \bibnamefont{Sklarz}} \bibnamefont{and}
  \bibinfo{author}{\bibfnamefont{D.~J.} \bibnamefont{Tannor}},
  \bibinfo{journal}{J. Chem. Phys.} \textbf{\bibinfo{volume}{322}},
  \bibinfo{pages}{87} (\bibinfo{year}{2006}).

\bibitem[{\citenamefont{Rabitz et~al.}(2005)\citenamefont{Rabitz, Hsieh, and
  Rosenthal}}]{Rabitz}
\bibinfo{author}{\bibfnamefont{H.}~\bibnamefont{Rabitz}},
  \bibinfo{author}{\bibfnamefont{M.}~\bibnamefont{Hsieh}}, \bibnamefont{and}
  \bibinfo{author}{\bibfnamefont{C.}~\bibnamefont{Rosenthal}},
  \bibinfo{journal}{Phys. Rev. A {\bf 72}, 052337}  (\bibinfo{year}{2005}).

\bibitem[{\citenamefont{Gordon and Rice}(1997)}]{Gord97}
\bibinfo{author}{\bibfnamefont{R.~J.} \bibnamefont{Gordon}} \bibnamefont{and}
  \bibinfo{author}{\bibfnamefont{S.~A.} \bibnamefont{Rice}},
  \bibinfo{journal}{Annu. Rev. Phys. Chem.} \textbf{\bibinfo{volume}{48}},
  \bibinfo{pages}{601} (\bibinfo{year}{1997}).

\bibitem[{\citenamefont{Dupont et~al.}(1995)\citenamefont{Dupont, Corkum, Liu,
  Buchanan, and Wasilewski}}]{Dupond95}
\bibinfo{author}{\bibfnamefont{E.}~\bibnamefont{Dupont}},
  \bibinfo{author}{\bibfnamefont{P.~B.} \bibnamefont{Corkum}},
  \bibinfo{author}{\bibfnamefont{H.~C.} \bibnamefont{Liu}},
  \bibinfo{author}{\bibfnamefont{M.}~\bibnamefont{Buchanan}}, \bibnamefont{and}
  \bibinfo{author}{\bibfnamefont{Z.~R.} \bibnamefont{Wasilewski}},
  \bibinfo{journal}{Phys. Rev. Lett.} \textbf{\bibinfo{volume}{74}},
  \bibinfo{pages}{3596} (\bibinfo{year}{1995}).

\bibitem[{\citenamefont{Hach\`{e} et~al.}(1997)\citenamefont{Hach\`{e},
  Kostoulas, Atanasov, Hughes, Sipe, , and van Driel}}]{vanDriel97}
\bibinfo{author}{\bibfnamefont{A.}~\bibnamefont{Hach\`{e}}},
  \bibinfo{author}{\bibfnamefont{Y.}~\bibnamefont{Kostoulas}},
  \bibinfo{author}{\bibfnamefont{R.}~\bibnamefont{Atanasov}},
  \bibinfo{author}{\bibfnamefont{J.~L.~P.} \bibnamefont{Hughes}},
  \bibinfo{author}{\bibfnamefont{J.~E.} \bibnamefont{Sipe}}, ,
  \bibnamefont{and} \bibinfo{author}{\bibfnamefont{H.~M.} \bibnamefont{van
  Driel}}, \bibinfo{journal}{Phys. Rev. Lett.} \textbf{\bibinfo{volume}{78}},
  \bibinfo{pages}{306} (\bibinfo{year}{1997}).

\bibitem[{\citenamefont{Bhat and Sipe}(2000)}]{Sipe00}
\bibinfo{author}{\bibfnamefont{R.~D.~R.} \bibnamefont{Bhat}} \bibnamefont{and}
  \bibinfo{author}{\bibfnamefont{J.~E.} \bibnamefont{Sipe}},
  \bibinfo{journal}{Phys. Rev. Lett.} \textbf{\bibinfo{volume}{85}},
  \bibinfo{pages}{5432} (\bibinfo{year}{2000}).

\bibitem[{\citenamefont{Stevens et~al.}(2003)\citenamefont{Stevens, Najmaie,
  Bhat, Sipe, van Driel, and Smirl}}]{Smirl03}
\bibinfo{author}{\bibfnamefont{M.~J.} \bibnamefont{Stevens}},
  \bibinfo{author}{\bibfnamefont{A.}~\bibnamefont{Najmaie}},
  \bibinfo{author}{\bibfnamefont{R.~D.~R.} \bibnamefont{Bhat}},
  \bibinfo{author}{\bibfnamefont{J.~E.} \bibnamefont{Sipe}},
  \bibinfo{author}{\bibfnamefont{H.~M.} \bibnamefont{van Driel}},
  \bibnamefont{and} \bibinfo{author}{\bibfnamefont{A.~L.} \bibnamefont{Smirl}},
  \bibinfo{journal}{J. Appl. Phys.} \textbf{\bibinfo{volume}{94}},
  \bibinfo{pages}{4999} (\bibinfo{year}{2003}).

\bibitem[{\citenamefont{Kerachian et~al.}(2004)\citenamefont{Kerachian, Nemec,
  van Driel, and Smirl}}]{Smirl04}
\bibinfo{author}{\bibfnamefont{Y.}~\bibnamefont{Kerachian}},
  \bibinfo{author}{\bibfnamefont{P.}~\bibnamefont{Nemec}},
  \bibinfo{author}{\bibfnamefont{H.~M.} \bibnamefont{van Driel}},
  \bibnamefont{and} \bibinfo{author}{\bibfnamefont{A.~L.} \bibnamefont{Smirl}},
  \bibinfo{journal}{J. Appl. Phys.} \textbf{\bibinfo{volume}{96}},
  \bibinfo{pages}{430} (\bibinfo{year}{2004}).

\bibitem[{\citenamefont{P\"{o}tz and Schroeder}(1999)}]{Poetz99}
\bibinfo{editor}{\bibfnamefont{W.}~\bibnamefont{P\"{o}tz}} \bibnamefont{and}
  \bibinfo{editor}{\bibfnamefont{W.~A.} \bibnamefont{Schroeder}}, eds.,
  \emph{\bibinfo{title}{Coherent Control in Atoms, Molecules, and
  Semiconductors}} (\bibinfo{publisher}{Kluwer, Dordrecht},
  \bibinfo{year}{1999}).

\bibitem[{\citenamefont{Hohenester and Stadler}(2004)}]{hohenester}
\bibinfo{author}{\bibfnamefont{U.}~\bibnamefont{Hohenester}} \bibnamefont{and}
  \bibinfo{author}{\bibfnamefont{G.}~\bibnamefont{Stadler}},
  \bibinfo{journal}{Phys. Rev. Lett.} \textbf{\bibinfo{volume}{92}},
  \bibinfo{pages}{196801} (\bibinfo{year}{2004}).

\bibitem[{\citenamefont{Romero et~al.}(2003)\citenamefont{Romero, Laverde, and
  Ardilla}}]{Romero}
\bibinfo{author}{\bibfnamefont{K.~M.~F.} \bibnamefont{Romero}},
  \bibinfo{author}{\bibfnamefont{G.~U.} \bibnamefont{Laverde}},
  \bibnamefont{and} \bibinfo{author}{\bibfnamefont{F.~T.}
  \bibnamefont{Ardilla}}, \bibinfo{journal}{J. Phys. A: Math. Gen.}
  \textbf{\bibinfo{volume}{36}}, \bibinfo{pages}{841} (\bibinfo{year}{2003}).

\bibitem[{\citenamefont{Emmanouilidou et~al.}(2000)\citenamefont{Emmanouilidou,
  Zhao, Ao, and Niu}}]{Zhao}
\bibinfo{author}{\bibfnamefont{A.}~\bibnamefont{Emmanouilidou}},
  \bibinfo{author}{\bibfnamefont{X.~G.} \bibnamefont{Zhao}},
  \bibinfo{author}{\bibfnamefont{P.}~\bibnamefont{Ao}}, \bibnamefont{and}
  \bibinfo{author}{\bibfnamefont{Q.}~\bibnamefont{Niu}},
  \bibinfo{journal}{Phys. Rev. Lett.} \textbf{\bibinfo{volume}{85}},
  \bibinfo{pages}{1626} (\bibinfo{year}{2000}).

\bibitem[{\citenamefont{Wenin and P\"otz}(2007)}]{Proc}
\bibinfo{author}{\bibfnamefont{M.}~\bibnamefont{Wenin}} \bibnamefont{and}
  \bibinfo{author}{\bibfnamefont{W.}~\bibnamefont{P\"otz}},
  \bibinfo{journal}{Proceedings of the 11th International Workshop on
  Computational Electronics (IWCE--11), May 25--27, 2006, Vienna, Austria.
  Journal of Computational Electronics, Vol. 6, Issue 1--3, pp. 271--274}
  (\bibinfo{year}{2007}).

\bibitem[{\citenamefont{Wu et~al.}(2003)\citenamefont{Wu, Pechen, Brif, and
  Rabitz}}]{Wu2}
\bibinfo{author}{\bibfnamefont{R.}~\bibnamefont{Wu}},
  \bibinfo{author}{\bibfnamefont{A.}~\bibnamefont{Pechen}},
  \bibinfo{author}{\bibfnamefont{C.}~\bibnamefont{Brif}}, \bibnamefont{and}
  \bibinfo{author}{\bibfnamefont{H.}~\bibnamefont{Rabitz}},
  \bibinfo{journal}{J. Phys. A: Math. Theor.} \textbf{\bibinfo{volume}{40}},
  \bibinfo{pages}{5681} (\bibinfo{year}{2003}).

\bibitem[{\citenamefont{Braun}(2000)}]{Braun}
\bibinfo{author}{\bibfnamefont{D.}~\bibnamefont{Braun}},
  \emph{\bibinfo{title}{Dissipative Quantum Chaos and Decoherence}}
  (\bibinfo{publisher}{Springer, Berlin, Auflage 1}, \bibinfo{year}{2000}).

\bibitem[{\citenamefont{Karasik et~al.}(2008)\citenamefont{Karasik, Marzlin,
  Sanders, and Whaley}}]{Karasik}
\bibinfo{author}{\bibfnamefont{R.~I.} \bibnamefont{Karasik}},
  \bibinfo{author}{\bibfnamefont{K.~P.} \bibnamefont{Marzlin}},
  \bibinfo{author}{\bibfnamefont{B.}~\bibnamefont{Sanders}}, \bibnamefont{and}
  \bibinfo{author}{\bibfnamefont{K.~B.} \bibnamefont{Whaley}},
  \bibinfo{journal}{Phys. Rev. A} \textbf{\bibinfo{volume}{77}},
  \bibinfo{pages}{052301} (\bibinfo{year}{2008}).

\bibitem[{\citenamefont{Lidar et~al.}(1998)\citenamefont{Lidar, Chuang, and
  Whaley}}]{Lidar}
\bibinfo{author}{\bibfnamefont{D.~A.} \bibnamefont{Lidar}},
  \bibinfo{author}{\bibfnamefont{I.~L.} \bibnamefont{Chuang}},
  \bibnamefont{and} \bibinfo{author}{\bibfnamefont{K.~B.}
  \bibnamefont{Whaley}}, \bibinfo{journal}{Phys. Rev. Lett. {\bf 81}, 12}
  (\bibinfo{year}{1998}).

\bibitem[{\citenamefont{Viola and Lloyd}(1998)}]{Viola}
\bibinfo{author}{\bibfnamefont{L.}~\bibnamefont{Viola}} \bibnamefont{and}
  \bibinfo{author}{\bibfnamefont{S.}~\bibnamefont{Lloyd}},
  \bibinfo{journal}{Phys. Rev. A {\bf 58}, 4}  (\bibinfo{year}{1998}).

\bibitem[{\citenamefont{Viola et~al.}(1999)\citenamefont{Viola, Knill, and
  Lloyd}}]{Viola1}
\bibinfo{author}{\bibfnamefont{L.}~\bibnamefont{Viola}},
  \bibinfo{author}{\bibfnamefont{E.}~\bibnamefont{Knill}}, \bibnamefont{and}
  \bibinfo{author}{\bibfnamefont{S.}~\bibnamefont{Lloyd}},
  \bibinfo{journal}{Phys. Rev. Lett. {\bf 82}, 12}  (\bibinfo{year}{1999}).

\bibitem[{\citenamefont{Gorden and Lidar}(2008)}]{Lidar2}
\bibinfo{author}{\bibfnamefont{G.}~\bibnamefont{Gorden}} \bibnamefont{and}
  \bibinfo{author}{\bibfnamefont{D.~A.} \bibnamefont{Lidar}},
  \bibinfo{journal}{Phys. Rev. Lett. {\bf 101}, 010403}
  (\bibinfo{year}{2008}).

\bibitem[{\citenamefont{Vitali and Tombesi}(1999)}]{Vitali}
\bibinfo{author}{\bibfnamefont{D.}~\bibnamefont{Vitali}} \bibnamefont{and}
  \bibinfo{author}{\bibfnamefont{P.}~\bibnamefont{Tombesi}},
  \bibinfo{journal}{Phys. Rev. A {\bf 59}, 4178}  (\bibinfo{year}{1999}).

\bibitem[{\citenamefont{Roloff et~al.}(2007)\citenamefont{Roloff, Wenin, and
  P\"{o}tz}}]{Rol07b}
\bibinfo{author}{\bibfnamefont{R.}~\bibnamefont{Roloff}},
  \bibinfo{author}{\bibfnamefont{M.}~\bibnamefont{Wenin}}, \bibnamefont{and}
  \bibinfo{author}{\bibfnamefont{W.}~\bibnamefont{P\"{o}tz}},
  \bibinfo{journal}{J. Comput. Electr.} \textbf{\bibinfo{volume}{8}},
  \bibinfo{pages}{29} (\bibinfo{year}{2009}).

\bibitem[{\citenamefont{Hu and P\"{o}tz}(1999)}]{Hu99}
\bibinfo{author}{\bibfnamefont{X.}~\bibnamefont{Hu}} \bibnamefont{and}
  \bibinfo{author}{\bibfnamefont{W.}~\bibnamefont{P\"{o}tz}},
  \bibinfo{journal}{Phys. Rev. Lett.} \textbf{\bibinfo{volume}{82}},
  \bibinfo{pages}{3116} (\bibinfo{year}{1999}).

\bibitem[{\citenamefont{Rossi and Kuhn}(2002)}]{Rossi02}
\bibinfo{author}{\bibfnamefont{F.}~\bibnamefont{Rossi}} \bibnamefont{and}
  \bibinfo{author}{\bibfnamefont{T.}~\bibnamefont{Kuhn}},
  \bibinfo{journal}{Rev. Mod. Phys.} \textbf{\bibinfo{volume}{74}},
  \bibinfo{pages}{895} (\bibinfo{year}{2002}).

\bibitem[{\citenamefont{Xua et~al.}(2004)\citenamefont{Xua, Yan, Ohtsuki,
  Fujimura, and Rabitz}}]{Xua04}
\bibinfo{author}{\bibfnamefont{R.}~\bibnamefont{Xua}},
  \bibinfo{author}{\bibfnamefont{Y.-J.} \bibnamefont{Yan}},
  \bibinfo{author}{\bibfnamefont{Y.}~\bibnamefont{Ohtsuki}},
  \bibinfo{author}{\bibfnamefont{Y.}~\bibnamefont{Fujimura}}, \bibnamefont{and}
  \bibinfo{author}{\bibfnamefont{H.}~\bibnamefont{Rabitz}},
  \bibinfo{journal}{J. Chem. Phys.} \textbf{\bibinfo{volume}{120}},
  \bibinfo{pages}{6600} (\bibinfo{year}{2004}).

\bibitem[{\citenamefont{P\"{o}tz et~al.}(2006)\citenamefont{P\"{o}tz,
  Goritschnig, and H.Jirari}}]{Poetz1}
\bibinfo{author}{\bibfnamefont{W.}~\bibnamefont{P\"{o}tz}},
  \bibinfo{author}{\bibfnamefont{A.}~\bibnamefont{Goritschnig}},
  \bibnamefont{and} \bibinfo{author}{\bibnamefont{H.Jirari}},
  \bibinfo{journal}{Proc. 5$^{th}$ MATHMOD, Feb. 8-10, ARGESIM Report no.~30,
  Vol. 1 p.~87, and Physical Modeling pp~9-1 to~9-10}  (\bibinfo{year}{2006}).

\bibitem[{\citenamefont{P\"otz}(2007)}]{WP}
\bibinfo{author}{\bibfnamefont{W.}~\bibnamefont{P\"otz}}, \bibinfo{journal}{J.
  Comp. Electronics} \textbf{\bibinfo{volume}{6}}, \bibinfo{pages}{171}
  (\bibinfo{year}{2007}).

\bibitem[{\citenamefont{Betts}(2001)}]{Betts}
\bibinfo{author}{\bibfnamefont{J.~T.} \bibnamefont{Betts}},
  \emph{\bibinfo{title}{Practical Methods for Optimal Control using Nonlinear
  Programming}} (\bibinfo{publisher}{SIAM, Philadelphia},
  \bibinfo{year}{2001}).

\bibitem[{\citenamefont{Shnirman et~al.}(1997)\citenamefont{Shnirman,
  Sch\"{o}n, and Hermon}}]{Shnir97}
\bibinfo{author}{\bibfnamefont{A.}~\bibnamefont{Shnirman}},
  \bibinfo{author}{\bibfnamefont{G.}~\bibnamefont{Sch\"{o}n}},
  \bibnamefont{and} \bibinfo{author}{\bibfnamefont{Z.}~\bibnamefont{Hermon}},
  \bibinfo{journal}{Phys. Rev. Lett.} \textbf{\bibinfo{volume}{79}},
  \bibinfo{pages}{2371} (\bibinfo{year}{1997}).

\bibitem[{\citenamefont{Cirac and Zoller}(1995)}]{Cir95}
\bibinfo{author}{\bibfnamefont{J.}~\bibnamefont{Cirac}} \bibnamefont{and}
  \bibinfo{author}{\bibfnamefont{P.}~\bibnamefont{Zoller}},
  \bibinfo{journal}{Phys. Rev. Letters} \textbf{\bibinfo{volume}{74}},
  \bibinfo{pages}{4091} (\bibinfo{year}{1995}).

\bibitem[{\citenamefont{Gershenfeld and Chuang}(1997)}]{Ger97}
\bibinfo{author}{\bibfnamefont{N.}~\bibnamefont{Gershenfeld}} \bibnamefont{and}
  \bibinfo{author}{\bibfnamefont{I.}~\bibnamefont{Chuang}},
  \bibinfo{journal}{Science} \textbf{\bibinfo{volume}{275}},
  \bibinfo{pages}{350} (\bibinfo{year}{1997}).

\bibitem[{\citenamefont{Knill et~al.}(2001)\citenamefont{Knill, Laflamme, and
  Milburn}}]{Knill01}
\bibinfo{author}{\bibfnamefont{E.}~\bibnamefont{Knill}},
  \bibinfo{author}{\bibfnamefont{R.}~\bibnamefont{Laflamme}}, \bibnamefont{and}
  \bibinfo{author}{\bibfnamefont{G.~J.} \bibnamefont{Milburn}},
  \bibinfo{journal}{Nature} \textbf{\bibinfo{volume}{409}}, \bibinfo{pages}{46}
  (\bibinfo{year}{2001}).

\bibitem[{\citenamefont{Elzerman et~al.}(2005)}]{Elz}
\bibinfo{author}{\bibfnamefont{J.}~\bibnamefont{Elzerman}}
  \bibnamefont{et~al.}, \emph{\bibinfo{title}{Semiconductor few-electron
  quantum dots as spin qubits}} (\bibinfo{publisher}{Springer},
  \bibinfo{year}{2005}).

\bibitem[{\citenamefont{Scully and Zubairy}(1997)}]{Scully}
\bibinfo{author}{\bibfnamefont{M.~O.} \bibnamefont{Scully}} \bibnamefont{and}
  \bibinfo{author}{\bibfnamefont{M.~S.} \bibnamefont{Zubairy}},
  \emph{\bibinfo{title}{Quantum Optics}} (\bibinfo{publisher}{Cambridge
  University Press}, \bibinfo{year}{1997}).

\bibitem[{\citenamefont{Taylor et~al.}(2006)\citenamefont{Taylor, Petta,
  Johnson, Yacoby, Marcus, and Lukin}}]{Tay06}
\bibinfo{author}{\bibfnamefont{J.}~\bibnamefont{Taylor}},
  \bibinfo{author}{\bibfnamefont{J.~R.} \bibnamefont{Petta}},
  \bibinfo{author}{\bibfnamefont{A.~C.} \bibnamefont{Johnson}},
  \bibinfo{author}{\bibfnamefont{A.}~\bibnamefont{Yacoby}},
  \bibinfo{author}{\bibfnamefont{C.~M.} \bibnamefont{Marcus}},
  \bibnamefont{and} \bibinfo{author}{\bibfnamefont{M.~D.} \bibnamefont{Lukin}},
  \bibinfo{journal}{arXiv:cond-mat/0602470 v1}  (\bibinfo{year}{2006}).

\bibitem[{\citenamefont{Golovach et~al.}(2004)\citenamefont{Golovach,
  Khaetskii, and Loss}}]{Gol04}
\bibinfo{author}{\bibfnamefont{V.}~\bibnamefont{Golovach}},
  \bibinfo{author}{\bibfnamefont{A.}~\bibnamefont{Khaetskii}},
  \bibnamefont{and} \bibinfo{author}{\bibfnamefont{D.}~\bibnamefont{Loss}},
  \bibinfo{journal}{Phys. Rev. Letters} \textbf{\bibinfo{volume}{93}},
  \bibinfo{pages}{016601} (\bibinfo{year}{2004}).

\bibitem[{\citenamefont{Mozyrsky et~al.}(2002)\citenamefont{Mozyrsky, Kogan,
  Gorshkov, and Berman}}]{Moz02}
\bibinfo{author}{\bibfnamefont{D.}~\bibnamefont{Mozyrsky}},
  \bibinfo{author}{\bibfnamefont{S.}~\bibnamefont{Kogan}},
  \bibinfo{author}{\bibfnamefont{V.~N.} \bibnamefont{Gorshkov}},
  \bibnamefont{and} \bibinfo{author}{\bibfnamefont{G.~P.}
  \bibnamefont{Berman}}, \bibinfo{journal}{Phys. Rev. B}
  \textbf{\bibinfo{volume}{65}}, \bibinfo{pages}{245213}
  (\bibinfo{year}{2002}).

\bibitem[{\citenamefont{Yu and Eberly}(2002)}]{Yu02}
\bibinfo{author}{\bibfnamefont{T.}~\bibnamefont{Yu}} \bibnamefont{and}
  \bibinfo{author}{\bibfnamefont{J.}~\bibnamefont{Eberly}},
  \bibinfo{journal}{Phys. Rev. B} \textbf{\bibinfo{volume}{66}},
  \bibinfo{pages}{193306} (\bibinfo{year}{2002}).

\bibitem[{\citenamefont{Hu and Sarma}(2006)}]{Hu06}
\bibinfo{author}{\bibfnamefont{X.}~\bibnamefont{Hu}} \bibnamefont{and}
  \bibinfo{author}{\bibfnamefont{S.~D.} \bibnamefont{Sarma}},
  \bibinfo{journal}{Phys. Rev. Letters} \textbf{\bibinfo{volume}{96}},
  \bibinfo{pages}{100501} (\bibinfo{year}{2006}).

\bibitem[{DEA()}]{DEA}
\urlprefix\url{http://www.icsi.berkeley.edu/~storn/devcpp.zip}.

\bibitem[{\citenamefont{Slichter}(1992)}]{slichter}
\bibinfo{author}{\bibfnamefont{C.~L.} \bibnamefont{Slichter}},
  \emph{\bibinfo{title}{Principles of Magnetic Resonance}}
  (\bibinfo{publisher}{Springer, Berlin}, \bibinfo{year}{1992}).

\bibitem[{\citenamefont{Brumer and Shapiro}(2003)}]{Brumer}
\bibinfo{author}{\bibfnamefont{P.~W.} \bibnamefont{Brumer}} \bibnamefont{and}
  \bibinfo{author}{\bibfnamefont{M.}~\bibnamefont{Shapiro}},
  \emph{\bibinfo{title}{Principles of the Quantum Control of Molecular
  Processes}} (\bibinfo{publisher}{Wiley-VCH, Berlin}, \bibinfo{year}{2003}).

\bibitem[{\citenamefont{Haroche and Raimond}(2006)}]{Haroche}
\bibinfo{author}{\bibfnamefont{S.}~\bibnamefont{Haroche}} \bibnamefont{and}
  \bibinfo{author}{\bibfnamefont{J.-M.} \bibnamefont{Raimond}},
  \emph{\bibinfo{title}{Exploring the Quantum - Atoms, Cavities, and Photons}}
  (\bibinfo{publisher}{Oxford University Press, New York},
  \bibinfo{year}{2006}).

\bibitem[{\citenamefont{Bushev et~al.}(2006)\citenamefont{Bushev, Rotter,
  Wilson, Dubin, Becher, Eschner, Blatt, Steixner, Rabl, and Zoller}}]{Blatt}
\bibinfo{author}{\bibfnamefont{P.}~\bibnamefont{Bushev}},
  \bibinfo{author}{\bibfnamefont{D.}~\bibnamefont{Rotter}},
  \bibinfo{author}{\bibfnamefont{A.}~\bibnamefont{Wilson}},
  \bibinfo{author}{\bibfnamefont{F.}~\bibnamefont{Dubin}},
  \bibinfo{author}{\bibfnamefont{C.}~\bibnamefont{Becher}},
  \bibinfo{author}{\bibfnamefont{J.}~\bibnamefont{Eschner}},
  \bibinfo{author}{\bibfnamefont{R.}~\bibnamefont{Blatt}},
  \bibinfo{author}{\bibfnamefont{V.}~\bibnamefont{Steixner}},
  \bibinfo{author}{\bibfnamefont{P.}~\bibnamefont{Rabl}}, \bibnamefont{and}
  \bibinfo{author}{\bibfnamefont{P.}~\bibnamefont{Zoller}},
  \bibinfo{journal}{Phys. Rev. Lett.} \textbf{\bibinfo{volume}{96}},
  \bibinfo{pages}{043003} (\bibinfo{year}{2006}).

\bibitem[{\citenamefont{Chu}(2002)}]{Chu}
\bibinfo{author}{\bibfnamefont{S.}~\bibnamefont{Chu}},
  \bibinfo{journal}{Nature} \textbf{\bibinfo{volume}{416}},
  \bibinfo{pages}{206} (\bibinfo{year}{2002}).

\bibitem[{\citenamefont{Vandersypen and Chuang}(2005)}]{NMR}
\bibinfo{author}{\bibfnamefont{L.~M.~K.} \bibnamefont{Vandersypen}}
  \bibnamefont{and} \bibinfo{author}{\bibfnamefont{I.~L.}
  \bibnamefont{Chuang}}, \bibinfo{journal}{Rev. Mod. Phys. {\bf 76} 1037 -
  1069}  (\bibinfo{year}{2005}).

\bibitem[{\citenamefont{Baltuska et~al.}(2003)\citenamefont{Baltuska, Udem,
  Uiberacker, Hentschel, Goulielmakis, Holzwarth, Yakovlev, Scrinzi, H\"ansch,
  and Krausz}}]{Krausz}
\bibinfo{author}{\bibfnamefont{A.}~\bibnamefont{Baltuska}},
  \bibinfo{author}{\bibfnamefont{T.}~\bibnamefont{Udem}},
  \bibinfo{author}{\bibfnamefont{M.}~\bibnamefont{Uiberacker}},
  \bibinfo{author}{\bibfnamefont{M.}~\bibnamefont{Hentschel}},
  \bibinfo{author}{\bibfnamefont{E.}~\bibnamefont{Goulielmakis}},
  \bibinfo{author}{\bibfnamefont{C.~G.~R.} \bibnamefont{Holzwarth}},
  \bibinfo{author}{\bibfnamefont{V.}~\bibnamefont{Yakovlev}},
  \bibinfo{author}{\bibfnamefont{A.}~\bibnamefont{Scrinzi}},
  \bibinfo{author}{\bibfnamefont{T.}~\bibnamefont{H\"ansch}}, \bibnamefont{and}
  \bibinfo{author}{\bibfnamefont{F.}~\bibnamefont{Krausz}},
  \bibinfo{journal}{Nature} \textbf{\bibinfo{volume}{421}},
  \bibinfo{pages}{611} (\bibinfo{year}{2003}).

\bibitem[{\citenamefont{Riebe et~al.}(2006)\citenamefont{Riebe, Kim, Schindler,
  Monz, Schmidt, K\"orber, H\"ansel, H\"affner, Roos, and Blatt}}]{Blatt2}
\bibinfo{author}{\bibfnamefont{M.}~\bibnamefont{Riebe}},
  \bibinfo{author}{\bibfnamefont{K.}~\bibnamefont{Kim}},
  \bibinfo{author}{\bibfnamefont{P.}~\bibnamefont{Schindler}},
  \bibinfo{author}{\bibfnamefont{T.}~\bibnamefont{Monz}},
  \bibinfo{author}{\bibfnamefont{P.~O.} \bibnamefont{Schmidt}},
  \bibinfo{author}{\bibfnamefont{T.~K.} \bibnamefont{K\"orber}},
  \bibinfo{author}{\bibfnamefont{W.}~\bibnamefont{H\"ansel}},
  \bibinfo{author}{\bibfnamefont{H.}~\bibnamefont{H\"affner}},
  \bibinfo{author}{\bibfnamefont{C.~F.} \bibnamefont{Roos}}, \bibnamefont{and}
  \bibinfo{author}{\bibfnamefont{R.}~\bibnamefont{Blatt}},
  \bibinfo{journal}{Phys. Rev. Lett.} \textbf{\bibinfo{volume}{97}},
  \bibinfo{pages}{220407} (\bibinfo{year}{2006}).

\bibitem[{\citenamefont{Nowack et~al.}(2007)\citenamefont{Nowack, Koppens,
  Nazarov, and Vandersypen}}]{Nowack}
\bibinfo{author}{\bibfnamefont{K.~C.} \bibnamefont{Nowack}},
  \bibinfo{author}{\bibfnamefont{F.~H.~L.} \bibnamefont{Koppens}},
  \bibinfo{author}{\bibfnamefont{Y.~V.} \bibnamefont{Nazarov}},
  \bibnamefont{and} \bibinfo{author}{\bibfnamefont{L.~M.~K.}
  \bibnamefont{Vandersypen}}, \bibinfo{journal}{Science}
  \textbf{\bibinfo{volume}{318}}, \bibinfo{pages}{5855} (\bibinfo{year}{2007}).

\bibitem[{\citenamefont{Wu et~al.}(2006)\citenamefont{Wu, Li, Duan, Steel, and
  Gammon}}]{Yanwen}
\bibinfo{author}{\bibfnamefont{Y.}~\bibnamefont{Wu}},
  \bibinfo{author}{\bibfnamefont{X.}~\bibnamefont{Li}},
  \bibinfo{author}{\bibfnamefont{L.}~\bibnamefont{Duan}},
  \bibinfo{author}{\bibfnamefont{D.}~\bibnamefont{Steel}}, \bibnamefont{and}
  \bibinfo{author}{\bibfnamefont{D.}~\bibnamefont{Gammon}},
  \bibinfo{journal}{Phys. Rev. Lett.} \textbf{\bibinfo{volume}{96}},
  \bibinfo{pages}{087402} (\bibinfo{year}{2006}).

\bibitem[{\citenamefont{Bartana et~al.}(2001)\citenamefont{Bartana, Kosloff,
  and Tannor}}]{Bartana}
\bibinfo{author}{\bibfnamefont{A.}~\bibnamefont{Bartana}},
  \bibinfo{author}{\bibfnamefont{R.}~\bibnamefont{Kosloff}}, \bibnamefont{and}
  \bibinfo{author}{\bibfnamefont{D.~J.} \bibnamefont{Tannor}},
  \bibinfo{journal}{Chem. Phys. {\bf 267}, 195}  (\bibinfo{year}{2001}).

\bibitem[{\citenamefont{Lindblad}(1976)}]{Lindblad76}
\bibinfo{author}{\bibfnamefont{G.}~\bibnamefont{Lindblad}},
  \bibinfo{journal}{Communications in Mathematical Physics}
  \textbf{\bibinfo{volume}{48}}, \bibinfo{pages}{119} (\bibinfo{year}{1976}).

\bibitem[{\citenamefont{Rabitz et~al.}(2004)\citenamefont{Rabitz, Hsieh, and
  Rosenthal}}]{Rabitz04}
\bibinfo{author}{\bibfnamefont{H.~A.} \bibnamefont{Rabitz}},
  \bibinfo{author}{\bibfnamefont{M.~M.} \bibnamefont{Hsieh}}, \bibnamefont{and}
  \bibinfo{author}{\bibfnamefont{C.~M.} \bibnamefont{Rosenthal}},
  \bibinfo{journal}{Science} \textbf{\bibinfo{volume}{303}},
  \bibinfo{pages}{1998} (\bibinfo{year}{2004}).

\bibitem[{\citenamefont{Astafiev et~al.}(2004)\citenamefont{Astafiev, Pashkin,
  Nakamura, Yamamoto, and Tsai}}]{Asta04}
\bibinfo{author}{\bibfnamefont{O.}~\bibnamefont{Astafiev}},
  \bibinfo{author}{\bibfnamefont{Y.}~\bibnamefont{Pashkin}},
  \bibinfo{author}{\bibfnamefont{Y.}~\bibnamefont{Nakamura}},
  \bibinfo{author}{\bibfnamefont{T.}~\bibnamefont{Yamamoto}}, \bibnamefont{and}
  \bibinfo{author}{\bibfnamefont{J.~S.} \bibnamefont{Tsai}},
  \bibinfo{journal}{Phys. Rev. Lett.} \textbf{\bibinfo{volume}{93}},
  \bibinfo{pages}{267007} (\bibinfo{year}{2004}).

\bibitem[{\citenamefont{Schriefl et~al.}(2006)\citenamefont{Schriefl, Makhlin,
  Shnirman, and Sch\"{o}n}}]{Schriefl06}
\bibinfo{author}{\bibfnamefont{J.}~\bibnamefont{Schriefl}},
  \bibinfo{author}{\bibfnamefont{Y.}~\bibnamefont{Makhlin}},
  \bibinfo{author}{\bibfnamefont{A.}~\bibnamefont{Shnirman}}, \bibnamefont{and}
  \bibinfo{author}{\bibfnamefont{G.}~\bibnamefont{Sch\"{o}n}},
  \bibinfo{journal}{New J. Phys.} \textbf{\bibinfo{volume}{8}},
  \bibinfo{pages}{1} (\bibinfo{year}{2006}).

\bibitem[{\citenamefont{Thorwart and H\"anggi}(2002)}]{Thorwart}
\bibinfo{author}{\bibfnamefont{M.}~\bibnamefont{Thorwart}} \bibnamefont{and}
  \bibinfo{author}{\bibfnamefont{P.}~\bibnamefont{H\"anggi}},
  \bibinfo{journal}{Phys. Rev. A} \textbf{\bibinfo{volume}{65}},
  \bibinfo{pages}{012309} (\bibinfo{year}{2002}).

\bibitem[{\citenamefont{Storcz and Wilhelm}(2003)}]{Storcz}
\bibinfo{author}{\bibfnamefont{M.~J.} \bibnamefont{Storcz}} \bibnamefont{and}
  \bibinfo{author}{\bibfnamefont{F.~K.} \bibnamefont{Wilhelm}},
  \bibinfo{journal}{Phys. Rev. A} \textbf{\bibinfo{volume}{67}},
  \bibinfo{pages}{042319} (\bibinfo{year}{2003}).

\bibitem[{\citenamefont{Thorwart et~al.}(2005)\citenamefont{Thorwart, Eckel,
  and Mucciolo}}]{Mucciolo}
\bibinfo{author}{\bibfnamefont{M.}~\bibnamefont{Thorwart}},
  \bibinfo{author}{\bibfnamefont{J.}~\bibnamefont{Eckel}}, \bibnamefont{and}
  \bibinfo{author}{\bibfnamefont{E.~R.} \bibnamefont{Mucciolo}},
  \bibinfo{journal}{Phys. Rev. B} \textbf{\bibinfo{volume}{72}},
  \bibinfo{pages}{235320} (\bibinfo{year}{2005}).

\bibitem[{\citenamefont{Pasini et~al.}(2008)\citenamefont{Pasini, Fischer,
  Karbach, and Uhrig}}]{Uhrig}
\bibinfo{author}{\bibfnamefont{S.}~\bibnamefont{Pasini}},
  \bibinfo{author}{\bibfnamefont{T.}~\bibnamefont{Fischer}},
  \bibinfo{author}{\bibfnamefont{P.}~\bibnamefont{Karbach}}, \bibnamefont{and}
  \bibinfo{author}{\bibfnamefont{G.~S.} \bibnamefont{Uhrig}},
  \bibinfo{journal}{Phys. Rev. A {\bf 77}, 032315}  (\bibinfo{year}{2008}).

\bibitem[{\citenamefont{Niskanen et~al.}(2003)\citenamefont{Niskanen,
  Vartiainen, and Salomaa}}]{Niskanen2}
\bibinfo{author}{\bibfnamefont{A.~O.} \bibnamefont{Niskanen}},
  \bibinfo{author}{\bibfnamefont{J.~J.} \bibnamefont{Vartiainen}},
  \bibnamefont{and} \bibinfo{author}{\bibfnamefont{M.~M.}
  \bibnamefont{Salomaa}}, \bibinfo{journal}{Phys. Rev. Lett.}
  \textbf{\bibinfo{volume}{90}}, \bibinfo{pages}{19} (\bibinfo{year}{2003}).

\bibitem[{\citenamefont{Havel}(2003)}]{Havel03}
\bibinfo{author}{\bibfnamefont{T.~F.} \bibnamefont{Havel}},
  \bibinfo{journal}{J. Math. Phys.} \textbf{\bibinfo{volume}{44}},
  \bibinfo{pages}{534} (\bibinfo{year}{2003}).

\bibitem[{\citenamefont{Altepeter et~al.}(2003)\citenamefont{Altepeter,
  Branning, Jeffrey, C.Wei, Kwiat, Thew, O'Brien, Nielsen, and
  A.G.White}}]{Alt03}
\bibinfo{author}{\bibfnamefont{J.~B.} \bibnamefont{Altepeter}},
  \bibinfo{author}{\bibfnamefont{D.}~\bibnamefont{Branning}},
  \bibinfo{author}{\bibfnamefont{E.}~\bibnamefont{Jeffrey}},
  \bibinfo{author}{\bibfnamefont{T.}~\bibnamefont{C.Wei}},
  \bibinfo{author}{\bibfnamefont{P.}~\bibnamefont{Kwiat}},
  \bibinfo{author}{\bibfnamefont{R.}~\bibnamefont{Thew}},
  \bibinfo{author}{\bibfnamefont{J.~L.} \bibnamefont{O'Brien}},
  \bibinfo{author}{\bibfnamefont{M.~A.} \bibnamefont{Nielsen}},
  \bibnamefont{and} \bibinfo{author}{\bibnamefont{A.G.White}},
  \bibinfo{journal}{Phys. Rev. Lett.} \textbf{\bibinfo{volume}{90}},
  \bibinfo{pages}{192601} (\bibinfo{year}{2003}).

\bibitem[{\citenamefont{Nielsen}(2002)}]{Niel02b}
\bibinfo{author}{\bibfnamefont{M.~A.} \bibnamefont{Nielsen}},
  \bibinfo{journal}{Phys. Lett. A} \textbf{\bibinfo{volume}{303}},
  \bibinfo{pages}{249} (\bibinfo{year}{2002}).

\end{thebibliography}
\end{document}